\documentclass[twocolumn,prb,floatfix,nofootinbib,superscriptaddress]{revtex4-2}

\usepackage[utf8]{inputenc}
\usepackage[T1]{fontenc} 
\usepackage[english]{babel}

\usepackage{amsmath}
\usepackage{multirow}
\usepackage{appendix}
\usepackage{float}

\usepackage{amsthm}
\usepackage{amssymb}
\usepackage[normalem]{ulem}
\usepackage{listings}
\usepackage[dvipsnames]{xcolor}
\usepackage{cancel}
\definecolor{codegreen}{rgb}{0,0.6,0}
\definecolor{codegray}{rgb}{0.5,0.5,0.5}
\definecolor{codepurple}{rgb}{0.58,0,0.82}
\definecolor{backcolour}{rgb}{0.95,0.95,0.92}

\lstdefinestyle{mystyle}{
    backgroundcolor=\color{backcolour},   
    commentstyle=\color{codegreen},
    keywordstyle=\color{magenta},
    numberstyle=\tiny\color{codegray},
    stringstyle=\color{codepurple},
    basicstyle=\ttfamily\footnotesize,
    breakatwhitespace=false,         
    breaklines=true,                 
    captionpos=b,                    
    keepspaces=true,                 
    numbersep=5pt,                  
    showspaces=false,                
    showstringspaces=false,
    showtabs=false,                  
    tabsize=1
}
\definecolor{dark-green}{RGB}{0, 128, 0}

\usepackage{enumitem}

\usepackage{dsfont}

\usepackage{graphicx}
\usepackage{color}

\usepackage{comment}
\usepackage{ stmaryrd }
\usepackage{physics}
\usepackage{tabularx}

\usepackage{hyperref}
\hypersetup{
    colorlinks,
    citecolor=black,
    filecolor=black,
    linkcolor=black,
    urlcolor=blue
}
\begin{document}

\title{Finite-state automata for exact matrix product operators of tight-binding Hamiltonians: fractals, quasicrystals, trees and hyperbolic lattices}
%Finite-state automata for exact matrix product operators of tight-binding Hamiltonians on fractals, quasicrystals, trees and hyperbolic lattices
%OR
%Generalized tensor networks for tight binding Hamiltonians from finite-state automata: fractals, quasiperiodic and hyperbolic lattices
%OR
%Generalized lattice tensor networks from finite-state automata: fractals, quasiperiodic and hyperbolic models

\author{Marta Brzezińska}
\affiliation{Quobly, 38000 Grenoble, France}
\author{Jeanne Colbois}
\affiliation{Univ. Grenoble Alpes, CNRS, Institut N\'eel, 38000 Grenoble, France}
\author{Lo\"ic Herviou}
\affiliation{Univ. Grenoble Alpes, CNRS, LPMMC, 38000 Grenoble, France}

\begin{abstract}
Inspired by the recent progress in the simulation of tight-binding Hamiltonians on large lattices using tensor networks, we introduce a systematic matrix product operator (MPO) construction for single-particle Hamiltonians on recursively structured lattices.
Taking advantage of this recursive structure, we encode the lattice geometry in a finite-state automaton and, adapting ideas from Abelian-symmetric tensor networks, obtain an exact and analytical MPO representation of the Hamiltonian, where the number of tensors grows logarithmically with the system size and the bond dimension is set by the number of automaton states.
Several examples demonstrate the generality of the framework: regular lattices, fractals, Cayley trees, hyperbolic lattices, and one- and two-dimensional Fibonacci quasicrystals.
Combined with the kernel polynomial method, these MPO representations enable large-scale calculations of spectral properties without explicitly constructing the full Hamiltonian.
This provides a unified route to the exploration of electronic properties across a broad class of lattices at exponentially large system sizes.
\end{abstract}

\maketitle

\section{Introduction}

Tight-binding Hamiltonians play an essential role as simple and versatile models of quantum matter. 
By offering a minimal setting for studying how lattice geometry, orbital structure, and hopping processes shape electronic properties, they form the cornerstone of our understanding of band structure, topology, and localization. 
Beyond regular lattices, tight-binding models defined on complex lattice geometries can exhibit remarkably rich physical behavior.
Among these, fractals, quasiperiodic lattices, or tree-like structures have historically played an important role in theoretical physics, often used as analytically tractable settings. 
Fractal lattices, characterized by (non-integer) Hausdorff dimension,
exhibit distinct electronic and topological properties~\cite{, PhysRevB.107.115424,  PhysRevB.100.155135, PhysRevB.105.L201301, PhysRevB.110.L241302, PhysRevB.98.205116, lage2026chiralsymmetrybreakingdegeneracykoch}, anomalous transport~\cite{PhysRevResearch.2.013044, PhysRevB.101.045413, PhysRevResearch.2.033063,Rojo-Francas2024}, and self-similar entanglement patterns~\cite{PhysRevResearch.6.043145}. 
Quasiperiodic chains, such as the Fibonacci chain, have served as paradigmatic models for studying critical states and localization phenomena~\cite{PhysRevB.35.1020, RevModPhys.93.045001, SciPostPhys.6.4.050}.
Cayley trees, finite rooted regular trees, are classic geometries of statistical mechanics on which many models are exactly tractable~\cite{OSTILLI20123417}. Together with Bethe lattices~\cite{zirnbauer_localization_1986-1,mirlin_distribution_1994-1, parisi_anderson_2019}, Cayley trees~\cite{monthus_anderson_2011,tikhonov_fractality_2016,sonner_multifractality_2017-1} play a particularly important role in the study of localization. 
Hyperbolic lattices are  regular tilings of negatively curved space which share with trees an exponential volume growth and an extensive boundary, and the vertices of a rooted hyperbolic lattice can be labeled by a tree with depth-dependent branching. They have been the focus of a renewed interest as platforms for exploring spectral properties in non-Euclidean geometries~\cite{doi:10.1126/sciadv.abe9170, PhysRevLett.129.246402, t8gg-xqls, PhysRevLett.125.053901, PhysRevB.106.155146}. 
Besides this theoretical motivation, the rise of quantum simulators has reignited interest in single- and many-body Hamiltonians on non-regular lattices, by giving access to their exotic quantum transport, topological and localization properties.
Advances in photonics enabled experimental evidence of anomalous quantum transport controlled by the fractal dimension and the different spectral properties of the Sierpiński fractals~\cite{Xu2021, doi:10.1126/science.abm2842}. 
Fractal geometries can also be experimentally realized in other devices from artificial electronic systems~\cite{Kempkes2019} to Rydberg arrays~\cite{verstraten2025controlsinglespinflipsrydberg}. 
Quasicrystals are, of course, a well-known solid state discovery~\cite{Shechtman1984}.
Ultracold atoms in optical lattices famously implement one- and two-dimensional quasicrystals, which can be used to capture delocalization-localization  transitions~\cite{Roati2008,schreiber_observation_2015,Bordia2017,yu_observing_2024}, both in interacting and non-interacting tight-binding models.
Photonics also enables their study, from topological properties in one dimension~\cite{verbin_observation_2013} to light localization in the two-dimensional Fibonacci lattice~\cite{boguslawski_light_2016}. 
Finally, hyperbolic lattices can be studied in circuit QED~\cite{kollar_hyperbolic_2019}, as well as in electrical circuits~\cite{Lenggenhager2022, Zhang2022,Chen2023, Zhang2023}, microwave networks~\cite{Chen2024}, and photonics~\cite{Huang2024, 10.1021/acsphotonics.4c01184}.
All these experiments motivate the development of numerical tools that can efficiently study tight-binding models on such geometries at large scale, exactly~\cite{Antao2026} or with well-chosen approximations~\cite{verstraten2026simgraphuniversalguidesymmetric}.
Among numerical methods, tensor networks have gradually been established as a powerful tool for the simulation of quantum many-body systems, ranging from ground states of frustrated magnets~\cite{PhysRevLett.104.187203, PhysRevB.79.195119, PhysRevLett.109.067201} or Hubbard models~\cite{PhysRevB.93.045116, r4q9-4yvj, PhysRevB.107.165112}, to finite-temperature simulation of fermionic systems~\cite{PhysRevB.106.195105}, and able to compete with near-term quantum processors~\cite{doi:10.1126/sciadv.adk4321, PhysRevResearch.6.013326}. 
The key to their success in capturing relevant states in exponentially large Hilbert spaces is that they provide an extremely efficient compression of many-body states whose entanglement is limited~\cite{RevModPhys.93.045003, Schoellwock2011}. 
The central observation is that such states can be efficiently approximated by a network of local tensors, one per site, connected through a limited virtual space, resulting in an exponential reduction in the memory and computational cost of storing and performing linear algebra operations on ansatz wavefunctions.
In one dimension, another key to the success of tensor networks in quantum many-body physics is that a Hamiltonian can be efficiently represented as a matrix product operator (MPO) through finite-state automata. 
Each term of a sum of local operators corresponds to a path through an automaton whose internal states label the virtual bond, so that the MPO bond dimension is simply the number of automaton states and is independent of system size~\cite{Crosswhite2008,McCulloch2007,Schoellwock2011}.
Over the last few years, it has become increasingly clear that spatial entanglement is not the only underlying structure in condensed matter physics which tensor networks can take advantage of~\cite{PhysRevLett.102.240603,PhysRevA.91.032306}. In particular, one-dimensional tensor networks known as tensor trains~\cite{Oseledets2011} or matrix product states (MPS)~\cite{vidal_efficient_2003,Schoellwock2011} also support the compression of multivariate functions and differential/integral operators~\cite{ReviewGarcia2024}. 
A particularly powerful variant, the quantics tensor train (QTT) representation of functions, uses one tensor per scale of an exponentially dense grid discretization of the space to accurately capture functions, in particular those with a strong separation of scales~\cite{khoromskij_odlog_2011-1,SciPostPhysLectNotes.133}.
Combined with the tensor cross interpolation (TCI) algorithm, which learns a tensor train from a small number of function evaluations~\cite{oseledets_tt-cross_2010,  Ritter2024, Núñez2025}, this enables the efficient evaluation of high-dimensional integrals, from Feynman diagrams~\cite{Nunez2022, Ishida2025} to Brillouin-zone sums~\cite{Ritter2024}. These tools also support the solution of partial differential equations, e.g., in fluid dynamics~\cite{Holscher2025, pisoni2026compressionsimulationsynthesisturbulent, Pham2026} or quantum molecular dynamics~\cite{gavrilyuk_quantized-tt-cayley_2011}, the efficient computation of Fourier transforms~\cite{dolgov_superfast_2012, SAVOSTYANOV20123215}, solving self-consistent equations for multivariate functions~\cite{Ritter2024, PhysRevResearch.7.023087, PhysRevX.13.021015, jx7h-lsqk, SciPostPhys.19.2.038}, and, most relevant here, the solution of single-particle lattice Hamiltonians with billions of sites~\cite{Antao2026, antao2026tensornetworksolversultralarge, Yitao2025}.
Several results in the literature suggest that an underlying separation of scales~\cite{qlrs-6f8t} or a sufficiently deep understanding of the lattice structure can guide the construction of an exact representation of single-particle Hamiltonians in the language of QTTs.
First, early in the origin of quantics tensor trains it was observed that the Laplace operator and multi-level Toeplitz matrices are easily expressed in this language~\cite{oseledets_approximation_2010,doi:10.1137/100820479, kazeev_multilevel_2013}. 
Second, Ref.~\cite{antao2026tensornetworksolversultralarge} builds tight-binding Hamiltonians on hypercubic grids as the sum of two MPOs: the first, corresponding to the on-site potential, is learned using quantics TCI, and the second, corresponding to the hopping term, is implemented using a shift~\cite{oseledets_approximation_2010,doi:10.1137/100820479,Antao2026} or the magic tensor~\cite{SciPostPhysLectNotes.133}. Other lattices embedded in the hypercubic one as well as modulated couplings are obtained by choosing shift distances and diagonal modulation MPO.
Finally, the indicator function of fractal lattices admits a very natural representation with QTT~\cite{gelss2018tensorgeneratedfractalsusingtensor}. This representation is not optimal, as it encodes zeros, but existing work without QTT suggests how to exploit the sparsity of fractals to obtain an efficient representation~\cite{Villalba-Diez2026}.

Inspired by these results, in the present work, we highlight the existence of a unifying framework for the representation of tight-binding Hamiltonians on a large family of lattices as low-rank MPOs.
This construction relies on finite-state automata with outputs: the lattice geometry is simply encoded in the automaton's transition function and the hopping terms in its output function. 
The bond dimension is determined by the number of automaton states, which is finite for a broad class of recursively structured lattices, including regular Bravais lattices, fractals, quasiperiodic chains, Cayley trees or hyperbolic lattices. 
The resulting MPO is, for the cases studied in this paper, obtained entirely analytically.
This enables the efficient evaluation of spectral properties and transport coefficients using the kernel polynomial method (KPM) and related techniques~\cite{RevModPhys.78.275,antao2026tensornetworksolversultralarge}.
We demonstrate this approach on a range of examples, from the square and triangular lattices to the Sierpiński carpet, the Fibonacci chain, and hyperbolic tilings, and we discuss its generalization to arbitrary substitution rules and finite-state geometries. \\

The rest of the paper is organized as follows. Section~\ref{sec:TNtightbinding} provides a reminder of tensor networks as weighted automata before introducing the general framework. 
In Section~\ref{sec:regularandfractal}, we represent tight binding Hamiltonians on both regular and fractal lattices by using a distance automaton. 
Section~\ref{sec:FibAuto} discusses the substitution-based automata applicable to quasiperiodic models on the 1D and 2D Fibonacci lattices.  
Finally, generalized Cayley trees and hyperbolic lattices are discussed in Section~\ref{sec:CayleyHyperbolic}.\\

Upon completing our work, we became aware of a preprint introducing an equivalent construction focusing on a family of one-dimensional quasicrystals with finite-state automata~\cite{kolehmainen2026onedimensionalquasicrystalstensornetworkfinitestate}.

\section{Tensor network representations of tight-binding Hamiltonians}
\label{sec:TNtightbinding}
\subsection{General formalism}\label{sec:GeneralFormalism}
In this section, we introduce a general framework to construct, analytically or algorithmically, non-interacting tight-binding Hamiltonians on arbitrary lattices in any dimension. 
We consider fermionic models\footnote{The same formalism applies to non-interacting bosonic models without Bogoliubov pairing terms.} on a lattice \(\Gamma\) with the Hamiltonian
\begin{equation}
    H = \sum_{\vec{r}, \vec{r}' \in \Gamma} \sum_{\sigma, \sigma'} c^\dagger_{\vec{r}, \sigma} \, h_{\vec{r}, \vec{r}', \sigma, \sigma'} \, c_{\vec{r}', \sigma'}, \label{eq:GenericHam}
\end{equation}
where \(\sigma\) and \(\sigma'\) label the different local degrees of freedom. Our goal is to find an efficient tensor network representation of \(h_{\vec{r}, \vec{r}', \sigma, \sigma'}\). For ease of notation, we consider a single orbital per site in the rest of the paper. The generalization to multi-orbital systems requires adding a single index labeling the orbital at one extremity of the tensor network representation, and in fact, it falls entirely within our framework. 
For fermionic Bogoliubov-de~Gennes Hamiltonians, the generalization will be similar and require doubling the number of indices to encode the pairing terms.\\

\paragraph*{Tensor network representations.} We start by reminding the usual formulation of matrix product states in quantum many-body physics. Tensor networks form a general representation of quantum states and operators in a product space. Let us consider the \(d^n\)-dimensional Hilbert space of \(n\) \(d\)-dimensional sites. The matrix product state (MPS) representation of a quantum wavefunction is a set of matrices \(A^{\sigma_j, j}\) of dimension \(\chi_{j-1} \times \chi_j\) for each site \(j\) and physical index \(\sigma_j\), such that
\begin{multline}
    \sum_{\sigma_1, \dots, \sigma_n} c_{\sigma_1, \dots, \sigma_n} \ket{\sigma_1, \dots, \sigma_n} 
    = \\
    \sum_{\sigma_1, \dots, \sigma_n} \left( \vec{l} \cdot A^{\sigma_1, 1} A^{\sigma_2, 2} \cdots A^{\sigma_n, n} \cdot \vec{r} \right) \ket{\sigma_1, \dots, \sigma_n},
\end{multline}
where \(\vec{l}\) is a row vector of dimension \(\chi_0\) and \(\vec{r}\) is a column vector of dimension \(\chi_n\). These boundary vectors can be absorbed into the definitions of \(A^{\sigma_1, 1}\) and \(A^{\sigma_n, n}\). 
The matrix product operator (MPO) representation of an operator is
\begin{equation}
    H_{\sigma_1, \dots, \sigma_n}^{\sigma_1', \dots, \sigma_n'} 
    = \vec{l} \cdot W^{\sigma_1, \sigma_1', 1} W^{\sigma_2, \sigma_2', 2} \cdots W^{\sigma_n, \sigma_n', n} \cdot \vec{r}.
\end{equation}
In the following, we denote by \(W_{v_{j-1},  \sigma_j, \sigma_j', v_j}\) the matrix elements of \(W^{\sigma_j, \sigma_j', j}\).
A tensor network decomposition is an efficient representation of states or operators if the bond dimensions \(\chi_j\) do not scale exponentially with the number of sites.\\

\paragraph*{Tensor networks as weighted automata.}
The bond indices \(v_j\) of an MPS or MPO can be interpreted as the states of a weighted finite automaton with output (WFAO)~\cite{BookAutomaton}.
A WFAO is a tuple \(\mathcal{A} = (V, X, \Delta, \delta, \lambda, v_0, F, \mathbb{K})\), where \(V\) is a finite set of states, \(X\) is a finite input alphabet, \(\Delta\) is a finite output alphabet, \(\delta : V \times X \times V \to \mathbb{K}\) is a weighted transition function into a semiring \(\mathbb{K}\) (e.g., \(\mathbb{C}\), \(\mathbb{R}_{\ge 0}\), or \(\mathbb{B} = \{0,1\}\)), \(\lambda : V \to \Delta\) is the output function, \(v_0 \in V\) is the initial state, and \(F \subseteq V\) is the set of accepting states. 
Given an input word \(x = \underline{x_1x_2 \dots x_n} \in X^n\), the automaton computes a formal power series
\begin{equation}
    f(x) = \sum_{\substack{v_1, \dots, v_{n-1} \in V \\ v_n \in F}} \delta(v_0, x_1, v_1)  \cdots \delta(v_{n-1}, x_n, v_n) \lambda(v_n),
\end{equation}
summing over all accepting paths weighted by the transition weights. 
The output of the automaton on the word \(x\) is then a weighted combination of the \(\lambda(v_n)\).
In the tensor network language, the bond states \(v_j\) map to the automaton states, the physical indices \(\sigma_j, \sigma_j'\) are the input symbols, the local tensors \(W_{v_{j-1}, \sigma_j, \sigma_j',v_j}^{}\) encode the transition weights, and the boundary vectors \(\vec{l}\) and \(\vec{r}\) fix the initial and accepting states~\cite{Crosswhite2008}. 
The contraction of the bonds is precisely the sum over accepting paths, and the semiring \(\mathbb{K}\) determines the arithmetic. 
Iteration-dependent transition functions ($\delta \equiv \delta_j$) can be incorporated at no additional cost.\\

\paragraph*{Requirements on the tight-binding Hamiltonian.}
To represent the tight-binding Hamiltonian in Eq.~\eqref{eq:GenericHam} as a tensor network, our general formalism requires three main ingredients. 
(i) We need an embedding of \(\Gamma\) into a product space \(\mathcal{H}\) of dimension \(d_1 \times d_2 \times \dots \times d_n\), such that each site is associated with a unique word \(\underline{x_1 x_2 \dots x_n}\), with \(x_j \in \{0, \dots, d_j - 1\}\). 

(ii) If the embedding is not bijective, we require the existence of a finite automaton distinguishing \(\Gamma\) from its complement in \(\mathcal{H}\), e.g., an automaton constructing the projector on $\Gamma$, so we can separate the physical states from the non-physical (ghost) states. 

(iii) \(h_{\vec{r}, \vec{r}'}\) also needs to be representable as such a WFAO.
In this work, we focus on two families of terms. 
First, terms of type \(h_{\vec{r}, \vec{r}'} = \delta_{\vec{r}, \vec{r}'} \mathbf{1}_{\vec{r} \in \Gamma'}\), where \(\Gamma'\) is a sublattice of \(\Gamma\), i.e., a chemical potential.
The representation of such an on-site term requires the existence of a finite automaton creating the projector on $\Gamma'$. 
Second, terms of type \(h_{\vec{r}, \vec{r}'}\) where \(h_{\vec{r}, \vec{r}'}\) is a function of the (signed) distance between \(\vec{r}\) and \(\vec{r}'\) on the lattice graph, with a finite range. 
The automaton generating this distance is the key to our construction.\\

\paragraph*{The distance automaton.}
More precisely, we consider an automaton such that, at step \(j\), the machine state \(v_j\) encodes the distance \(\lambda(v_j)\) between \(\underline{x_1 \dots x_j}\) and \(\underline{y_1 \dots y_j}\), with the corresponding tensor representation
\begin{equation}
    W^\mathrm{dis}_{v_{j-1}, x_j, y_j, v_j} = 
    \begin{cases}
        1 & \text{if } v_j = \delta(v_{j-1}, x_j, y_j), \\
        0 & \text{otherwise},
    \end{cases} \label{eq:DistanceTensor}
\end{equation}
where \(\delta\) updates the distance between step $j-1$ and $j$. We note that this tensor is normally a deterministic automaton. In the following, we take inspiration from symmetric tensor networks and explicitly label the virtual dimensions by \(v_j\)~\cite{PhysRevA.82.050301}. 
The connection with symmetric tensor networks will be made explicit in Sec.~\ref{sec:additiveautomata}. 
The tensor \(W^\mathrm{dis}_{v_{j-1}, x_j, y_j, v_j}\), along with the left boundary condition \(v_0 = \bra{0}\), defines an MPO that carries the distance between \(x\) and \(y\) on its virtual dimension. 
We generate the appropriate hopping terms by selecting the right boundary vector of the MPO
\begin{equation}
    \sum_{v_n} h_{v_n} \ket{v_n}. \label{eq:boundarycondition}
\end{equation}
While the number of states \(v_j\) in principle grows with the system size, the backpropagation of the right boundary condition often generates a finite sub-automaton with a very limited number of internal states. This is the explicit condition for the efficiency of the tensor network representation: the bond dimension of the resulting MPO is determined by the number of states that survive the backpropagation, and remains finite if the automaton is sufficiently constrained.\\

In Sec.~\ref{sec:additiveautomata}, we present a class of lattices that admit such an efficient representation. 
Remarkably, it encompasses a wide variety of models, ranging from regular Bravais lattices to fractals or quasiperiodic models.
We note that this formalism is a natural extension of the magic tensor introduced in Ref.~\cite{SciPostPhysLectNotes.133}.

\subsection{Additive distance automata}\label{sec:additiveautomata}
We consider an $n$th-iteration lattice embedded in a $d$-dimensional space and denote by $\vec{u}(\underline{x_1 \dots x_n})$ the position of a site on the lattice, with
\begin{equation}
    \vec{u}(\underline{x_1 \dots x_n}) = \sum_{j=1}^n \vec{u}_{x_j, j}. \label{eq:position}
\end{equation}
The integers $x_j$ label possible moves at each generation $j$, and we assume $x \neq y \implies \vec{u}_{x, j} \neq \vec{u}_{y, j}$.

The distance tensor associated with this position encoding, following Eq.~\eqref{eq:DistanceTensor}, is
\begin{equation}
    W^\mathrm{dis}_{v_{j-1}, x_j, y_j, v_j} = 
    \begin{cases}
        1 & \text{if } v_j = v_{j-1} + \vec{u}_{y_j, j} - \vec{u}_{x_j, j}, \\
        0 & \text{otherwise},
    \end{cases} \label{eq:distance_tensor_linear}
\end{equation}
and $v_j$ indeed tracks the accumulated displacement between the two sites after $j$ generations (with the left boundary condition $v_0 = \vec{0} $).
The hopping terms are generated by selecting the right boundary vector in Eq.~\eqref{eq:boundarycondition}.
This tensor in fact encodes the displacement as an Abelian symmetry. 
In a standard Abelian-symmetric MPS or MPO, each physical and virtual index carries a charge (here $u_j$ and $v_j$, respectively), and tensor elements are non-zero only if the charges are conserved.
The main difference is that the \emph{charge} here (i.e., the displacement) is not a conserved quantum number of the Hamiltonian, but a bookkeeping device. 

We follow a coarse-to-fine decomposition, i.e., the distance between sites differing only on the last digit $n$ corresponds to the lattice spacing. 
By enforcing a scale-separation condition
\begin{equation}
    \forall j, \ \forall \vec{x}, \qquad 
    \left\| \sum_{k=j+1}^n \vec{u}_{x_k, k} \right\| < \left\| \vec{u}_{x_j, j} \right\|
    \quad \text{if } \vec{u}_{x_j, j} \neq \vec{0}, \label{eq:construction}
\end{equation}
we ensure that tight-binding Hamiltonians with finite range have a finite bond dimension. 
Indeed, the condition in Eq.~\eqref{eq:construction} guarantees that the displacement accumulated at generation $j$ dominates all subsequent contributions, so that backpropagating a right boundary condition with a finite range only populates a finite number of virtual states.
The resulting MPO is therefore U($1$)$^d$-labeled, with a finite bond dimension and virtual charges in a finite subset of $\mathbb{R}^d$.

\subsection{Spectral densities from the kernel polynomial method}
Once an MPO representation of the Hamiltonian is constructed, the single-particle spectral properties can be obtained without diagonalizing \(H\). 
We employ the KPM~\cite{RevModPhys.78.275}, an expansion technique using Chebyshev polynomials particularly well suited to tensor networks~\cite{Holzner_chebyshev_2011, antao2026tensornetworksolversultralarge, kolehmainen2026onedimensionalquasicrystalstensornetworkfinitestate}.\\

For a single-particle Hamiltonian \(H\), the retarded Green's function at energy \(E\) is
\begin{equation}
    G(E) = \lim_{\eta \to 0^+} \frac{1}{E + \mathrm{i}\eta - H}.
\end{equation}
The \emph{density of states} (DOS) is defined as
\begin{equation}
    \rho(E) = -\frac{1}{\pi} \operatorname{Im} \operatorname{Tr} G(E)
    = \sum_{n} \delta(E - E_n),
\end{equation}
where \(\{E_n\}\) are the eigenvalues of \(H\). The \emph{local density of states} (LDOS) at site \(i\) is the diagonal matrix element of the spectral operator,
\begin{equation}
    \rho_i(E) = \bra{i} \delta(E - H) \ket{i}
    = \sum_{n} |\langle i \vert n \rangle |^2 \, \delta(E - E_n),
\end{equation}
where \(\ket{n}\) denotes the eigenstate with energy \(E_n\). The LDOS is the central quantity for spatially resolved spectroscopy.\\

KPM approximates \(\rho(E)\) and \(\rho_i(E)\) by expanding the \(\delta\)-function in Chebyshev polynomials of the first kind.
Details of the formalism are kept in App.~\ref{app:Chebyshev}.
We apply recursively \(H\) on itself (for the DOS) or on an MPS representing $\vert i \rangle$ (for the LDOS) to compute Chebyshev polynomials, followed by a compression step.
This avoids constructing the full sparse matrix and preserves the memory advantages of the tensor network representation.
If the bond dimensions of the MPO or the MPS remain under control during iterations, the DOS and LDOS can be computed in a time quasilinear in $n$, and therefore (poly)logarithmic in the number of physical sites.\\

Throughout the rest of the paper, we use \texttt{ITensors.jl}~\cite{ITensor, ITensor-r0.3} as a backend for tensor computations.
Computations are performed using standard compression schemes after every iteration of the Chebyshev polynomials, with a low cut-off of $10^{-14}$.
We denote by $M$ the order of the Chebyshev expansion.
Details on the convergence and numerical parameters used for the different examples can be found in the relevant sections and in App.~\ref{app:convergence}.

\section{Regular and fractal lattices}
\label{sec:regularandfractal}
In this section, we apply the formalism developed in Sec.~\ref{sec:additiveautomata} to regular two-dimensional lattices and then extend the construction to self-similar lattices, i.e., fractals.

\subsection{2D square and triangular lattices} \label{sec:regularlattice}
We start with a two-dimensional square lattice with unit lattice spacing. 
In this construction, the elementary square lattice consists of four sites with positions
\begin{equation}
\begin{aligned}
    \vec{u}_{1, 1} = (0, 0), \quad \vec{u}_{2, 1} = (0, 1), \\
    \quad \vec{u}_{3, 1} = (1, 0), \quad \vec{u}_{4, 1} = (1, 1).
\end{aligned}
\end{equation}
The additive process introduced in Sec.~\ref{sec:additiveautomata} corresponds to choosing the displacement vectors 
\begin{equation}
    \vec{u}_{x, j} = 2^{n-j} \vec{u}_{x, 1}.
\end{equation}
This is the standard decomposition for a \(2^n \times 2^n\) square lattice: a point at position \((x, y)\), where \(x = \underline{x_1 \dots x_n}\) and \(y = \underline{y_1 \dots y_n}\), is labeled by the word \(\underline{x_1 y_1 \dots x_n y_n}\).

We consider the tight-binding model with open boundary conditions,
\begin{equation}
    H_\mathrm{Square}^\mathrm{NN} = -\sum_{\vec{r} \in \Gamma} \left( t_x c^\dagger_{\vec{r}+\vec{e}_x} c_{\vec{r}} + t_y c^\dagger_{\vec{r}+\vec{e}_y} c_{\vec{r}} + \mathrm{h.c.} \right),
\end{equation}
where $t_{x/y}$ are direction-dependent hoppings. 
The corresponding MPO is obtained by imposing on the distance tensor the right boundary condition 
\begin{equation}
\begin{aligned}
   \vec{r}^\square=  -t_x \ket{(1, 0)} - t_y \ket{(0, 1)} \\
   - t_x^* \ket{(-1, 0)} - t_y^* \ket{(0, -1)}. 
   \end{aligned}
   \label{eq:square_bc}
\end{equation}
The propagation of the constraints ensures that the nearest-neighbor (NN) Hamiltonian is represented by a \(5\)-dimensional MPO, where the allowed virtual bond vectors satisfy
\begin{equation}
    \vec{v}_j \in \left\{ \vec{0}, \pm 2^{n-j} \vec{e}_x, \pm 2^{n-j} \vec{e}_y \right\}.
\end{equation}
To include next-NN contributions, we can modify the right boundary condition and the allowed values of \(\vec{v}_{n-1}\), while leaving the rest of the MPO unchanged. 
For practical purposes, we can split the scaling basis and work on a \(2n\)-site tensor network with a two-dimensional Hilbert space, via QR or SVD decompositions of the local tensors \(W\). 
This recovers a more conventional tensor decomposition of the square lattice, with physical legs alternating between the \(x\) and \(y\) directions~\cite{antao2026tensornetworksolversultralarge}.
Finally, periodic boundary conditions also fit the formalism of Sec.~\ref{sec:additiveautomata}. 
For NN hopping, the MPO corresponding to the additional hopping terms is obtained by choosing the right boundary conditions \(\pm \ket{(2^n-1, 0)}\) and \(\pm \ket{(0, 2^n-1)}\), which yields
\begin{equation}
    \vec{v}_i = \pm \vec{e}_x \sum_{j=1}^i 2^{n-j}
    \quad \text{and} \quad
    \vec{v}_i = \pm \vec{e}_y \sum_{j=1}^i 2^{n-j}.
\end{equation}

The triangular lattice takes a similar form. 
The lattice is constructed by the same \(\vec{u}\) vectors as for the square lattice, now interpreted as coordinates in the non-orthogonal basis \(\mathcal{A} = (\vec{e}_x, \frac{1}{2} \vec{e}_x + \frac{\sqrt{3}}{2} \vec{e}_y)\). 
NN Hamiltonians are generated by adding the vectors \((1, -1)\) and \((-1, 1)\) to the right boundary condition of Eq.~\eqref{eq:square_bc}: 
\begin{equation}
    \vec{r}^\triangle = \vec{r}^\square - t_3 \vert (1, -1) \rangle - t_3^* \vert (-1, 1)\rangle.
\end{equation}

\subsection{Fractal lattices}
This framework naturally lends itself to the description of fractals due to their scale invariance. 
In this section, we discuss the Sierpiński carpet as a direct extension of the square lattice construction. 
Several complementary examples are in App.~\ref{app:fractal}.

\begin{figure}[H]
    \centering
    \includegraphics[width=\linewidth]{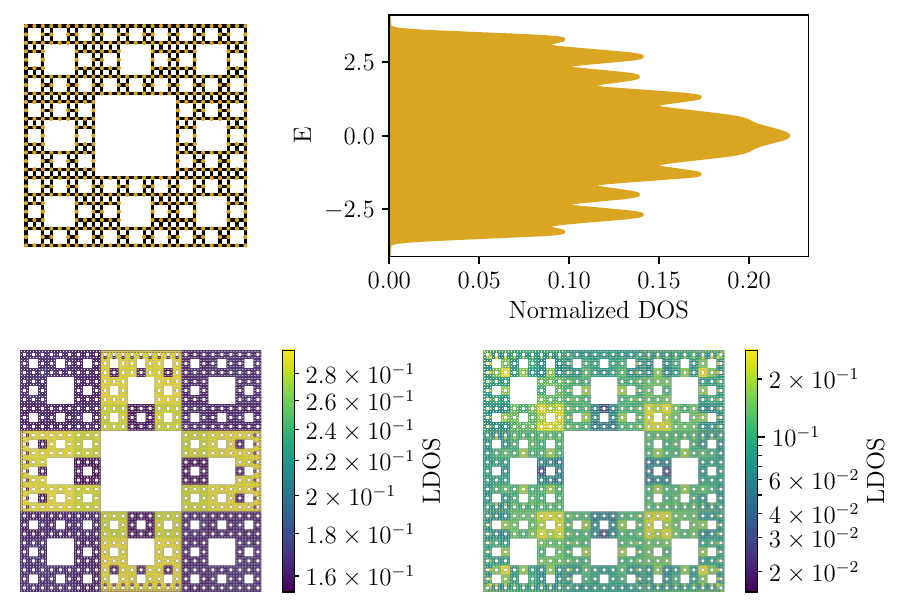}
    \caption{Top left: Sierpi\'nski carpet at generation $n = 3$. Top right: normalized DOS calculated from a Chebyshev expansion up to moment $M = 100$ and $n = 6$, using a NN Hamiltonian on the carpet (with Uunit hopping). Bottom: Representative LDOS maps at energies $E = 0$ (left) and $E = 2.5$ (right), computed with $M = 55$, showing similarly extended spatial distributions, albeit localized in different regions of the carpet.}
    \label{fig:Carpet}
\end{figure}

The Sierpiński carpet is constructed by repeatedly subdividing a $3^n \times 3^n$ square into a \(3 \times 3\) grid and removing the central block, leaving eight smaller copies at every iteration (see Fig.~\ref{fig:Carpet}). 
It is therefore natural to introduce an \(8\)-dimensional Hilbert space labeling the eight allowed positions in the unit cell, namely,
\begin{equation}
\begin{aligned}
    \vec{u}_{1, 1} = (0, 0), \quad \vec{u}_{2, 1} = (1, 0), \quad \vec{u}_{3, 1} = (2, 0),  \\
     \vec{u}_{4, 1} = (0, 1), \quad  \vec{u}_{5, 1} = (2, 1), \quad \vec{u}_{6, 1} = (0, 2),  \\
    \vec{u}_{7, 1} = (1, 2), \quad \vec{u}_{8, 1} = (2, 2),
\end{aligned}
\end{equation}
and
\begin{equation}
    \vec{u}_{x, j} = 3^{n-j} \vec{u}_{x, 1}.
    \label{eq:SC-scaling}
\end{equation}
The NN Hamiltonian is defined using the same right boundary conditions as in Eq.~\eqref{eq:square_bc}, resulting in a similar MPO of bond dimension \(5\).\\

In Fig.~\ref{fig:Carpet}, we show, as an illustration, the total DOS and the local density of states at $E=0$ and $E=2.5$ for an $n=6$ Sierpi\'nski carpet ($2^{18}$ sites) with NN hopping.
The extended nature of the modes is directly reflected in the local DOS, which remains spatially distributed without pronounced localization peaks.
This translates into a fast quadratic increase in the bond dimension of the MPOs during the KPM iterations.
Including disorder, or a suitably chosen magnetic field, may help localize the wavefunctions and therefore alleviate this difficulty.

\section{Tensor network representations of substitution quasicrystals}
Quasicrystals are aperiodic structures that can exhibit long-range order, characterized by diffraction spectra with non-crystallographic symmetry~\cite{Shechtman1984}. 
A prominent class of quasicrystals is generated by substitution rules: a finite set of tiles or atoms is repeatedly replaced according to a fixed rule, producing a self-similar, non-periodic arrangement. 
They are consequently a natural fit for our finite automaton framework.
In this section, we show how our tensor network formalism can be used to study substitutional quasicrystals through the example of the Fibonacci chain~\cite{PhysRevB.35.1020, PhysRevB.40.7413, PhysRevB.93.205153, Suto1989, RevModPhys.93.045001} and one of its two-dimensional extensions~\cite{ILAN01012004,LIFSHITZ2002186}.

\subsection{Automaton representation of the Fibonacci chain}\label{sec:FibAuto}
The Fibonacci chain consists of two types of atoms, \(0\) and \(1\), whose arrangement is generated by the substitution rule
\begin{equation}
    0 \rightarrow 01, \qquad 1 \rightarrow 0. \label{eq:FibSub}
\end{equation}
Starting from a site \(0\) at the $0$th generation, the \(n\)th generation of the chain has length \(L = F_{n+1}\), where \(F_n\) is the \(n\)th Fibonacci number (with \(F_0 = F_1 = 1\), \(F_2 = 2\)).
We remind that $F_n$ grows approximately as $\varphi^n$, where $\varphi$ is the golden ratio $\frac{1 + \sqrt{5}}{2}$.\\

Crucially, the Fibonacci chain follows the formalism of Sec.~\ref{sec:additiveautomata}. 
Each site index \(i \in \{0, \dots, L-1\}\) is uniquely expressed in the Zeckendorf representation~\cite{Zeckendorf1972},
\begin{equation}
    i = \sum_{j=1}^{n} x_j F_{n+1-j}, \qquad x_j \in \{0, 1\}, \label{eq:positionFibo}
\end{equation}
with the constraint that no two consecutive \(x_j\) are equal to \(1\). 
Note that the unusual form of Eq.~\eqref{eq:positionFibo} is due to our coarse-to-fine convention. 
The Zeckendorf representation therefore naturally embeds the Fibonacci chain into a \(2^n\)-dimensional Hilbert space, with a large (exponential) number of ghost states.
Contrary to the previous examples, to study the spectral properties of the Fibonacci chain, we need to project our these unphysical states.
We denote by $\sigma^\alpha$ the Pauli operators on the tensor space, and by $p_0$ (resp. $p_1$) the projector on the state $\vert 0 \rangle$ (resp.  $\vert 1 \rangle$).
The projector onto the physical subspace is a two-dimensional MPO defined by
\begin{equation}
    W_\mathrm{Fib}^{\mathrm{proj}} = 
    \begin{pmatrix}
        p_0 & p_1 \\
        p_0 & 0 
    \end{pmatrix}, \label{eq:FibProj}
\end{equation}
with left boundary condition \((1, 0)\) and right boundary condition \((1, 1)\). \\

The tight-binding Hamiltonian on the Fibonacci chain is
\begin{equation}
    H_\mathrm{Fib} = \sum_i \varepsilon_{\sigma_i} c_i^\dagger c_i + \sum_{\langle i,j \rangle} t_{\sigma_i, \sigma_j} c_i^\dagger c_j,
\end{equation}
where \(\varepsilon_{\sigma_i}\) are on-site potentials depending on the atomic type and \(t_{\sigma_i, \sigma_j}\) are species-dependent hopping amplitudes.
For simplicity, we consider real hopping terms.\\

The on-site term is a single-site diagonal operator acting on the last physical leg, \(\mathrm{diag}(\varepsilon_0, \varepsilon_1)\), since the nature of the site is entirely encoded in \(x_n\).
For the hopping term, Eq.~\eqref{eq:positionFibo} matches the additive distance-tensor formalism with \(u(0, j) = 0\) and \(u(1, j) = F_{n + 1-j}\), so the NN uniform hopping is obtained by fixing the right boundary vector of the MPO to \(\ket{\pm 1}\). 
The scale-separation condition in Eq.~\eqref{eq:construction} is respected within the physical states and the MPO dimension remains bounded. 
We proceed below to an explicit construction.
Let \(\underline{x_1 \dots x_n}\) and \(\underline{y_1 \dots y_n}\) be two points of the chain in their Zeckendorf representation, and let \(x_j\) and \(y_j\) be the first differing digits, with \(x_j = 0\) and \(y_j = 1\). Then \(x\) lies immediately to the left of \(y\) iff \(x_{j+1+2k} = 1\) for all \(k\) such that \(j+1+2k \le n\), and \(y_i = 0\) for all \(i > j\). 
Consequently, a (symmetric) hopping Hamiltonian in the Zeckendorf basis reads
\begin{equation}
    H_\mathrm{Fib}^{\mathrm{hop}} = \sum_{j=1}^{n} \sigma_j^+ \bigotimes_{k=1}^{n-j} \hat{o}_{k, j+k} + \mathrm{h.c.},
\end{equation}
where
\begin{equation}
    \hat{o}_{k, l} = 
    \begin{cases}
        p_{0, l} & \text{if } k \text{ is even}, \\[2mm]
        \sigma_l^- & \text{if } k \text{ is odd}.
    \end{cases}
\end{equation}
The corresponding MPO has bond dimension \(5\):
\begin{equation}
    W_\mathrm{Fib}^{\mathrm{hop}} = 
    \begin{pmatrix}
        \mathrm{Id} & \sigma^+ & 0 & \sigma^- & 0 \\
        0 & 0 & \sigma^- & 0 & 0 \\
        0 & p_0 & 0 & 0 & 0 \\
        0 & 0 & 0 & 0 & \sigma^+ \\
        0 & 0 & 0 & p_0 & 0
    \end{pmatrix},
\end{equation}
with left boundary condition \((1, 0, 0, 0, 0)\) and right boundary condition \((0, -t, -t, -t, -t)\).
As written, this hopping term connects physical and ghost states.
We address this problem, either by projecting the Hamiltonian on the physical Hilbert space using the projector defined in Eq.~\eqref{eq:FibProj}, or by increasing its bond dimension by $1$ and directly incorporating the projection within the MPO.
Similarly, an anisotropic hopping $t_{00} \neq t_{01}$ can be introduced at no cost in bond dimension by adding the relevant amplitudes in the last tensor.
The explicit tensors can be found in App.~\ref{app:Fib}. \\

\begin{figure}[H]
    \centering
    \includegraphics[width=0.48\linewidth]{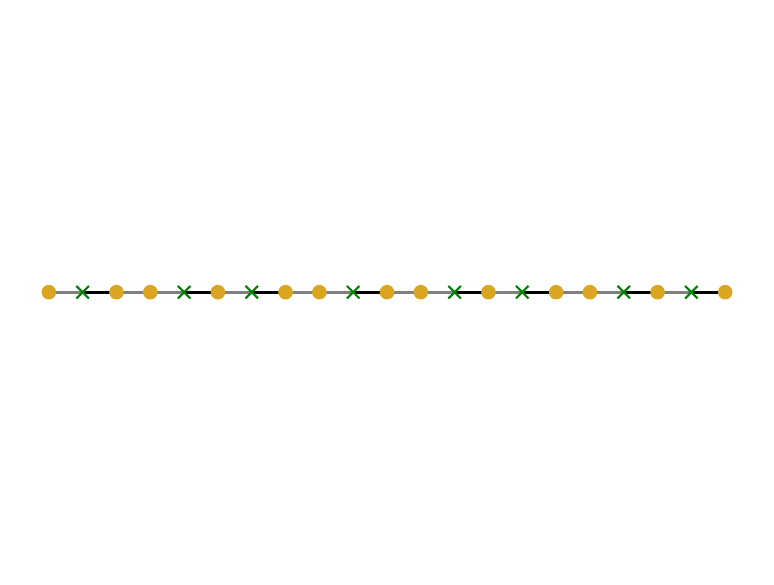}\\
    \vspace*{-1cm}
    \includegraphics[width=\columnwidth]{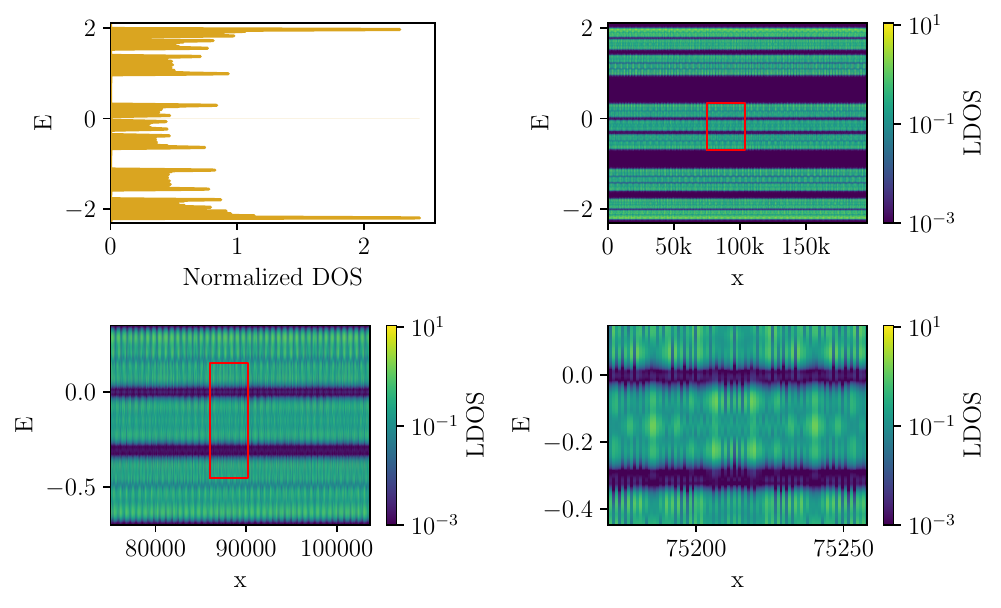}
    \caption{Top: Fibonacci chain at iteration $n = 6$. The two sites types, corresponding to $0$ (gold circle) and $1$ (green cross) in Eq.~\eqref{eq:FibSub}, are connected by two types of links, with $t_{00} = t_{10}$ (gray) and $t_{01}$ (black). This representation makes explicit the correspondence between the patterns of hopping amplitudes and on-site potentials.
    Center left: Normalized DOS for $n=25$ generation, calculated from a Chebyshev expansion up to moment $M = 1000$. We set $\epsilon_1 = -\epsilon_0 = 0.5$ and $t_{\alpha, \beta} = 1$.
    Center right and bottom: Energy-resolved LDOS ($M = 500$) along the full chain, together with zoomed-in views of the central region of the chain around $E=0$, highlighting the strongly inhomogeneous and quasiperiodic spatial structure of the spectral weight.}
    \label{fig:Fibonacci}
\end{figure}

As an application of the construction, we study a chain with $\varepsilon_1 = - \varepsilon_0 = 0.5$ and uniform hopping $t_{\alpha, \beta} =1$, breaking chiral symmetry.
In Fig.~\ref{fig:Fibonacci}, we show the DOS ($M = 1000$) and local DOS ($M=500$) for a finite chain at iteration $n = 25$ ($196\,418$ sites).
At this low value of disordered potential, the chain remains weakly localized, yet, the two spectra present the characteristic recursive patterns of quasi-periodic models.
The reduced moment $M$ used for the LDOS is purely practical: we do not perform any coarse-graining in our evaluation of the LDOS, and therefore explicitly compute the LDOS for the $196\, 418$ states.
Significant numerical improvement can be obtained either by taking a partial trace over the smaller scales, coarse-graining the LDOS, therefore evaluating an exponentially smaller number of states.
Alternatively, by saving the Chebyshev MPOs, each LDOS can then be computed in parallel at large scale.

The construction extends directly to any substitution rule that admits a finite-state description.
Prominent examples include the metallic-mean quasicrystals (e.g., silver-mean and bronze-mean chains)~\cite{RevModPhys.93.045001, 7rhl-z7f3} and the higher-order Fibonacci generalizations such as the \(k\)-bonacci chains~\cite{PhysRevB.108.104204}.
Ref.~\cite{kolehmainen2026onedimensionalquasicrystalstensornetworkfinitestate} provides the explicit MPO construction of the silver-mean and the Tribonacci chains.
In each case, the substitution rule defines a finite automaton, and the associated numeration system provides the analogue of the Zeckendorf representation, allowing the Hamiltonian to be encoded as an MPO within the same framework.

\subsection{Limits of the cut-and-project construction for the Fibonacci chain}
The cut-and-project method is a standard approach for the construction of quasiperiodic models.
In the case of the Fibonacci chain, it consists in introducing a $\mathbb{Z}^2$ lattice.
Then we select a non-rational line with slope $\frac{1}{\varphi}$, e.g., passing by $(0, 0)$ and keep the points of the lattice within a predefined width $\frac{1 + \varphi}{\sqrt{2 + \varphi}}$.
The position of the conserved points projected on the line determines their position in a one-dimensional chain, with a quasiperiodic alternance of long ($\varphi$) and short (1) distance between the neighboring sites.
Points that were NN in the $x$ direction in $\mathbb{Z}^2$ have a long separation, while those that were NN in the $y$ are separated by a short distance.

At first, the cut-and-project method seems perfectly adapted to our tensor construction, as all properties depend on the position of the points in the larger space, which is easy to calculate through tensor computations.
Indeed, we can represent the square lattice as in Sec.~\ref{sec:regularlattice}, and the Hamiltonian is the projection of the anisotropic $t_x \neq t_y$ NN hopping of the square lattice.
The difficulty resides in the encoding of the projector onto the physical space.
The binary representation in $\mathbb{Z}^2$ does not offer a simple finite-state machine to compute the distance to the projected subspace, due to the irrationality of $\varphi$.
The path forward is then to translate the binary representation into a more convenient representation, such as the Zeckendorf basis, using another automaton.
While technically possible, it therefore appears more efficient to directly work in that basis as in Sec. \ref{sec:FibAuto}.
We expect a similar difficulty with all constructions based on the cut-and-project method given that quasiperiodicity relies on the irrationality of the projection.

\subsection{Two-dimensional Fibonacci lattice}
Our substitution-based automaton can be applied to two-dimensional quasicrystals. 
A technical difficulty arises, however, for the position-based schemes discussed in Sec.~\ref{sec:additiveautomata}: rotations at every step are frequent in two-dimensional quasiperiodic tilings, and encoding them requires tracking the local orientation in addition to the displacement. 
This increases the bond dimension and obscures the connection with the one-dimensional construction.
Instead, we focus here on the two-dimensional Fibonacci model, for which the orientation is fixed. 
General two-dimensional quasiperiodic lattices are left for future works. \\

The square two-dimensional Fibonacci tiling is the Cartesian product of two one-dimensional Fibonacci chains, with tile lengths \(1\) and \(\varphi \) along each axis~\cite{PhysRevB.41.4314, LIFSHITZ2002186, Even-DarMandel21022006}. 
It can alternatively be seen as a finite automaton consisting of three different tiles: two squares of side $1$ and $\varphi$ ($S$ and $L$), as well as a rectangle ($R$) \(1 \times \varphi \) in the horizontal and vertical directions, subject to the substitution rule
\begin{equation}
    S \rightarrow L, \ R \rightarrow L + R, \ L \rightarrow L + S + 2R.
\end{equation}
Given the structure of the lattice, we can decompose both the $x$ and $y$ coordinates of a lattice point in their Zeckendorf representation $\underline{x_1\dots x_n}$ and $\underline{y_1\dots y_n}$, leading to an embedding in $\mathbb{Z}^2$.
The projector on the physical subspace is therefore simply $\mathcal{P}_\mathrm{Fib2} = \mathcal{P}_{\mathrm{Fib}, x} \otimes \mathcal{P}_{\mathrm{Fib}, y}$, with $\mathcal{P}_{\mathrm{Fib}, \alpha}$ the projector defined in Sec.~\ref{sec:FibAuto} applied on the $\alpha$ bits.
The type of each site is now parametrized by $(x_n, y_n)$, with 
\begin{equation}
    x_n = y_n = 0 \Leftrightarrow L,\ x_n = y_n = 1 \Leftrightarrow S,\ R \text{ else.}
\end{equation}
We consider the NN tight-binding model of the form 
\begin{multline}
    H_\mathrm{Fib2}= \sum_{x,y} \varepsilon_{x_n, y_n} \, c^\dagger_{x,y} c_{x,y} \\
    + \sum_{\langle (x,y), (x', y')\rangle_x} \left( t^x_{x_n,x'_n} \, c^\dagger_{x,y} c_{x',y'}+ \mathrm{h.c.} \right) \\
     + \sum_{\langle (x,y), (x', y')\rangle_y} \left( t^y_{y_n,y'_n} \, c^\dagger_{x,y} c_{x',y'}+ \mathrm{h.c.} \right).
\end{multline}
In particular, the hopping contribution takes the form
\begin{equation}
    H_\mathrm{Fib2}^\mathrm{hop} = H_{\mathrm{Fib}, x}^\mathrm{hop} \otimes \mathrm{Id} +\mathrm{Id} \otimes H_{\mathrm{Fib}, y}^\mathrm{hop}.
\end{equation}
The chemical potential is a simple two-site operator acting on $x_n$ and $y_n$.\\

The tiling can be encoded as a one-dimensional MPO in two ways: by interleaving the \(x\) and \(y\) coordinates, as in our study of the square lattice or fractals, or by concatenating them ($\underline{x_1\dots x_n y_1\dots y_n}$ or $\underline{x_1\dots x_n y_n\dots y_1}$).
In the interleaved representation, we preserve the scale ordering.
The resulting hopping MPO and projector MPO have a bond dimension twice larger than in the one-dimensional model.
The chemical potential is a two-site operator at the end of the chain.
Conversely, in the concatenated representation, the MPOs have a significantly reduced dimension: the bond dimension of the hopping MPO is only one larger than in one dimension, and the projector remains of bond dimension $2$.
In both cases, the bond dimension of the operators remain manageable and practical for the KPM scheme.\\

\begin{figure}[H]
    \centering
    \includegraphics[width=\linewidth]{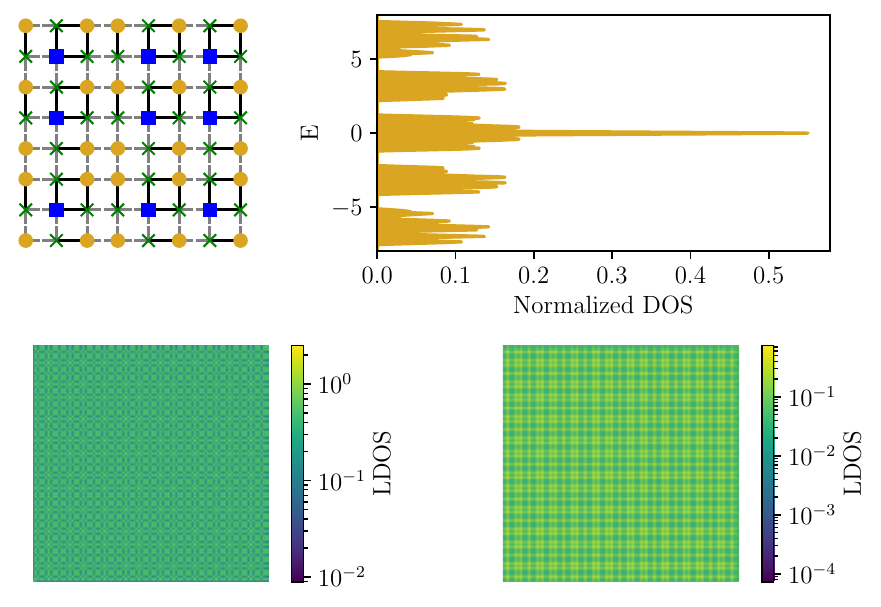}
    \caption{Top left: Two-dimensional Fibonacci lattice generated from the one-dimensional Fibonacci chain, at iteration $5$.
    The four types of sites indicated by different colors and symbols. The hoppings follow the same convention as in Fig.~\ref{fig:Fibonacci}.
    Top right: Normalized DOS obtained for a Fibonacci square at generation $n = 12$ ($142\,129$ sites), for $\varepsilon_{\alpha, \beta} = 0$ and $t_{01} = 3$, $t_{00} = t_{10} = 1$, using the KPM expansion up to $M = 1000$.
    Bottom: the corresponding LDOS at $E=0$ (left) and $E= 3$ (right) maps, revealing characteristic square-like spatial patterns, in a quasiperiodic ordering, with $M = 500$.}
    \label{fig:Fibonacci2D}
\end{figure}

Using the concatenated representation, we study a 2D Fibonacci square at iteration $n =12$ ($377 \times 377$ sites) with $\varepsilon_{\alpha\beta} = 0$, and $t_{0, 0} = t_{1, 0} = 1$, $t_{0, 1} = 3$ in both directions.
The DOS in Fig.~\ref{fig:Fibonacci2D}, obtained using a Chebyshev expansion up to order $M = 1000$ shows the predicted five sub-bands~\citep{Fu1993}: in contrast to the one-dimensional Fibonacci chain, the square here remains gapless.
The corresponding LDOS (with $M = 500$) is delocalized, matching the expected critical behavior, though the quasiperiodic patterns are apparent in the parameter range we show.

\section{Cayley trees and hyperbolic lattices}
\label{sec:CayleyHyperbolic}
Cayley trees are rooted, loopless graphs in which every vertex at a given distance from the root has the same number of branches~\cite{OSTILLI20123417}. 
They play a central role in statistical mechanics as exactly solvable models of phase transitions.
They correspond to truncated Bethe lattices, an infinite-dimensional limit of Euclidean lattices.
The core difference is that a finite Cayley tree has a boundary that remains macroscopically important~\cite{OSTILLI20123417,tikhonov_fractality_2016,sonner_multifractality_2017-1}.
Their loop-free structure makes them amenable to recursive and cavity methods, and they are the natural setting for the Bethe approximation and its generalizations. 
Hyperbolic lattices are tilings of the hyperbolic plane (with negative curvature).
For example, the \(\{p,q\}\)  hyperbolic lattices correspond to regular tilings of the hyperbolic plane, in which \(q\) regular \(p\)-gons meet at each vertex. 
Tight-binding models on hyperbolic lattices tend to consider sites localized on the vertices of the \(p\)-gons~\cite{doi:10.1126/sciadv.abe9170, PhysRevLett.129.246402, t8gg-xqls, PhysRevLett.125.053901, PhysRevB.106.155146}.
Without loss of generality, we will consider lattices with a vertex at the origin of the curved plane and of the Poincar\'e disk.

These two structures are intimately connected.
The Bethe lattice has hyperbolic character and can be naturally embedded in hyperbolic space~\cite{refId0, PhysRevE.47.4582}.
Conversely, the vertices of a \(\{p,q\}\) hyperbolic lattice form a generalized Cayley tree, with a coordination number varying with the distance to the origin.

In this section, we exploit this relationship to extend our tensor-network formalism to both geometries.

\subsection{Generalized Cayley trees}\label{sec:Cayley}
We define a generalized Cayley tree as a rooted graph parametrized by a sequence \((m_1, \dots, m_n)\) of positive integers, such that every vertex at distance \(k\) from the root has exactly \(m_{k+1}\) branches connecting it to vertices at distance \(k+1\). 
The nodes are parametrized by the unique path connecting them to the root, which is realized by a finite automaton with the following encoding: each site is labeled by the input \(\underline{x_1 \dots x_n}\) with \(x_k \in \{0, \dots, m_k\}\), subject to the grammatical rule \(x_j = 0 \implies x_{j'} = 0\) for all \(j' > j\). 
Here, \(x_j \neq 0\) indicates which descendant at distance \(j\) the path passes through, while \(x_j = 0\) indicates that the path has terminated. 
This provides the first requirement of Sec.~\ref{sec:GeneralFormalism}: a natural embedding of the Cayley tree into a \(\prod_{j=1}^n (m_j + 1)\)-dimensional Hilbert space.
The ghost states correspond to the invalid paths, which include a sequence $\underline{\dots0x_j\dots}$ with $x_j \neq 0$.
The separation between ghost and physical states can be enforced through the projector
\begin{equation}
    W^{\mathrm{proj}}_\mathrm{Cay} = \begin{pmatrix}
        \mathrm{Id} - | 0 \rangle\langle 0\vert  &  | 0 \rangle\langle 0\vert \\
        0 &  | 0 \rangle\langle 0\vert
    \end{pmatrix}
\end{equation}
with left and right boundary condition $(1, 1)$.\\

A Cayley tree cannot be embedded in a Euclidean vector space while respecting the graph distance, and therefore does not follow the additive construction of Eq.~\eqref{eq:construction}. 
Yet, the distance tensor admits a simple definition:
\begin{equation}
    \left(W_\mathrm{Cay}^\mathrm{dis}\right)_{v_{j-1}, x_j, y_j, v_j} = 1 
    \quad \text{if} \quad 
    v_j = v_{j-1} + \delta(v_{j-1}, x_j, y_j),
\end{equation}
where the increment is
\begin{equation}
    \delta(v_{j-1}, x_j, y_j) = 
    \begin{cases}
        0 & \text{if } x_j = y_j = 0 \\
          & \text{ or } (x_j = y_j \text{ and } v_{j-1} = 0), \\
        1 & \text{else if } x_j = 0 \text{ or } y_j = 0, \\
        2 & \text{otherwise}.
    \end{cases}
\end{equation}
With the left boundary condition \(v_0 = 0\), the state \(v_j\) tracks the partial graph distance between the sites \(x\) and \(y\). 
The NN Hamiltonian is obtained by enforcing the right boundary condition \(v_n = 1\). 
Since \(v_j\) is non-decreasing and bounded by \(v_n = 1\), the resulting MPO has bond dimension \(2\), with \(v_j \in \{0,1\}\) for all \(1 \le j < n\):
\begin{equation}
     W^\mathrm{NN}_\mathrm{Cay} = \begin{pmatrix}
        \mathrm{Id} - p_0 & \tau^+ + \tau^- \\
        0 & p_0 
     \end{pmatrix}, \label{eq:TenCay}
\end{equation}
where $p_0$ is the projector on $\vert 0 \rangle$ and $\tau^+ = \sum\limits_{x = 1}^q \vert x \rangle \langle 0 \vert$.
The left boundary condition is $(1, 0)$ and the right is $(0, 1)$.
The term $\mathrm{Id} - p_0$, instead of $\mathrm{Id}$ ensures that physical and ghost states are not connected.\\

\begin{figure}
    \centering
     \includegraphics[width=\linewidth]{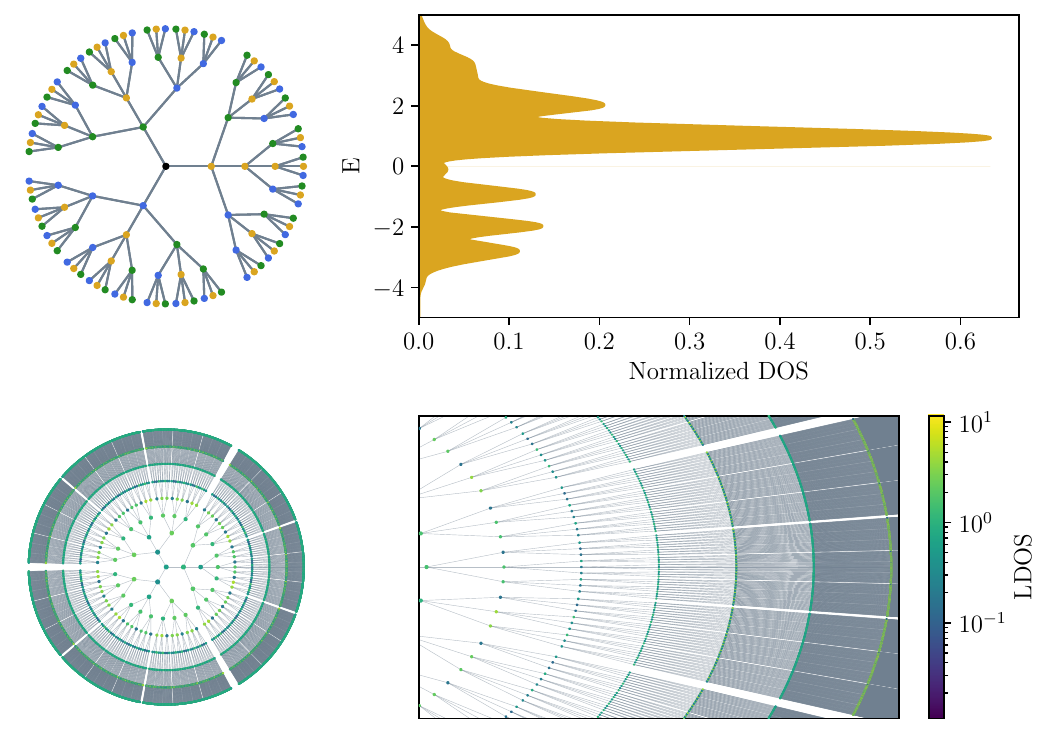}
    \caption{Top left: the $3$-Cayley tree with unit NN hopping at iteration $n = 4$. At each depth, nodes of the same color have the same on-site potential. The central node is in black, with zero chemical-potential. Top right: DOS ($M = 140$) for a $3$-Cayley tree at iteration $10$, in the presence of disorder defined in Eq.~\eqref{eq:DisCay}, with disorder strength $W = 3$. Bottom: the corresponding local density of states at $E=-2$, with a zoom on iterations $3$ to $9$.}
    \label{fig:Cayley}
\end{figure}

In Fig.~\ref{fig:Cayley}, we study the spectral properties of a disordered tight-binding model on a $3$-Cayley tree at iteration $n = 10$, with a chemical potential following a correlated disorder:
\begin{equation}
    H^\mathrm{Dis}_\mathrm{Cay} =\sum\limits_{\vec{r}} \varepsilon_{m_{\vec{r}}, d_{\vec{r}}}  c^\dagger_{\vec{r}} c_{\vec{r}}, \label{eq:DisCay}
 \end{equation}
$\varepsilon_{m_{\vec{r}}, d_{\vec{r}}}$ depends only on the distance to the origin $ d_{\vec{r}}$, and the label $m \in [1, q]$ as shown in the top panel of the figure, and is taken uniformly in $[-W, W]$, with $W = 3$.
This operator admits a straightforward MPO representation.
In the bottom panels of Fig.~\ref{fig:Cayley}, the $C_3$ rotation symmetry is broken at very short distances but is restored quickly going towards the leaves. By contrast, the mirror symmetry stays broken.

\subsection{Hyperbolic lattices}\label{sec:Hyperbolic}

\begin{figure}
    \centering
    \includegraphics[width=\linewidth]{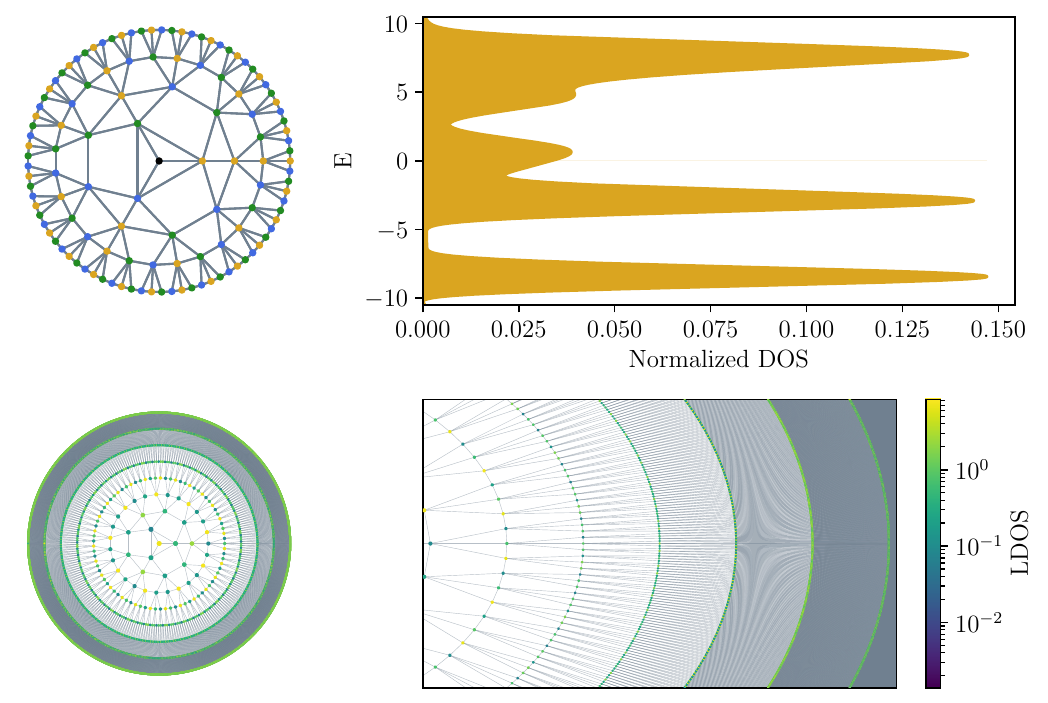}
    \caption{Top left: Graphical representation of an hyperbolic lattice following the convention of Fig.~\ref{fig:Cayley}. Top right: the DOS ($M=68$) for an hyperbolic lattice with unit NN hopping and  disorder defined in Eq.~\eqref{eq:DisCay} with disorder strength $W = 9$. Bottom: the corresponding LDOS.}
    \label{fig:HyperbolicLattice}
\end{figure}
As discussed previously, the vertices of the $\{p, q\}$ lattices only form generalized Cayley trees.
For simplicity, we will work in this section with a more convenient class of hyperbolic models, whose nodes are exactly described by a $q$-Cayley tree, as depicted in Fig.~\ref{fig:HyperbolicLattice}.
The construction we develop can be adapted to the more general case, at the price of a varying local physical dimension of the tensors.

We start with a standard Cayley tree where every node has $q$ children, labeled as in Sec.~\ref{sec:Cayley}.
The first descendants labeled by $x_1 \in [1, q]$ of the root nodes will also be considered to be neighbours on the hyperbolic lattice if $ x_1 = y_{1} \pm 1 \text{ mod } q$.
By convention, we will say that $x_10\dots$ is to the left of $y_10\dots$ if  $x_1 = y_{1} + 1 \text{ mod } q$ (assuming $q > 2$).
For further generation $j > 1$, two nodes $\underline{\dots x_j0\dots}$ and $\underline{\dots y_j0\dots}$ ($x_j, y_j\neq 0$) will be neighbours in the following cases:
\begin{enumerate}
    \item If they have the same parent ($x_k=y_k\ \forall k \leq j-1$) and  $x_j = y_{j}+1$.
    \item If their parents are neighbours, $\underline{\dots x_{j-1}0\dots}$ is to the left of $\underline{\dots y_{j-1}0\dots}$,  $x_j = 1$ and $y_j = q$.
\end{enumerate}
In all cases, $\underline{\dots x_j0\dots}$ will be to the left of $\underline{\dots y_j0\dots}$.
It only requires a minor modification in the definition of the distance tensor to track whether the paths belong to neighbouring branches.
The resulting MPO, following Eq.~\eqref{eq:TenCay}, has bond dimension $4$: 
\begin{equation}
     W^\mathrm{NN}_\mathrm{CayHyp} = \begin{pmatrix}
        \mathrm{Id} - p_0 & \tau^+ + \tau^- & \delta_{x = y+1 > 1} & \delta_{x +1 = y > 1} \\
        0 & p_0  & 0 & 0 \\
        0 & p_0 & \delta_{x, 1} \delta_{y, q} & 0 \\
        0 & p_0 & 0 &\delta_{x, q} \delta_{y, 1} 
     \end{pmatrix}.
\end{equation}
Note that for the first tensor site, the $\delta$ symbols on the first line should be understood modulo $q$.
The left boundary condition is $(1, 0, 0, 0)$ and the right-boundary condition is $(0, 1, 1, 1)$.

In Fig.~\ref{fig:HyperbolicLattice}, we show the global and local DOS for the hyperbolic lattice constructed from a $3$-Cayley tree at generation $n = 10$, in the presence of the disordered potential defined in Eq.~\eqref{eq:DisCay} with disorder strength $W = 9$.

\section{Conclusion}
In this work, we introduced a systematic tensor-network construction for single-particle Hamiltonians defined on recursively generated lattices.
By decomposing the lattice structure into the local degrees of freedom associated with each level of the recursive construction, the lattice connectivity can be encoded through a finite set of virtual states.
A central feature of this approach is that the number of tensors is determined by the number of recursion levels (or generations) rather than by the total number of physical sites.
When the required set of virtual states remains finite and independent of generation, this leads to an exact representation of the Hamiltonian whose storage grows only logarithmically with the physical system size.
The resulting MPOs can be directly combined with tensor-network algorithms for studying systems far beyond the sizes accessible through an explicit representation of the Hamiltonian. 
Their integration with standard numerical methods such as KPM provides a direct route to obtaining spectral properties.
We demonstrated the versatility of the framework across several classes of recursively structured lattices, including Cayley trees, hyperbolic lattices, self-similar fractal lattices, and substitutional quasicrystals. 
Despite their different connectivity properties, these systems can be described within the same tensor-network framework by adapting the local recursive rules and the corresponding finite-state transitions.
We note that the link to Abelian-symmetric tensors is only a convenient labeling trick that can be relaxed in more involved models.
The compact representation of the Hamiltonian alone does not guarantee a logarithmic computational cost, since the bond dimensions of the intermediate tensor-network representations may increase with the expansion order and system size beyond numerical capacity.
The incorporation of the crystal symmetries, such as the $D_4$ symmetry for the Sierpiński carpet, may provide a path to reduce the numerical costs of the simulations, as well as provide a direct access to symmetry-resolved observables.
A particularly interesting direction for future work is the extension of the present framework to intrinsically two-dimensional quasicrystals, such as the Penrose or Ammann–Beenker tilings.
Such structures require generalizing the displacement-only recursion to more general affine transformations, potentially enlarging the set of virtual states that must be tracked and, consequently, increasing the MPO bond dimension.
Conveniently, the existence of simple substitution rules essentially guarantees the existence of a bounded tensor representation.
Whether the distance formalism is the most favorable approach remains to be seen.
In parallel, adding long-range hopping terms calls for a systematic investigation of how the range and structure of the hoppings affect the number of virtual states.
Similarly to the standard encoding of decaying-exponentials in the MPO of interacting models, it is likely to require abandoning the strict distance-based framework, to fully exploit the WFAO structure of tensor networks.

\acknowledgements
The authors thank Peru d'Ornellas and Yuriel Núñez Fernández for useful discussions.
LH acknowledges the support of the ANR JCJC ANR-25-CE30-2205-01.
\appendix

\section{Chebyshev expansion of the DOS}\label{app:Chebyshev}
The Chebyshev polynomials of the first kind, \(T_m(x)\) are defined on \(x \in [-1,1]\) by the recursion
\begin{equation}
    T_0(x) = 1, \qquad T_1(x) = x,
\end{equation}
\begin{equation}
     T_{m+1}(x) = 2x  T_m(x) - T_{m-1}(x).
\end{equation}
The Hamiltonian must first be rescaled so that its spectrum lies within the interval \([-1,1]\). Introducing the rescaled operator
\begin{equation}
    \tilde{H} = \frac{H - b}{a},
    \qquad
    a = \frac{E_{\max} - E_{\min}}{2(1 - \epsilon)},
    \qquad
    b = \frac{E_{\max} + E_{\min}}{2},
\end{equation}
where \(\epsilon\) is a small safety margin, the spectral function is expanded as
\begin{equation}
    \rho(E) = \frac{1}{\pi a \sqrt{1 - \tilde{E}^2}}
    \left[ \mu_0 + 2 \sum_{m=1}^{N_c} \mu_m T_m(\tilde{E}) \right],
    \label{eq:kpm_dos}
\end{equation}
and the LDOS as
\begin{equation}
    \rho_i(E) = \frac{1}{\pi a \sqrt{1 - \tilde{E}^2}}
    \left[ \mu_0^{(i)} + 2 \sum_{m=1}^{N_c} \mu_m^{(i)} T_m(\tilde{E}) \right],
    \label{eq:kpm_ldos}
\end{equation}
where \(N_c\) is the truncation order and the Chebyshev moments are
\begin{equation}
    \mu_m = \operatorname{Tr} T_m(\tilde{H}),
    \qquad
    \mu_m^{(i)} = \bra{i} T_m(\tilde{H}) \ket{i}.
\end{equation}
The moments are computed recursively using the same three-term recurrence as the polynomials themselves. 
For the LDOS, one can start from the local vector \(\ket{i}\); for the total DOS, one either traces over a full basis or uses the stochastic trace estimator \(\mu_m \approx \frac{1}{R} \sum_{r=1}^{R} \langle{v_r} \vert {T_m(\tilde{H})} \vert{v_r} \rangle \) with \(R\) random vectors, which converges as \(1/\sqrt{R}\).\\

The finite truncation of the Chebyshev series in Eqs.~\eqref{eq:kpm_dos}–\eqref{eq:kpm_ldos} produces Gibbs oscillations that must be damped by a kernel \(g_m\). The standard choice is the Jackson kernel,
\begin{equation}
    g_m = \frac{(N_c - m + 1)\cos\frac{\pi m}{N_c + 1} + \sin\frac{\pi m}{N_c + 1}\cot\frac{\pi}{N_c + 1}}{N_c + 1},
\end{equation}
which enforces positivity of the reconstructed spectral density and yields the optimal uniform approximation for a given truncation order. With the kernel, the moments are replaced by \(g_m \mu_m\) (and \(g_m \mu_m^{(i)}\) for the LDOS). The effective energy resolution is set by the broadening parameter \(\eta = a \pi / N_c\), so increasing \(N_c\) improves the resolution at the cost of additional matrix-vector multiplications.

\section{Additional examples}
\subsection{Fractal lattices} \label{app:fractal}
\paragraph{Fractal dimension}
The fractal dimension $d_f$ appears naturally in our quantics encoding.
Let $d$ be the local physical dimension, $d = |  \{ \vec{u}_{x, 1}\}| $, and $b$ the scale factor that appears in the recursive construction for uniformly self-similar fractals

\begin{equation}
    \vec{u}_{x,j} = b^{n-j} \vec{u}_{x,1}.
    \label{eq:general-recursive}
\end{equation}

Then $d_f = \log d / \log b$.
\paragraph{The Sierpi\'nski gasket}
The Sierpiński gasket is constructed by starting from an equilateral triangle and iteratively removing the central inverted triangle from each remaining solid triangle.
We work in the basis $\mathcal{A} = (\vec{e}_x, \frac{1}{2}\vec{e}_x + \frac{\sqrt{3}}{2}\vec{e}_y)$ of the underlying triangular lattice. 
At the first generation, the elementary gasket consists of three sites forming an elementary triangle,
\begin{equation}
    \vec{u}_{1,1} = (0,0), \qquad
    \vec{u}_{2,1} = (1,0), \qquad
    \vec{u}_{3,1} = (0,1),
\end{equation}
where the coordinates are understood in the $\mathcal{A}$ basis. The recursive construction follows Eq.~\eqref{eq:general-recursive} with $b =2$.
Thus, a site at generation $n$ is uniquely labeled by the word $\underline{x_1 \dots x_n}$ with $x_j \in \{1,2,3\}$, and the total Hilbert space has dimension $\dim \mathcal{H} = 3^n$.

The NN tight-binding Hamiltonian on the gasket takes the form
\begin{equation}
    H = -\sum_{\langle \vec{r}, \vec{r}' \rangle} t_{\vec{r}, \vec{r}'} \, c^\dagger_{\vec{r}} c_{\vec{r}'} + \mathrm{h.c.},
\end{equation}
where the sum runs over pairs of sites connected by an edge of the triangular lattice. In the present formalism, the Hamiltonian is obtained by imposing the right boundary condition
\begin{equation}
\begin{aligned}
   -t_1 \ket{(1,0)} - t_2 \ket{(0,1)} - t_3 \ket{(-1,1)}  \\ 
    - t_1^* \ket{(-1,0)} - t_2^* \ket{(0,-1)} - t_3^* \ket{(1,-1)},
\end{aligned}
\end{equation}
where $t_1$, $t_2$, and $t_3$ denote the hopping amplitudes along the three directions of the triangular lattice. 
The propagation of constraints yields an MPO of bond dimension $7$, with allowed virtual bond vectors
\begin{equation}
    \vec{v}_i \in \left\{ \vec{0}, \pm 2^{n-i} \vec{e}_x, \pm 2^{n-i} \vec{e}_y, \pm 2^{n-i} (\vec{e}_x - \vec{e}_y) \right\}.
\end{equation}

Similarly to the Sierpiński carpet, it is naturally embedded in the Hilbert space of the full triangular lattice introduced in the main text by adding the extra state $\vec{u}_{4,1} = (1,1)$.
The projector onto the gasket subspace is $\mathcal{P}_G = P_G^{\otimes n}$ with $P_G = \mathrm{Id} - \ket{4}\bra{4}$.
In fact, the scale $b=2$ in the construction of the Sierpiński gasket originates from the base-$2$ self-similarity of Pascal's triangle modulo $2$.
By Lucas' theorem, this structure generalizes to Pascal's triangle modulo any prime $p$, suggesting that our MPO framework extends to this family of fractals by taking $b=p$.
\paragraph{Cantor dust}
The two-dimensional Cantor dust is the Cartesian product of two one-dimensional Cantor sets. 
At each generation, each direction is divided into three segments, with only the first and third retained.
Their Cartesian product therefore retains the four corners of the resulting $(3\times3)$ grid, corresponding to the displacement vectors

\begin{equation}
\begin{aligned}
    \vec{u}_{1,1} = (0,0), \qquad & \vec{u}_{2,1} = (2,0), \\
    \vec{u}_{3,1} = (0,2), \qquad & \vec{u}_{4,1} = (2,2).
\end{aligned}
\end{equation}

For the Cantor dust, the scale factor is $b=3$.
\paragraph{Vicsek cross}
The cross form of the Vicsek fractal is generated by dividing a square into a $3\times3$ grid, while keeping the central cell together with the four cells sharing an edge with it.
The $ \{ \vec{u}_{x, 1} \}$ vectors are
\begin{equation}
\begin{aligned}
    \vec{u}_{1,1} = (1,0), \qquad & \vec{u}_{2,1} = (0,1), \qquad \vec{u}_{3,1} = (1,1), \\ 
    \vec{u}_{4,1} = (2,1), \qquad & \vec{u}_{5,1} = (1,2). 
\end{aligned}
\end{equation}

Here, the scale $b$ is also 3.
\paragraph{Unfolded representation of the Koch curve} 
In its conventional planar representation, each line segment in the Koch curve is replaced at every generation by four segments of one third of its length, with relative rotations of $\pm\pi/3$.
These rotations make Eq.~\eqref{eq:general-recursive} insufficient and call for a generalization of the recursive construction to affine transformations.
We instead consider an unfolded representation of the Koch curve, in which the four segments generated at each iteration are arranged along a one-dimensional chain while preserving their recursive ordering.
This representation does not preserve the two-dimensional embedding, and consequently the fractal dimension, of the Koch curve, but retains its hierarchical recursive structure in a form directly compatible with our MPO construction.

The displacement vectors are

\begin{equation}
\begin{aligned}
    \vec{u}_{1,1} = 0, \qquad & \vec{u}_{2,1} = 1, \\
    \vec{u}_{3,1} = 2, \qquad & \vec{u}_{4,1} = 3.
\end{aligned}
\end{equation}

and $b = 4$.

\subsection{Quasiperiodic models} \label{app:Fib}
\paragraph{Complementary tensor representations for the Fibonacci chain}
As discussed in the main text, we can modify the hopping terms to ensure that ghost space is no longer connected with physical space.
It is enough to ensure that the first $\sigma^\pm$ follows a $\vert 0 \rangle$.
The resulting tensor is of dimension $6$:
\begin{equation}
    \tilde{W}_\mathrm{Fibonacci}^{\mathrm{hopping}} = 
    \begin{pmatrix}
        \mathrm{Id} &p_0 & 0 & 0 & 0  & 0  \\
         0  &0 & \sigma^+ & 0 & \sigma^- & 0 \\
         0  &0 & 0 & \sigma^- & 0 & 0 \\
         0  &0 & p_0 & 0 & 0 & 0 \\
         0  &0 & 0 & 0 & 0 & \sigma^+ \\
         0  &0 & 0 & 0 & p_0 & 0
    \end{pmatrix},
\end{equation}
with left boundary condition $(1, 1, 0, \dots)$.
We use this form of the hopping in our simulations.
Anisotropic hopping is ensured by replacing $p_0 \rightarrow -t_{00} p_0$,  $\sigma^{\pm} \rightarrow -t_{01} \sigma^{\pm}$ (on the first line) and $\sigma^{\pm} \rightarrow -t_{10} \sigma^{\pm}$ (on the other line) in the last tensor, with right boundary condition $(0, 0, 1, 1, 1, 1)$.
Here $t_{\alpha \beta}$ describes hopping between $\alpha \beta$ orbitals, reading left to right.

\section{Bond dimension evolution} \label{app:convergence}
\begin{figure}
    \centering
    \includegraphics[width=\linewidth]{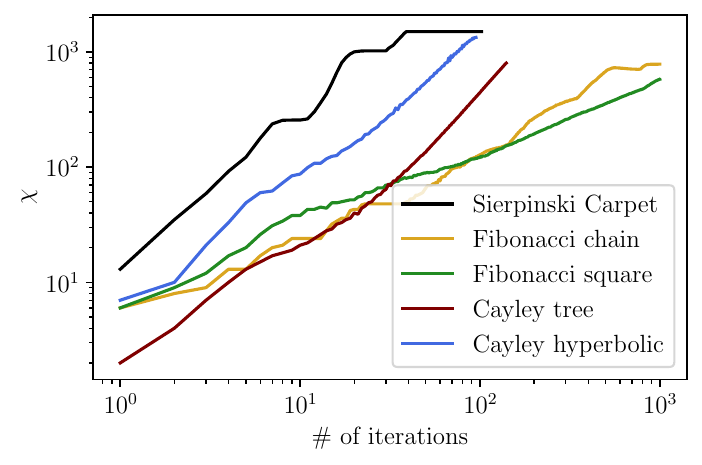}
    \caption{Evolution of the bond dimension of the Chebyshev MPO as function of the iteration, for the different examples in this paper.
 }
    \label{fig:bonddimensions}
\end{figure}

Fig.~\ref{fig:bonddimensions} summarizes the evolution of the bond dimensions of the Chebyshev MPOs as a function of the number of iterations.
All models follow the same iterative procedure, with a truncation of the Chebyshev MPO at each iteration, with a cutoff at $10^{-14}$.
Despite similar original MPO bond dimensions, the growth of the MPO bond dimension differ significantly.
In general, the more extended wavefunctions tend to also lead to a more complex representation of the Chebyshev MPOs.
The effective local Hilbert space, even after splitting sites to recover effective two-level systems, appears to be a major issue.
A careful analysis and understanding of the reasons for these various behaviours is left for future work.

\bibliography{bibliography}

%apsrev4-2.bst 2019-01-14 (MD) hand-edited version of apsrev4-1.bst
%Control: key (0)
%Control: author (8) initials jnrlst
%Control: editor formatted (1) identically to author
%Control: production of article title (0) allowed
%Control: page (0) single
%Control: year (1) truncated
%Control: production of eprint (0) enabled
\begin{thebibliography}{110}%
\makeatletter
\providecommand \@ifxundefined [1]{%
 \@ifx{#1\undefined}
}%
\providecommand \@ifnum [1]{%
 \ifnum #1\expandafter \@firstoftwo
 \else \expandafter \@secondoftwo
 \fi
}%
\providecommand \@ifx [1]{%
 \ifx #1\expandafter \@firstoftwo
 \else \expandafter \@secondoftwo
 \fi
}%
\providecommand \natexlab [1]{#1}%
\providecommand \enquote  [1]{``#1''}%
\providecommand \bibnamefont  [1]{#1}%
\providecommand \bibfnamefont [1]{#1}%
\providecommand \citenamefont [1]{#1}%
\providecommand \href@noop [0]{\@secondoftwo}%
\providecommand \href [0]{\begingroup \@sanitize@url \@href}%
\providecommand \@href[1]{\@@startlink{#1}\@@href}%
\providecommand \@@href[1]{\endgroup#1\@@endlink}%
\providecommand \@sanitize@url [0]{\catcode `\\12\catcode `\$12\catcode `\&12\catcode `\#12\catcode `\^12\catcode `\_12\catcode `\%12\relax}%
\providecommand \@@startlink[1]{}%
\providecommand \@@endlink[0]{}%
\providecommand \url  [0]{\begingroup\@sanitize@url \@url }%
\providecommand \@url [1]{\endgroup\@href {#1}{\urlprefix }}%
\providecommand \urlprefix  [0]{URL }%
\providecommand \Eprint [0]{\href }%
\providecommand \doibase [0]{https://doi.org/}%
\providecommand \selectlanguage [0]{\@gobble}%
\providecommand \bibinfo  [0]{\@secondoftwo}%
\providecommand \bibfield  [0]{\@secondoftwo}%
\providecommand \translation [1]{[#1]}%
\providecommand \BibitemOpen [0]{}%
\providecommand \bibitemStop [0]{}%
\providecommand \bibitemNoStop [0]{.\EOS\space}%
\providecommand \EOS [0]{\spacefactor3000\relax}%
\providecommand \BibitemShut  [1]{\csname bibitem#1\endcsname}%
\let\auto@bib@innerbib\@empty
%</preamble>
\bibitem [{\citenamefont {Yao}\ \emph {et~al.}(2023)\citenamefont {Yao}, \citenamefont {Yang}, \citenamefont {Iliasov}, \citenamefont {Katsnelson},\ and\ \citenamefont {Yuan}}]{PhysRevB.107.115424}%
  \BibitemOpen
  \bibfield  {author} {\bibinfo {author} {\bibfnamefont {Q.}~\bibnamefont {Yao}}, \bibinfo {author} {\bibfnamefont {X.}~\bibnamefont {Yang}}, \bibinfo {author} {\bibfnamefont {A.~A.}\ \bibnamefont {Iliasov}}, \bibinfo {author} {\bibfnamefont {M.~I.}\ \bibnamefont {Katsnelson}},\ and\ \bibinfo {author} {\bibfnamefont {S.}~\bibnamefont {Yuan}},\ }\bibfield  {title} {\bibinfo {title} {Energy-level statistics in planar fractal tight-binding models},\ }\href {https://doi.org/10.1103/PhysRevB.107.115424} {\bibfield  {journal} {\bibinfo  {journal} {Phys. Rev. B}\ }\textbf {\bibinfo {volume} {107}},\ \bibinfo {pages} {115424} (\bibinfo {year} {2023})}\BibitemShut {NoStop}%
\bibitem [{\citenamefont {Pai}\ and\ \citenamefont {Prem}(2019)}]{PhysRevB.100.155135}%
  \BibitemOpen
  \bibfield  {author} {\bibinfo {author} {\bibfnamefont {S.}~\bibnamefont {Pai}}\ and\ \bibinfo {author} {\bibfnamefont {A.}~\bibnamefont {Prem}},\ }\bibfield  {title} {\bibinfo {title} {Topological states on fractal lattices},\ }\href {https://doi.org/10.1103/PhysRevB.100.155135} {\bibfield  {journal} {\bibinfo  {journal} {Phys. Rev. B}\ }\textbf {\bibinfo {volume} {100}},\ \bibinfo {pages} {155135} (\bibinfo {year} {2019})}\BibitemShut {NoStop}%
\bibitem [{\citenamefont {Manna}\ \emph {et~al.}(2022)\citenamefont {Manna}, \citenamefont {Nandy},\ and\ \citenamefont {Roy}}]{PhysRevB.105.L201301}%
  \BibitemOpen
  \bibfield  {author} {\bibinfo {author} {\bibfnamefont {S.}~\bibnamefont {Manna}}, \bibinfo {author} {\bibfnamefont {S.}~\bibnamefont {Nandy}},\ and\ \bibinfo {author} {\bibfnamefont {B.}~\bibnamefont {Roy}},\ }\bibfield  {title} {\bibinfo {title} {Higher-order topological phases on fractal lattices},\ }\href {https://doi.org/10.1103/PhysRevB.105.L201301} {\bibfield  {journal} {\bibinfo  {journal} {Phys. Rev. B}\ }\textbf {\bibinfo {volume} {105}},\ \bibinfo {pages} {L201301} (\bibinfo {year} {2022})}\BibitemShut {NoStop}%
\bibitem [{\citenamefont {Salib}\ \emph {et~al.}(2024)\citenamefont {Salib}, \citenamefont {Mains},\ and\ \citenamefont {Roy}}]{PhysRevB.110.L241302}%
  \BibitemOpen
  \bibfield  {author} {\bibinfo {author} {\bibfnamefont {D.~J.}\ \bibnamefont {Salib}}, \bibinfo {author} {\bibfnamefont {A.~J.}\ \bibnamefont {Mains}},\ and\ \bibinfo {author} {\bibfnamefont {B.}~\bibnamefont {Roy}},\ }\bibfield  {title} {\bibinfo {title} {Topological insulators on fractal lattices: A general principle of construction},\ }\href {https://doi.org/10.1103/PhysRevB.110.L241302} {\bibfield  {journal} {\bibinfo  {journal} {Phys. Rev. B}\ }\textbf {\bibinfo {volume} {110}},\ \bibinfo {pages} {L241302} (\bibinfo {year} {2024})}\BibitemShut {NoStop}%
\bibitem [{\citenamefont {Brzezi\ifmmode~\acute{n}\else \'{n}\fi{}ska}\ \emph {et~al.}(2018)\citenamefont {Brzezi\ifmmode~\acute{n}\else \'{n}\fi{}ska}, \citenamefont {Cook},\ and\ \citenamefont {Neupert}}]{PhysRevB.98.205116}%
  \BibitemOpen
  \bibfield  {author} {\bibinfo {author} {\bibfnamefont {M.}~\bibnamefont {Brzezi\ifmmode~\acute{n}\else \'{n}\fi{}ska}}, \bibinfo {author} {\bibfnamefont {A.~M.}\ \bibnamefont {Cook}},\ and\ \bibinfo {author} {\bibfnamefont {T.}~\bibnamefont {Neupert}},\ }\bibfield  {title} {\bibinfo {title} {Topology in the sierpi\ifmmode \acute{n}\else \'{n}\fi{}ski-hofstadter problem},\ }\href {https://doi.org/10.1103/PhysRevB.98.205116} {\bibfield  {journal} {\bibinfo  {journal} {Phys. Rev. B}\ }\textbf {\bibinfo {volume} {98}},\ \bibinfo {pages} {205116} (\bibinfo {year} {2018})}\BibitemShut {NoStop}%
\bibitem [{\citenamefont {Lage}\ and\ \citenamefont {Latgé}(2026)}]{lage2026chiralsymmetrybreakingdegeneracykoch}%
  \BibitemOpen
  \bibfield  {author} {\bibinfo {author} {\bibfnamefont {L.~L.}\ \bibnamefont {Lage}}\ and\ \bibinfo {author} {\bibfnamefont {A.}~\bibnamefont {Latgé}},\ }\href {https://arxiv.org/abs/2608.30007} {\bibinfo {title} {Chiral-symmetry breaking and degeneracy in koch fractal geometries}} (\bibinfo {year} {2026}),\ \Eprint {https://arxiv.org/abs/2608.30007} {arXiv:2608.30007} \BibitemShut {NoStop}%
\bibitem [{\citenamefont {Fremling}\ \emph {et~al.}(2020)\citenamefont {Fremling}, \citenamefont {van Hooft}, \citenamefont {Smith},\ and\ \citenamefont {Fritz}}]{PhysRevResearch.2.013044}%
  \BibitemOpen
  \bibfield  {author} {\bibinfo {author} {\bibfnamefont {M.}~\bibnamefont {Fremling}}, \bibinfo {author} {\bibfnamefont {M.}~\bibnamefont {van Hooft}}, \bibinfo {author} {\bibfnamefont {C.~M.}\ \bibnamefont {Smith}},\ and\ \bibinfo {author} {\bibfnamefont {L.}~\bibnamefont {Fritz}},\ }\bibfield  {title} {\bibinfo {title} {Existence of robust edge currents in sierpi\ifmmode \acute{n}\else \'{n}\fi{}ski fractals},\ }\href {https://doi.org/10.1103/PhysRevResearch.2.013044} {\bibfield  {journal} {\bibinfo  {journal} {Phys. Rev. Res.}\ }\textbf {\bibinfo {volume} {2}},\ \bibinfo {pages} {013044} (\bibinfo {year} {2020})}\BibitemShut {NoStop}%
\bibitem [{\citenamefont {Iliasov}\ \emph {et~al.}(2020)\citenamefont {Iliasov}, \citenamefont {Katsnelson},\ and\ \citenamefont {Yuan}}]{PhysRevB.101.045413}%
  \BibitemOpen
  \bibfield  {author} {\bibinfo {author} {\bibfnamefont {A.~A.}\ \bibnamefont {Iliasov}}, \bibinfo {author} {\bibfnamefont {M.~I.}\ \bibnamefont {Katsnelson}},\ and\ \bibinfo {author} {\bibfnamefont {S.}~\bibnamefont {Yuan}},\ }\bibfield  {title} {\bibinfo {title} {Hall conductivity of a sierpi\ifmmode \acute{n}\else \'{n}\fi{}ski carpet},\ }\href {https://doi.org/10.1103/PhysRevB.101.045413} {\bibfield  {journal} {\bibinfo  {journal} {Phys. Rev. B}\ }\textbf {\bibinfo {volume} {101}},\ \bibinfo {pages} {045413} (\bibinfo {year} {2020})}\BibitemShut {NoStop}%
\bibitem [{\citenamefont {Bouzerar}\ and\ \citenamefont {Mayou}(2020)}]{PhysRevResearch.2.033063}%
  \BibitemOpen
  \bibfield  {author} {\bibinfo {author} {\bibfnamefont {G.}~\bibnamefont {Bouzerar}}\ and\ \bibinfo {author} {\bibfnamefont {D.}~\bibnamefont {Mayou}},\ }\bibfield  {title} {\bibinfo {title} {Quantum transport in self-similar graphene carpets},\ }\href {https://doi.org/10.1103/PhysRevResearch.2.033063} {\bibfield  {journal} {\bibinfo  {journal} {Phys. Rev. Res.}\ }\textbf {\bibinfo {volume} {2}},\ \bibinfo {pages} {033063} (\bibinfo {year} {2020})}\BibitemShut {NoStop}%
\bibitem [{\citenamefont {Rojo-Franc{\`a}s}\ \emph {et~al.}(2024)\citenamefont {Rojo-Franc{\`a}s}, \citenamefont {Pansari}, \citenamefont {Bhattacharya}, \citenamefont {Juli{\'a}-D{\'i}az},\ and\ \citenamefont {Grass}}]{Rojo-Francas2024}%
  \BibitemOpen
  \bibfield  {author} {\bibinfo {author} {\bibfnamefont {A.}~\bibnamefont {Rojo-Franc{\`a}s}}, \bibinfo {author} {\bibfnamefont {P.}~\bibnamefont {Pansari}}, \bibinfo {author} {\bibfnamefont {U.}~\bibnamefont {Bhattacharya}}, \bibinfo {author} {\bibfnamefont {B.}~\bibnamefont {Juli{\'a}-D{\'i}az}},\ and\ \bibinfo {author} {\bibfnamefont {T.}~\bibnamefont {Grass}},\ }\bibfield  {title} {\bibinfo {title} {Anomalous quantum transport in fractal lattices},\ }\href {https://doi.org/10.1038/s42005-024-01747-x} {\bibfield  {journal} {\bibinfo  {journal} {Communications Physics}\ }\textbf {\bibinfo {volume} {7}},\ \bibinfo {pages} {259} (\bibinfo {year} {2024})}\BibitemShut {NoStop}%
\bibitem [{\citenamefont {Zhou}\ and\ \citenamefont {Ye}(2024)}]{PhysRevResearch.6.043145}%
  \BibitemOpen
  \bibfield  {author} {\bibinfo {author} {\bibfnamefont {Y.}~\bibnamefont {Zhou}}\ and\ \bibinfo {author} {\bibfnamefont {P.}~\bibnamefont {Ye}},\ }\bibfield  {title} {\bibinfo {title} {Entanglement fractalization},\ }\href {https://doi.org/10.1103/PhysRevResearch.6.043145} {\bibfield  {journal} {\bibinfo  {journal} {Phys. Rev. Res.}\ }\textbf {\bibinfo {volume} {6}},\ \bibinfo {pages} {043145} (\bibinfo {year} {2024})}\BibitemShut {NoStop}%
\bibitem [{\citenamefont {Kohmoto}\ \emph {et~al.}(1987)\citenamefont {Kohmoto}, \citenamefont {Sutherland},\ and\ \citenamefont {Tang}}]{PhysRevB.35.1020}%
  \BibitemOpen
  \bibfield  {author} {\bibinfo {author} {\bibfnamefont {M.}~\bibnamefont {Kohmoto}}, \bibinfo {author} {\bibfnamefont {B.}~\bibnamefont {Sutherland}},\ and\ \bibinfo {author} {\bibfnamefont {C.}~\bibnamefont {Tang}},\ }\bibfield  {title} {\bibinfo {title} {Critical wave functions and a cantor-set spectrum of a one-dimensional quasicrystal model},\ }\href {https://doi.org/10.1103/PhysRevB.35.1020} {\bibfield  {journal} {\bibinfo  {journal} {Phys. Rev. B}\ }\textbf {\bibinfo {volume} {35}},\ \bibinfo {pages} {1020} (\bibinfo {year} {1987})}\BibitemShut {NoStop}%
\bibitem [{\citenamefont {Jagannathan}(2021)}]{RevModPhys.93.045001}%
  \BibitemOpen
  \bibfield  {author} {\bibinfo {author} {\bibfnamefont {A.}~\bibnamefont {Jagannathan}},\ }\bibfield  {title} {\bibinfo {title} {The fibonacci quasicrystal: Case study of hidden dimensions and multifractality},\ }\href {https://doi.org/10.1103/RevModPhys.93.045001} {\bibfield  {journal} {\bibinfo  {journal} {Rev. Mod. Phys.}\ }\textbf {\bibinfo {volume} {93}},\ \bibinfo {pages} {045001} (\bibinfo {year} {2021})}\BibitemShut {NoStop}%
\bibitem [{\citenamefont {Macé}\ \emph {et~al.}(2019)\citenamefont {Macé}, \citenamefont {Laflorencie},\ and\ \citenamefont {Alet}}]{SciPostPhys.6.4.050}%
  \BibitemOpen
  \bibfield  {author} {\bibinfo {author} {\bibfnamefont {N.}~\bibnamefont {Macé}}, \bibinfo {author} {\bibfnamefont {N.}~\bibnamefont {Laflorencie}},\ and\ \bibinfo {author} {\bibfnamefont {F.}~\bibnamefont {Alet}},\ }\bibfield  {title} {\bibinfo {title} {Many-body localization in a quasiperiodic fibonacci chain},\ }\href {https://doi.org/10.21468/SciPostPhys.6.4.050} {\bibfield  {journal} {\bibinfo  {journal} {SciPost Phys.}\ }\textbf {\bibinfo {volume} {6}},\ \bibinfo {pages} {050} (\bibinfo {year} {2019})}\BibitemShut {NoStop}%
\bibitem [{\citenamefont {Ostilli}(2012)}]{OSTILLI20123417}%
  \BibitemOpen
  \bibfield  {author} {\bibinfo {author} {\bibfnamefont {M.}~\bibnamefont {Ostilli}},\ }\bibfield  {title} {\bibinfo {title} {Cayley trees and bethe lattices: A concise analysis for mathematicians and physicists},\ }\href {https://doi.org/https://doi.org/10.1016/j.physa.2012.01.038} {\bibfield  {journal} {\bibinfo  {journal} {Physica A: Statistical Mechanics and its Applications}\ }\textbf {\bibinfo {volume} {391}},\ \bibinfo {pages} {3417} (\bibinfo {year} {2012})}\BibitemShut {NoStop}%
\bibitem [{\citenamefont {Zirnbauer}(1986)}]{zirnbauer_localization_1986-1}%
  \BibitemOpen
  \bibfield  {author} {\bibinfo {author} {\bibfnamefont {M.~R.}\ \bibnamefont {Zirnbauer}},\ }\bibfield  {title} {\bibinfo {title} {Localization transition on the {{Bethe}} lattice},\ }\href {https://doi.org/10.1103/PhysRevB.34.6394} {\bibfield  {journal} {\bibinfo  {journal} {Physical Review B}\ }\textbf {\bibinfo {volume} {34}},\ \bibinfo {pages} {6394} (\bibinfo {year} {1986})}\BibitemShut {NoStop}%
\bibitem [{\citenamefont {Mirlin}\ and\ \citenamefont {Fyodorov}(1994)}]{mirlin_distribution_1994-1}%
  \BibitemOpen
  \bibfield  {author} {\bibinfo {author} {\bibfnamefont {A.~D.}\ \bibnamefont {Mirlin}}\ and\ \bibinfo {author} {\bibfnamefont {Y.~V.}\ \bibnamefont {Fyodorov}},\ }\bibfield  {title} {\bibinfo {title} {Distribution of local densities of states, order parameter function, and critical behavior near the {{Anderson}} transition},\ }\href {https://doi.org/10.1103/PhysRevLett.72.526} {\bibfield  {journal} {\bibinfo  {journal} {Physical Review Letters}\ }\textbf {\bibinfo {volume} {72}},\ \bibinfo {pages} {526} (\bibinfo {year} {1994})}\BibitemShut {NoStop}%
\bibitem [{\citenamefont {Parisi}\ \emph {et~al.}(2019)\citenamefont {Parisi}, \citenamefont {Pascazio}, \citenamefont {Pietracaprina}, \citenamefont {Ros},\ and\ \citenamefont {Scardicchio}}]{parisi_anderson_2019}%
  \BibitemOpen
  \bibfield  {author} {\bibinfo {author} {\bibfnamefont {G.}~\bibnamefont {Parisi}}, \bibinfo {author} {\bibfnamefont {S.}~\bibnamefont {Pascazio}}, \bibinfo {author} {\bibfnamefont {F.}~\bibnamefont {Pietracaprina}}, \bibinfo {author} {\bibfnamefont {V.}~\bibnamefont {Ros}},\ and\ \bibinfo {author} {\bibfnamefont {A.}~\bibnamefont {Scardicchio}},\ }\bibfield  {title} {\bibinfo {title} {Anderson transition on the {{Bethe}} lattice: An approach with real energies},\ }\href {https://doi.org/10.1088/1751-8121/ab56e8} {\bibfield  {journal} {\bibinfo  {journal} {Journal of Physics A: Mathematical and Theoretical}\ }\textbf {\bibinfo {volume} {53}},\ \bibinfo {pages} {014003} (\bibinfo {year} {2019})}\BibitemShut {NoStop}%
\bibitem [{\citenamefont {Monthus}\ and\ \citenamefont {Garel}(2011)}]{monthus_anderson_2011}%
  \BibitemOpen
  \bibfield  {author} {\bibinfo {author} {\bibfnamefont {C.}~\bibnamefont {Monthus}}\ and\ \bibinfo {author} {\bibfnamefont {T.}~\bibnamefont {Garel}},\ }\bibfield  {title} {\bibinfo {title} {Anderson localization on the {{Cayley}} tree: Multifractal statistics of the transmission at criticality and off criticality},\ }\href {https://doi.org/10.1088/1751-8113/44/14/145001} {\bibfield  {journal} {\bibinfo  {journal} {Journal of Physics A: Mathematical and Theoretical}\ }\textbf {\bibinfo {volume} {44}},\ \bibinfo {pages} {145001} (\bibinfo {year} {2011})}\BibitemShut {NoStop}%
\bibitem [{\citenamefont {Tikhonov}\ and\ \citenamefont {Mirlin}(2016)}]{tikhonov_fractality_2016}%
  \BibitemOpen
  \bibfield  {author} {\bibinfo {author} {\bibfnamefont {K.~S.}\ \bibnamefont {Tikhonov}}\ and\ \bibinfo {author} {\bibfnamefont {A.~D.}\ \bibnamefont {Mirlin}},\ }\bibfield  {title} {\bibinfo {title} {Fractality of wave functions on a {{Cayley}} tree: {{Difference}} between tree and locally treelike graph without boundary},\ }\href {https://doi.org/10.1103/PhysRevB.94.184203} {\bibfield  {journal} {\bibinfo  {journal} {Physical Review B}\ }\textbf {\bibinfo {volume} {94}},\ \bibinfo {pages} {184203} (\bibinfo {year} {2016})}\BibitemShut {NoStop}%
\bibitem [{\citenamefont {Sonner}\ \emph {et~al.}(2017)\citenamefont {Sonner}, \citenamefont {Tikhonov},\ and\ \citenamefont {Mirlin}}]{sonner_multifractality_2017-1}%
  \BibitemOpen
  \bibfield  {author} {\bibinfo {author} {\bibfnamefont {M.}~\bibnamefont {Sonner}}, \bibinfo {author} {\bibfnamefont {K.~S.}\ \bibnamefont {Tikhonov}},\ and\ \bibinfo {author} {\bibfnamefont {A.~D.}\ \bibnamefont {Mirlin}},\ }\bibfield  {title} {\bibinfo {title} {Multifractality of wave functions on a {{Cayley}} tree: {{From}} root to leaves},\ }\href {https://doi.org/10.1103/PhysRevB.96.214204} {\bibfield  {journal} {\bibinfo  {journal} {Physical Review B}\ }\textbf {\bibinfo {volume} {96}},\ \bibinfo {pages} {214204} (\bibinfo {year} {2017})}\BibitemShut {NoStop}%
\bibitem [{\citenamefont {Maciejko}\ and\ \citenamefont {Rayan}(2021)}]{doi:10.1126/sciadv.abe9170}%
  \BibitemOpen
  \bibfield  {author} {\bibinfo {author} {\bibfnamefont {J.}~\bibnamefont {Maciejko}}\ and\ \bibinfo {author} {\bibfnamefont {S.}~\bibnamefont {Rayan}},\ }\bibfield  {title} {\bibinfo {title} {Hyperbolic band theory},\ }\href {https://doi.org/10.1126/sciadv.abe9170} {\bibfield  {journal} {\bibinfo  {journal} {Science Advances}\ }\textbf {\bibinfo {volume} {7}},\ \bibinfo {pages} {eabe9170} (\bibinfo {year} {2021})}\BibitemShut {NoStop}%
\bibitem [{\citenamefont {Urwyler}\ \emph {et~al.}(2022)\citenamefont {Urwyler}, \citenamefont {Lenggenhager}, \citenamefont {Boettcher}, \citenamefont {Thomale}, \citenamefont {Neupert},\ and\ \citenamefont {Bzdu\ifmmode~\check{s}\else \v{s}\fi{}ek}}]{PhysRevLett.129.246402}%
  \BibitemOpen
  \bibfield  {author} {\bibinfo {author} {\bibfnamefont {D.~M.}\ \bibnamefont {Urwyler}}, \bibinfo {author} {\bibfnamefont {P.~M.}\ \bibnamefont {Lenggenhager}}, \bibinfo {author} {\bibfnamefont {I.}~\bibnamefont {Boettcher}}, \bibinfo {author} {\bibfnamefont {R.}~\bibnamefont {Thomale}}, \bibinfo {author} {\bibfnamefont {T.}~\bibnamefont {Neupert}},\ and\ \bibinfo {author} {\bibfnamefont {T.~c.~v.}\ \bibnamefont {Bzdu\ifmmode~\check{s}\else \v{s}\fi{}ek}},\ }\bibfield  {title} {\bibinfo {title} {Hyperbolic topological band insulators},\ }\href {https://doi.org/10.1103/PhysRevLett.129.246402} {\bibfield  {journal} {\bibinfo  {journal} {Phys. Rev. Lett.}\ }\textbf {\bibinfo {volume} {129}},\ \bibinfo {pages} {246402} (\bibinfo {year} {2022})}\BibitemShut {NoStop}%
\bibitem [{\citenamefont {Duss}\ \emph {et~al.}(2026)\citenamefont {Duss}, \citenamefont {Iliasov},\ and\ \citenamefont {Bzdu\ifmmode~\check{s}\else \v{s}\fi{}ek}}]{t8gg-xqls}%
  \BibitemOpen
  \bibfield  {author} {\bibinfo {author} {\bibfnamefont {W.~P.}\ \bibnamefont {Duss}}, \bibinfo {author} {\bibfnamefont {A.}~\bibnamefont {Iliasov}},\ and\ \bibinfo {author} {\bibfnamefont {T.~c.~v.}\ \bibnamefont {Bzdu\ifmmode~\check{s}\else \v{s}\fi{}ek}},\ }\bibfield  {title} {\bibinfo {title} {Topological states and flat bands in exactly solvable decorated cayley trees},\ }\href {https://doi.org/10.1103/t8gg-xqls} {\bibfield  {journal} {\bibinfo  {journal} {Phys. Rev. Res.}\ }\textbf {\bibinfo {volume} {8}},\ \bibinfo {pages} {033036} (\bibinfo {year} {2026})}\BibitemShut {NoStop}%
\bibitem [{\citenamefont {Yu}\ \emph {et~al.}(2020)\citenamefont {Yu}, \citenamefont {Piao},\ and\ \citenamefont {Park}}]{PhysRevLett.125.053901}%
  \BibitemOpen
  \bibfield  {author} {\bibinfo {author} {\bibfnamefont {S.}~\bibnamefont {Yu}}, \bibinfo {author} {\bibfnamefont {X.}~\bibnamefont {Piao}},\ and\ \bibinfo {author} {\bibfnamefont {N.}~\bibnamefont {Park}},\ }\bibfield  {title} {\bibinfo {title} {Topological hyperbolic lattices},\ }\href {https://doi.org/10.1103/PhysRevLett.125.053901} {\bibfield  {journal} {\bibinfo  {journal} {Phys. Rev. Lett.}\ }\textbf {\bibinfo {volume} {125}},\ \bibinfo {pages} {053901} (\bibinfo {year} {2020})}\BibitemShut {NoStop}%
\bibitem [{\citenamefont {Bzdu\ifmmode~\check{s}\else \v{s}\fi{}ek}\ and\ \citenamefont {Maciejko}(2022)}]{PhysRevB.106.155146}%
  \BibitemOpen
  \bibfield  {author} {\bibinfo {author} {\bibfnamefont {T.~c.~v.}\ \bibnamefont {Bzdu\ifmmode~\check{s}\else \v{s}\fi{}ek}}\ and\ \bibinfo {author} {\bibfnamefont {J.}~\bibnamefont {Maciejko}},\ }\bibfield  {title} {\bibinfo {title} {Flat bands and band-touching from real-space topology in hyperbolic lattices},\ }\href {https://doi.org/10.1103/PhysRevB.106.155146} {\bibfield  {journal} {\bibinfo  {journal} {Phys. Rev. B}\ }\textbf {\bibinfo {volume} {106}},\ \bibinfo {pages} {155146} (\bibinfo {year} {2022})}\BibitemShut {NoStop}%
\bibitem [{\citenamefont {Xu}\ \emph {et~al.}(2021)\citenamefont {Xu}, \citenamefont {Wang}, \citenamefont {Chen}, \citenamefont {Smith},\ and\ \citenamefont {Jin}}]{Xu2021}%
  \BibitemOpen
  \bibfield  {author} {\bibinfo {author} {\bibfnamefont {X.-Y.}\ \bibnamefont {Xu}}, \bibinfo {author} {\bibfnamefont {X.-W.}\ \bibnamefont {Wang}}, \bibinfo {author} {\bibfnamefont {D.-Y.}\ \bibnamefont {Chen}}, \bibinfo {author} {\bibfnamefont {C.~M.}\ \bibnamefont {Smith}},\ and\ \bibinfo {author} {\bibfnamefont {X.-M.}\ \bibnamefont {Jin}},\ }\bibfield  {title} {\bibinfo {title} {Quantum transport in fractal networks},\ }\href {https://doi.org/10.1038/s41566-021-00845-4} {\bibfield  {journal} {\bibinfo  {journal} {Nature Photonics}\ }\textbf {\bibinfo {volume} {15}},\ \bibinfo {pages} {703} (\bibinfo {year} {2021})}\BibitemShut {NoStop}%
\bibitem [{\citenamefont {Biesenthal}\ \emph {et~al.}(2022)\citenamefont {Biesenthal}, \citenamefont {Maczewsky}, \citenamefont {Yang}, \citenamefont {Kremer}, \citenamefont {Segev}, \citenamefont {Szameit},\ and\ \citenamefont {Heinrich}}]{doi:10.1126/science.abm2842}%
  \BibitemOpen
  \bibfield  {author} {\bibinfo {author} {\bibfnamefont {T.}~\bibnamefont {Biesenthal}}, \bibinfo {author} {\bibfnamefont {L.~J.}\ \bibnamefont {Maczewsky}}, \bibinfo {author} {\bibfnamefont {Z.}~\bibnamefont {Yang}}, \bibinfo {author} {\bibfnamefont {M.}~\bibnamefont {Kremer}}, \bibinfo {author} {\bibfnamefont {M.}~\bibnamefont {Segev}}, \bibinfo {author} {\bibfnamefont {A.}~\bibnamefont {Szameit}},\ and\ \bibinfo {author} {\bibfnamefont {M.}~\bibnamefont {Heinrich}},\ }\bibfield  {title} {\bibinfo {title} {Fractal photonic topological insulators},\ }\href {https://doi.org/10.1126/science.abm2842} {\bibfield  {journal} {\bibinfo  {journal} {Science}\ }\textbf {\bibinfo {volume} {376}},\ \bibinfo {pages} {1114} (\bibinfo {year} {2022})}\BibitemShut {NoStop}%
\bibitem [{\citenamefont {Kempkes}\ \emph {et~al.}(2019)\citenamefont {Kempkes}, \citenamefont {Slot}, \citenamefont {Freeney}, \citenamefont {Zevenhuizen}, \citenamefont {Vanmaekelbergh}, \citenamefont {Swart},\ and\ \citenamefont {Smith}}]{Kempkes2019}%
  \BibitemOpen
  \bibfield  {author} {\bibinfo {author} {\bibfnamefont {S.~N.}\ \bibnamefont {Kempkes}}, \bibinfo {author} {\bibfnamefont {M.~R.}\ \bibnamefont {Slot}}, \bibinfo {author} {\bibfnamefont {S.~E.}\ \bibnamefont {Freeney}}, \bibinfo {author} {\bibfnamefont {S.~J.~M.}\ \bibnamefont {Zevenhuizen}}, \bibinfo {author} {\bibfnamefont {D.}~\bibnamefont {Vanmaekelbergh}}, \bibinfo {author} {\bibfnamefont {I.}~\bibnamefont {Swart}},\ and\ \bibinfo {author} {\bibfnamefont {C.~M.}\ \bibnamefont {Smith}},\ }\bibfield  {title} {\bibinfo {title} {Design and characterization of electrons in a fractal geometry},\ }\href {https://doi.org/10.1038/s41567-018-0328-0} {\bibfield  {journal} {\bibinfo  {journal} {Nature Physics}\ }\textbf {\bibinfo {volume} {15}},\ \bibinfo {pages} {127} (\bibinfo {year} {2019})}\BibitemShut {NoStop}%
\bibitem [{\citenamefont {Verstraten}\ \emph {et~al.}(2025)\citenamefont {Verstraten}, \citenamefont {Knottnerus}, \citenamefont {Tseng}, \citenamefont {Urech}, \citenamefont {do~Espirito~Santo}, \citenamefont {Zampronio}, \citenamefont {Schreck}, \citenamefont {Spreeuw},\ and\ \citenamefont {Smith}}]{verstraten2025controlsinglespinflipsrydberg}%
  \BibitemOpen
  \bibfield  {author} {\bibinfo {author} {\bibfnamefont {R.~C.}\ \bibnamefont {Verstraten}}, \bibinfo {author} {\bibfnamefont {I.~H.~A.}\ \bibnamefont {Knottnerus}}, \bibinfo {author} {\bibfnamefont {Y.~C.}\ \bibnamefont {Tseng}}, \bibinfo {author} {\bibfnamefont {A.}~\bibnamefont {Urech}}, \bibinfo {author} {\bibfnamefont {T.~S.}\ \bibnamefont {do~Espirito~Santo}}, \bibinfo {author} {\bibfnamefont {V.}~\bibnamefont {Zampronio}}, \bibinfo {author} {\bibfnamefont {F.}~\bibnamefont {Schreck}}, \bibinfo {author} {\bibfnamefont {R.~J.~C.}\ \bibnamefont {Spreeuw}},\ and\ \bibinfo {author} {\bibfnamefont {C.~M.}\ \bibnamefont {Smith}},\ }\href {https://arxiv.org/abs/2509.03514} {\bibinfo {title} {Control of single spin-flips in a rydberg atomic fractal}} (\bibinfo {year} {2025}),\ \Eprint {https://arxiv.org/abs/2509.03514} {arXiv:2509.03514} \BibitemShut {NoStop}%
\bibitem [{\citenamefont {Shechtman}\ \emph {et~al.}(1984)\citenamefont {Shechtman}, \citenamefont {Blech}, \citenamefont {Gratias},\ and\ \citenamefont {Cahn}}]{Shechtman1984}%
  \BibitemOpen
  \bibfield  {author} {\bibinfo {author} {\bibfnamefont {D.}~\bibnamefont {Shechtman}}, \bibinfo {author} {\bibfnamefont {I.}~\bibnamefont {Blech}}, \bibinfo {author} {\bibfnamefont {D.}~\bibnamefont {Gratias}},\ and\ \bibinfo {author} {\bibfnamefont {J.~W.}\ \bibnamefont {Cahn}},\ }\bibfield  {title} {\bibinfo {title} {Metallic phase with long-range orientational order and no translational symmetry},\ }\href {https://doi.org/10.1103/PhysRevLett.53.1951} {\bibfield  {journal} {\bibinfo  {journal} {Phys. Rev. Lett.}\ }\textbf {\bibinfo {volume} {53}},\ \bibinfo {pages} {1951} (\bibinfo {year} {1984})}\BibitemShut {NoStop}%
\bibitem [{\citenamefont {Roati}\ \emph {et~al.}(2008)\citenamefont {Roati}, \citenamefont {D'Errico}, \citenamefont {Fallani}, \citenamefont {Fattori}, \citenamefont {Fort}, \citenamefont {Zaccanti}, \citenamefont {Modugno}, \citenamefont {Modugno},\ and\ \citenamefont {Inguscio}}]{Roati2008}%
  \BibitemOpen
  \bibfield  {author} {\bibinfo {author} {\bibfnamefont {G.}~\bibnamefont {Roati}}, \bibinfo {author} {\bibfnamefont {C.}~\bibnamefont {D'Errico}}, \bibinfo {author} {\bibfnamefont {L.}~\bibnamefont {Fallani}}, \bibinfo {author} {\bibfnamefont {M.}~\bibnamefont {Fattori}}, \bibinfo {author} {\bibfnamefont {C.}~\bibnamefont {Fort}}, \bibinfo {author} {\bibfnamefont {M.}~\bibnamefont {Zaccanti}}, \bibinfo {author} {\bibfnamefont {G.}~\bibnamefont {Modugno}}, \bibinfo {author} {\bibfnamefont {M.}~\bibnamefont {Modugno}},\ and\ \bibinfo {author} {\bibfnamefont {M.}~\bibnamefont {Inguscio}},\ }\bibfield  {title} {\bibinfo {title} {Anderson localization of a non-interacting bose--einstein condensate},\ }\href {https://doi.org/10.1038/nature07071} {\bibfield  {journal} {\bibinfo  {journal} {Nature}\ }\textbf {\bibinfo {volume} {453}},\ \bibinfo {pages} {895} (\bibinfo {year} {2008})}\BibitemShut {NoStop}%
\bibitem [{\citenamefont {Schreiber}\ \emph {et~al.}(2015)\citenamefont {Schreiber}, \citenamefont {Hodgman}, \citenamefont {Bordia}, \citenamefont {L{\"u}schen}, \citenamefont {Fischer}, \citenamefont {Vosk}, \citenamefont {Altman}, \citenamefont {Schneider},\ and\ \citenamefont {Bloch}}]{schreiber_observation_2015}%
  \BibitemOpen
  \bibfield  {author} {\bibinfo {author} {\bibfnamefont {M.}~\bibnamefont {Schreiber}}, \bibinfo {author} {\bibfnamefont {S.~S.}\ \bibnamefont {Hodgman}}, \bibinfo {author} {\bibfnamefont {P.}~\bibnamefont {Bordia}}, \bibinfo {author} {\bibfnamefont {H.~P.}\ \bibnamefont {L{\"u}schen}}, \bibinfo {author} {\bibfnamefont {M.~H.}\ \bibnamefont {Fischer}}, \bibinfo {author} {\bibfnamefont {R.}~\bibnamefont {Vosk}}, \bibinfo {author} {\bibfnamefont {E.}~\bibnamefont {Altman}}, \bibinfo {author} {\bibfnamefont {U.}~\bibnamefont {Schneider}},\ and\ \bibinfo {author} {\bibfnamefont {I.}~\bibnamefont {Bloch}},\ }\bibfield  {title} {\bibinfo {title} {Observation of many-body localization of interacting fermions in a quasirandom optical lattice},\ }\href {https://doi.org/10.1126/science.aaa7432} {\bibfield  {journal} {\bibinfo  {journal} {Science}\ }\textbf {\bibinfo {volume} {349}},\ \bibinfo {pages} {842} (\bibinfo {year} {2015})}\BibitemShut {NoStop}%
\bibitem [{\citenamefont {Bordia}\ \emph {et~al.}(2017)\citenamefont {Bordia}, \citenamefont {L\"uschen}, \citenamefont {Scherg}, \citenamefont {Gopalakrishnan}, \citenamefont {Knap}, \citenamefont {Schneider},\ and\ \citenamefont {Bloch}}]{Bordia2017}%
  \BibitemOpen
  \bibfield  {author} {\bibinfo {author} {\bibfnamefont {P.}~\bibnamefont {Bordia}}, \bibinfo {author} {\bibfnamefont {H.}~\bibnamefont {L\"uschen}}, \bibinfo {author} {\bibfnamefont {S.}~\bibnamefont {Scherg}}, \bibinfo {author} {\bibfnamefont {S.}~\bibnamefont {Gopalakrishnan}}, \bibinfo {author} {\bibfnamefont {M.}~\bibnamefont {Knap}}, \bibinfo {author} {\bibfnamefont {U.}~\bibnamefont {Schneider}},\ and\ \bibinfo {author} {\bibfnamefont {I.}~\bibnamefont {Bloch}},\ }\bibfield  {title} {\bibinfo {title} {Probing slow relaxation and many-body localization in two-dimensional quasiperiodic systems},\ }\href {https://doi.org/10.1103/PhysRevX.7.041047} {\bibfield  {journal} {\bibinfo  {journal} {Phys. Rev. X}\ }\textbf {\bibinfo {volume} {7}},\ \bibinfo {pages} {041047} (\bibinfo {year} {2017})}\BibitemShut {NoStop}%
\bibitem [{\citenamefont {Yu}\ \emph {et~al.}(2024)\citenamefont {Yu}, \citenamefont {Bhave}, \citenamefont {Reeve}, \citenamefont {Song},\ and\ \citenamefont {Schneider}}]{yu_observing_2024}%
  \BibitemOpen
  \bibfield  {author} {\bibinfo {author} {\bibfnamefont {J.-C.}\ \bibnamefont {Yu}}, \bibinfo {author} {\bibfnamefont {S.}~\bibnamefont {Bhave}}, \bibinfo {author} {\bibfnamefont {L.}~\bibnamefont {Reeve}}, \bibinfo {author} {\bibfnamefont {B.}~\bibnamefont {Song}},\ and\ \bibinfo {author} {\bibfnamefont {U.}~\bibnamefont {Schneider}},\ }\bibfield  {title} {\bibinfo {title} {Observing the two-dimensional {{Bose}} glass in an optical quasicrystal},\ }\href {https://doi.org/10.1038/s41586-024-07875-2} {\bibfield  {journal} {\bibinfo  {journal} {Nature}\ }\textbf {\bibinfo {volume} {633}},\ \bibinfo {pages} {338} (\bibinfo {year} {2024})}\BibitemShut {NoStop}%
\bibitem [{\citenamefont {Verbin}\ \emph {et~al.}(2013)\citenamefont {Verbin}, \citenamefont {Zilberberg}, \citenamefont {Kraus}, \citenamefont {Lahini},\ and\ \citenamefont {Silberberg}}]{verbin_observation_2013}%
  \BibitemOpen
  \bibfield  {author} {\bibinfo {author} {\bibfnamefont {M.}~\bibnamefont {Verbin}}, \bibinfo {author} {\bibfnamefont {O.}~\bibnamefont {Zilberberg}}, \bibinfo {author} {\bibfnamefont {Y.~E.}\ \bibnamefont {Kraus}}, \bibinfo {author} {\bibfnamefont {Y.}~\bibnamefont {Lahini}},\ and\ \bibinfo {author} {\bibfnamefont {Y.}~\bibnamefont {Silberberg}},\ }\bibfield  {title} {\bibinfo {title} {Observation of topological phase transitions in photonic quasicrystals},\ }\href {https://doi.org/10.1103/PhysRevLett.110.076403} {\bibfield  {journal} {\bibinfo  {journal} {Phys. Rev. Lett.}\ }\textbf {\bibinfo {volume} {110}},\ \bibinfo {pages} {076403} (\bibinfo {year} {2013})}\BibitemShut {NoStop}%
\bibitem [{\citenamefont {Boguslawski}\ \emph {et~al.}(2016)\citenamefont {Boguslawski}, \citenamefont {Lu{\v c}i{\'c}}, \citenamefont {Diebel}, \citenamefont {Timotijevi{\'c}}, \citenamefont {Denz},\ and\ \citenamefont {Savi{\'c}}}]{boguslawski_light_2016}%
  \BibitemOpen
  \bibfield  {author} {\bibinfo {author} {\bibfnamefont {M.}~\bibnamefont {Boguslawski}}, \bibinfo {author} {\bibfnamefont {N.~M.}\ \bibnamefont {Lu{\v c}i{\'c}}}, \bibinfo {author} {\bibfnamefont {F.}~\bibnamefont {Diebel}}, \bibinfo {author} {\bibfnamefont {D.~V.}\ \bibnamefont {Timotijevi{\'c}}}, \bibinfo {author} {\bibfnamefont {C.}~\bibnamefont {Denz}},\ and\ \bibinfo {author} {\bibfnamefont {D.~M.~J.}\ \bibnamefont {Savi{\'c}}},\ }\bibfield  {title} {\bibinfo {title} {Light localization in optically induced deterministic aperiodic {{Fibonacci}} lattices},\ }\href {https://doi.org/10.1364/OPTICA.3.000711} {\bibfield  {journal} {\bibinfo  {journal} {Optica}\ }\textbf {\bibinfo {volume} {3}},\ \bibinfo {pages} {711} (\bibinfo {year} {2016})}\BibitemShut {NoStop}%
\bibitem [{\citenamefont {Koll{\'a}r}\ \emph {et~al.}(2019)\citenamefont {Koll{\'a}r}, \citenamefont {Fitzpatrick},\ and\ \citenamefont {Houck}}]{kollar_hyperbolic_2019}%
  \BibitemOpen
  \bibfield  {author} {\bibinfo {author} {\bibfnamefont {A.~J.}\ \bibnamefont {Koll{\'a}r}}, \bibinfo {author} {\bibfnamefont {M.}~\bibnamefont {Fitzpatrick}},\ and\ \bibinfo {author} {\bibfnamefont {A.~A.}\ \bibnamefont {Houck}},\ }\bibfield  {title} {\bibinfo {title} {Hyperbolic lattices in circuit quantum electrodynamics},\ }\href {https://doi.org/10.1038/s41586-019-1348-3} {\bibfield  {journal} {\bibinfo  {journal} {Nature}\ }\textbf {\bibinfo {volume} {571}},\ \bibinfo {pages} {45} (\bibinfo {year} {2019})}\BibitemShut {NoStop}%
\bibitem [{\citenamefont {Lenggenhager}\ \emph {et~al.}(2022)\citenamefont {Lenggenhager}, \citenamefont {Stegmaier}, \citenamefont {Upreti}, \citenamefont {Hofmann}, \citenamefont {Helbig}, \citenamefont {Vollhardt}, \citenamefont {Greiter}, \citenamefont {Lee}, \citenamefont {Imhof}, \citenamefont {Brand}, \citenamefont {Kie{\ss}ling}, \citenamefont {Boettcher}, \citenamefont {Neupert}, \citenamefont {Thomale},\ and\ \citenamefont {Bzdu{\v{s}}ek}}]{Lenggenhager2022}%
  \BibitemOpen
  \bibfield  {author} {\bibinfo {author} {\bibfnamefont {P.~M.}\ \bibnamefont {Lenggenhager}}, \bibinfo {author} {\bibfnamefont {A.}~\bibnamefont {Stegmaier}}, \bibinfo {author} {\bibfnamefont {L.~K.}\ \bibnamefont {Upreti}}, \bibinfo {author} {\bibfnamefont {T.}~\bibnamefont {Hofmann}}, \bibinfo {author} {\bibfnamefont {T.}~\bibnamefont {Helbig}}, \bibinfo {author} {\bibfnamefont {A.}~\bibnamefont {Vollhardt}}, \bibinfo {author} {\bibfnamefont {M.}~\bibnamefont {Greiter}}, \bibinfo {author} {\bibfnamefont {C.~H.}\ \bibnamefont {Lee}}, \bibinfo {author} {\bibfnamefont {S.}~\bibnamefont {Imhof}}, \bibinfo {author} {\bibfnamefont {H.}~\bibnamefont {Brand}}, \bibinfo {author} {\bibfnamefont {T.}~\bibnamefont {Kie{\ss}ling}}, \bibinfo {author} {\bibfnamefont {I.}~\bibnamefont {Boettcher}}, \bibinfo {author} {\bibfnamefont {T.}~\bibnamefont {Neupert}}, \bibinfo {author} {\bibfnamefont {R.}~\bibnamefont {Thomale}},\ and\ \bibinfo {author} {\bibfnamefont {T.}~\bibnamefont {Bzdu{\v{s}}ek}},\ }\bibfield  {title}
  {\bibinfo {title} {Simulating hyperbolic space on a circuit board},\ }\href {https://doi.org/10.1038/s41467-022-32042-4} {\bibfield  {journal} {\bibinfo  {journal} {Nature Communications}\ }\textbf {\bibinfo {volume} {13}},\ \bibinfo {pages} {4373} (\bibinfo {year} {2022})}\BibitemShut {NoStop}%
\bibitem [{\citenamefont {Zhang}\ \emph {et~al.}(2022)\citenamefont {Zhang}, \citenamefont {Yuan}, \citenamefont {Sun}, \citenamefont {Sun},\ and\ \citenamefont {Zhang}}]{Zhang2022}%
  \BibitemOpen
  \bibfield  {author} {\bibinfo {author} {\bibfnamefont {W.}~\bibnamefont {Zhang}}, \bibinfo {author} {\bibfnamefont {H.}~\bibnamefont {Yuan}}, \bibinfo {author} {\bibfnamefont {N.}~\bibnamefont {Sun}}, \bibinfo {author} {\bibfnamefont {H.}~\bibnamefont {Sun}},\ and\ \bibinfo {author} {\bibfnamefont {X.}~\bibnamefont {Zhang}},\ }\bibfield  {title} {\bibinfo {title} {Observation of novel topological states in hyperbolic lattices},\ }\href {https://doi.org/10.1038/s41467-022-30631-x} {\bibfield  {journal} {\bibinfo  {journal} {Nature Communications}\ }\textbf {\bibinfo {volume} {13}},\ \bibinfo {pages} {2937} (\bibinfo {year} {2022})}\BibitemShut {NoStop}%
\bibitem [{\citenamefont {Chen}\ \emph {et~al.}(2023)\citenamefont {Chen}, \citenamefont {Brand}, \citenamefont {Helbig}, \citenamefont {Hofmann}, \citenamefont {Imhof}, \citenamefont {Fritzsche}, \citenamefont {Kie{\ss}ling}, \citenamefont {Stegmaier}, \citenamefont {Upreti}, \citenamefont {Neupert}, \citenamefont {Bzdu{\v{s}}ek}, \citenamefont {Greiter}, \citenamefont {Thomale},\ and\ \citenamefont {Boettcher}}]{Chen2023}%
  \BibitemOpen
  \bibfield  {author} {\bibinfo {author} {\bibfnamefont {A.}~\bibnamefont {Chen}}, \bibinfo {author} {\bibfnamefont {H.}~\bibnamefont {Brand}}, \bibinfo {author} {\bibfnamefont {T.}~\bibnamefont {Helbig}}, \bibinfo {author} {\bibfnamefont {T.}~\bibnamefont {Hofmann}}, \bibinfo {author} {\bibfnamefont {S.}~\bibnamefont {Imhof}}, \bibinfo {author} {\bibfnamefont {A.}~\bibnamefont {Fritzsche}}, \bibinfo {author} {\bibfnamefont {T.}~\bibnamefont {Kie{\ss}ling}}, \bibinfo {author} {\bibfnamefont {A.}~\bibnamefont {Stegmaier}}, \bibinfo {author} {\bibfnamefont {L.~K.}\ \bibnamefont {Upreti}}, \bibinfo {author} {\bibfnamefont {T.}~\bibnamefont {Neupert}}, \bibinfo {author} {\bibfnamefont {T.}~\bibnamefont {Bzdu{\v{s}}ek}}, \bibinfo {author} {\bibfnamefont {M.}~\bibnamefont {Greiter}}, \bibinfo {author} {\bibfnamefont {R.}~\bibnamefont {Thomale}},\ and\ \bibinfo {author} {\bibfnamefont {I.}~\bibnamefont {Boettcher}},\ }\bibfield  {title} {\bibinfo {title} {Hyperbolic matter in electrical circuits with tunable complex
  phases},\ }\href {https://doi.org/10.1038/s41467-023-36359-6} {\bibfield  {journal} {\bibinfo  {journal} {Nature Communications}\ }\textbf {\bibinfo {volume} {14}},\ \bibinfo {pages} {622} (\bibinfo {year} {2023})}\BibitemShut {NoStop}%
\bibitem [{\citenamefont {Zhang}\ \emph {et~al.}(2023)\citenamefont {Zhang}, \citenamefont {Di}, \citenamefont {Zheng}, \citenamefont {Sun},\ and\ \citenamefont {Zhang}}]{Zhang2023}%
  \BibitemOpen
  \bibfield  {author} {\bibinfo {author} {\bibfnamefont {W.}~\bibnamefont {Zhang}}, \bibinfo {author} {\bibfnamefont {F.}~\bibnamefont {Di}}, \bibinfo {author} {\bibfnamefont {X.}~\bibnamefont {Zheng}}, \bibinfo {author} {\bibfnamefont {H.}~\bibnamefont {Sun}},\ and\ \bibinfo {author} {\bibfnamefont {X.}~\bibnamefont {Zhang}},\ }\bibfield  {title} {\bibinfo {title} {Hyperbolic band topology with non-trivial second chern numbers},\ }\href {https://doi.org/10.1038/s41467-023-36767-8} {\bibfield  {journal} {\bibinfo  {journal} {Nature Communications}\ }\textbf {\bibinfo {volume} {14}},\ \bibinfo {pages} {1083} (\bibinfo {year} {2023})}\BibitemShut {NoStop}%
\bibitem [{\citenamefont {Chen}\ \emph {et~al.}(2024)\citenamefont {Chen}, \citenamefont {Zhang}, \citenamefont {Qin}, \citenamefont {Bossart}, \citenamefont {Yang}, \citenamefont {Chen},\ and\ \citenamefont {Fleury}}]{Chen2024}%
  \BibitemOpen
  \bibfield  {author} {\bibinfo {author} {\bibfnamefont {Q.}~\bibnamefont {Chen}}, \bibinfo {author} {\bibfnamefont {Z.}~\bibnamefont {Zhang}}, \bibinfo {author} {\bibfnamefont {H.}~\bibnamefont {Qin}}, \bibinfo {author} {\bibfnamefont {A.}~\bibnamefont {Bossart}}, \bibinfo {author} {\bibfnamefont {Y.}~\bibnamefont {Yang}}, \bibinfo {author} {\bibfnamefont {H.}~\bibnamefont {Chen}},\ and\ \bibinfo {author} {\bibfnamefont {R.}~\bibnamefont {Fleury}},\ }\bibfield  {title} {\bibinfo {title} {Anomalous and chern topological waves in hyperbolic networks},\ }\href {https://doi.org/10.1038/s41467-024-46551-x} {\bibfield  {journal} {\bibinfo  {journal} {Nature Communications}\ }\textbf {\bibinfo {volume} {15}},\ \bibinfo {pages} {2293} (\bibinfo {year} {2024})}\BibitemShut {NoStop}%
\bibitem [{\citenamefont {Huang}\ \emph {et~al.}(2024)\citenamefont {Huang}, \citenamefont {He}, \citenamefont {Zhang}, \citenamefont {Zhang}, \citenamefont {Liu}, \citenamefont {Feng}, \citenamefont {Liu}, \citenamefont {Cui}, \citenamefont {Huang}, \citenamefont {Zhang},\ and\ \citenamefont {Zhang}}]{Huang2024}%
  \BibitemOpen
  \bibfield  {author} {\bibinfo {author} {\bibfnamefont {L.}~\bibnamefont {Huang}}, \bibinfo {author} {\bibfnamefont {L.}~\bibnamefont {He}}, \bibinfo {author} {\bibfnamefont {W.}~\bibnamefont {Zhang}}, \bibinfo {author} {\bibfnamefont {H.}~\bibnamefont {Zhang}}, \bibinfo {author} {\bibfnamefont {D.}~\bibnamefont {Liu}}, \bibinfo {author} {\bibfnamefont {X.}~\bibnamefont {Feng}}, \bibinfo {author} {\bibfnamefont {F.}~\bibnamefont {Liu}}, \bibinfo {author} {\bibfnamefont {K.}~\bibnamefont {Cui}}, \bibinfo {author} {\bibfnamefont {Y.}~\bibnamefont {Huang}}, \bibinfo {author} {\bibfnamefont {W.}~\bibnamefont {Zhang}},\ and\ \bibinfo {author} {\bibfnamefont {X.}~\bibnamefont {Zhang}},\ }\bibfield  {title} {\bibinfo {title} {Hyperbolic photonic topological insulators},\ }\href {https://doi.org/10.1038/s41467-024-46035-y} {\bibfield  {journal} {\bibinfo  {journal} {Nature Communications}\ }\textbf {\bibinfo {volume} {15}},\ \bibinfo {pages} {1647} (\bibinfo {year} {2024})}\BibitemShut {NoStop}%
\bibitem [{\citenamefont {Park}\ \emph {et~al.}(2024)\citenamefont {Park}, \citenamefont {Piao},\ and\ \citenamefont {Yu}}]{10.1021/acsphotonics.4c01184}%
  \BibitemOpen
  \bibfield  {author} {\bibinfo {author} {\bibfnamefont {H.}~\bibnamefont {Park}}, \bibinfo {author} {\bibfnamefont {X.}~\bibnamefont {Piao}},\ and\ \bibinfo {author} {\bibfnamefont {S.}~\bibnamefont {Yu}},\ }\bibfield  {title} {\bibinfo {title} {Scalable and programmable emulation of photonic hyperbolic lattices},\ }\href {https://doi.org/10.1021/acsphotonics.4c01184} {\bibfield  {journal} {\bibinfo  {journal} {ACS Photonics}\ }\textbf {\bibinfo {volume} {11}},\ \bibinfo {pages} {3890} (\bibinfo {year} {2024})}\BibitemShut {NoStop}%
\bibitem [{\citenamefont {Ant\~ao}\ \emph {et~al.}(2026)\citenamefont {Ant\~ao}, \citenamefont {Sun}, \citenamefont {Fumega},\ and\ \citenamefont {Lado}}]{Antao2026}%
  \BibitemOpen
  \bibfield  {author} {\bibinfo {author} {\bibfnamefont {T.~V.~C.}\ \bibnamefont {Ant\~ao}}, \bibinfo {author} {\bibfnamefont {Y.}~\bibnamefont {Sun}}, \bibinfo {author} {\bibfnamefont {A.~O.}\ \bibnamefont {Fumega}},\ and\ \bibinfo {author} {\bibfnamefont {J.~L.}\ \bibnamefont {Lado}},\ }\bibfield  {title} {\bibinfo {title} {Tensor network method for real-space topology in quasicrystal chern mosaics},\ }\href {https://doi.org/10.1103/hhdf-xpwg} {\bibfield  {journal} {\bibinfo  {journal} {Phys. Rev. Lett.}\ }\textbf {\bibinfo {volume} {136}},\ \bibinfo {pages} {156601} (\bibinfo {year} {2026})}\BibitemShut {NoStop}%
\bibitem [{\citenamefont {Verstraten}\ and\ \citenamefont {Smith}(2026)}]{verstraten2026simgraphuniversalguidesymmetric}%
  \BibitemOpen
  \bibfield  {author} {\bibinfo {author} {\bibfnamefont {R.~C.}\ \bibnamefont {Verstraten}}\ and\ \bibinfo {author} {\bibfnamefont {C.~M.}\ \bibnamefont {Smith}},\ }\href {https://arxiv.org/abs/2609.09996} {\bibinfo {title} {Sim-graph: A universal guide to symmetric interactions}} (\bibinfo {year} {2026}),\ \Eprint {https://arxiv.org/abs/2609.09996} {arXiv:2609.09996} \BibitemShut {NoStop}%
\bibitem [{\citenamefont {Evenbly}\ and\ \citenamefont {Vidal}(2010)}]{PhysRevLett.104.187203}%
  \BibitemOpen
  \bibfield  {author} {\bibinfo {author} {\bibfnamefont {G.}~\bibnamefont {Evenbly}}\ and\ \bibinfo {author} {\bibfnamefont {G.}~\bibnamefont {Vidal}},\ }\bibfield  {title} {\bibinfo {title} {Frustrated antiferromagnets with entanglement renormalization: Ground state of the spin-$\frac{1}{2}$ heisenberg model on a kagome lattice},\ }\href {https://doi.org/10.1103/PhysRevLett.104.187203} {\bibfield  {journal} {\bibinfo  {journal} {Phys. Rev. Lett.}\ }\textbf {\bibinfo {volume} {104}},\ \bibinfo {pages} {187203} (\bibinfo {year} {2010})}\BibitemShut {NoStop}%
\bibitem [{\citenamefont {Murg}\ \emph {et~al.}(2009)\citenamefont {Murg}, \citenamefont {Verstraete},\ and\ \citenamefont {Cirac}}]{PhysRevB.79.195119}%
  \BibitemOpen
  \bibfield  {author} {\bibinfo {author} {\bibfnamefont {V.}~\bibnamefont {Murg}}, \bibinfo {author} {\bibfnamefont {F.}~\bibnamefont {Verstraete}},\ and\ \bibinfo {author} {\bibfnamefont {J.~I.}\ \bibnamefont {Cirac}},\ }\bibfield  {title} {\bibinfo {title} {Exploring frustrated spin systems using projected entangled pair states},\ }\href {https://doi.org/10.1103/PhysRevB.79.195119} {\bibfield  {journal} {\bibinfo  {journal} {Phys. Rev. B}\ }\textbf {\bibinfo {volume} {79}},\ \bibinfo {pages} {195119} (\bibinfo {year} {2009})}\BibitemShut {NoStop}%
\bibitem [{\citenamefont {Depenbrock}\ \emph {et~al.}(2012)\citenamefont {Depenbrock}, \citenamefont {McCulloch},\ and\ \citenamefont {Schollw\"ock}}]{PhysRevLett.109.067201}%
  \BibitemOpen
  \bibfield  {author} {\bibinfo {author} {\bibfnamefont {S.}~\bibnamefont {Depenbrock}}, \bibinfo {author} {\bibfnamefont {I.~P.}\ \bibnamefont {McCulloch}},\ and\ \bibinfo {author} {\bibfnamefont {U.}~\bibnamefont {Schollw\"ock}},\ }\bibfield  {title} {\bibinfo {title} {Nature of the spin-liquid ground state of the $s=1/2$ heisenberg model on the kagome lattice},\ }\href {https://doi.org/10.1103/PhysRevLett.109.067201} {\bibfield  {journal} {\bibinfo  {journal} {Phys. Rev. Lett.}\ }\textbf {\bibinfo {volume} {109}},\ \bibinfo {pages} {067201} (\bibinfo {year} {2012})}\BibitemShut {NoStop}%
\bibitem [{\citenamefont {Corboz}(2016)}]{PhysRevB.93.045116}%
  \BibitemOpen
  \bibfield  {author} {\bibinfo {author} {\bibfnamefont {P.}~\bibnamefont {Corboz}},\ }\bibfield  {title} {\bibinfo {title} {Improved energy extrapolation with infinite projected entangled-pair states applied to the two-dimensional hubbard model},\ }\href {https://doi.org/10.1103/PhysRevB.93.045116} {\bibfield  {journal} {\bibinfo  {journal} {Phys. Rev. B}\ }\textbf {\bibinfo {volume} {93}},\ \bibinfo {pages} {045116} (\bibinfo {year} {2016})}\BibitemShut {NoStop}%
\bibitem [{\citenamefont {Liu}\ \emph {et~al.}(2025)\citenamefont {Liu}, \citenamefont {Zhai}, \citenamefont {Peng}, \citenamefont {Gu},\ and\ \citenamefont {Chan}}]{r4q9-4yvj}%
  \BibitemOpen
  \bibfield  {author} {\bibinfo {author} {\bibfnamefont {W.-Y.}\ \bibnamefont {Liu}}, \bibinfo {author} {\bibfnamefont {H.}~\bibnamefont {Zhai}}, \bibinfo {author} {\bibfnamefont {R.}~\bibnamefont {Peng}}, \bibinfo {author} {\bibfnamefont {Z.-C.}\ \bibnamefont {Gu}},\ and\ \bibinfo {author} {\bibfnamefont {G.~K.-L.}\ \bibnamefont {Chan}},\ }\bibfield  {title} {\bibinfo {title} {Accurate simulation of the hubbard model with finite fermionic projected entangled pair states},\ }\href {https://doi.org/10.1103/r4q9-4yvj} {\bibfield  {journal} {\bibinfo  {journal} {Phys. Rev. Lett.}\ }\textbf {\bibinfo {volume} {134}},\ \bibinfo {pages} {256502} (\bibinfo {year} {2025})}\BibitemShut {NoStop}%
\bibitem [{\citenamefont {Scheb}\ and\ \citenamefont {Noack}(2023)}]{PhysRevB.107.165112}%
  \BibitemOpen
  \bibfield  {author} {\bibinfo {author} {\bibfnamefont {M.}~\bibnamefont {Scheb}}\ and\ \bibinfo {author} {\bibfnamefont {R.~M.}\ \bibnamefont {Noack}},\ }\bibfield  {title} {\bibinfo {title} {Finite projected entangled pair states for the hubbard model},\ }\href {https://doi.org/10.1103/PhysRevB.107.165112} {\bibfield  {journal} {\bibinfo  {journal} {Phys. Rev. B}\ }\textbf {\bibinfo {volume} {107}},\ \bibinfo {pages} {165112} (\bibinfo {year} {2023})}\BibitemShut {NoStop}%
\bibitem [{\citenamefont {Sinha}\ \emph {et~al.}(2022)\citenamefont {Sinha}, \citenamefont {Rams}, \citenamefont {Czarnik},\ and\ \citenamefont {Dziarmaga}}]{PhysRevB.106.195105}%
  \BibitemOpen
  \bibfield  {author} {\bibinfo {author} {\bibfnamefont {A.}~\bibnamefont {Sinha}}, \bibinfo {author} {\bibfnamefont {M.~M.}\ \bibnamefont {Rams}}, \bibinfo {author} {\bibfnamefont {P.}~\bibnamefont {Czarnik}},\ and\ \bibinfo {author} {\bibfnamefont {J.}~\bibnamefont {Dziarmaga}},\ }\bibfield  {title} {\bibinfo {title} {Finite-temperature tensor network study of the hubbard model on an infinite square lattice},\ }\href {https://doi.org/10.1103/PhysRevB.106.195105} {\bibfield  {journal} {\bibinfo  {journal} {Phys. Rev. B}\ }\textbf {\bibinfo {volume} {106}},\ \bibinfo {pages} {195105} (\bibinfo {year} {2022})}\BibitemShut {NoStop}%
\bibitem [{\citenamefont {Begušić}\ \emph {et~al.}(2024)\citenamefont {Begušić}, \citenamefont {Gray},\ and\ \citenamefont {Chan}}]{doi:10.1126/sciadv.adk4321}%
  \BibitemOpen
  \bibfield  {author} {\bibinfo {author} {\bibfnamefont {T.}~\bibnamefont {Begušić}}, \bibinfo {author} {\bibfnamefont {J.}~\bibnamefont {Gray}},\ and\ \bibinfo {author} {\bibfnamefont {G.~K.-L.}\ \bibnamefont {Chan}},\ }\bibfield  {title} {\bibinfo {title} {Fast and converged classical simulations of evidence for the utility of quantum computing before fault tolerance},\ }\href {https://doi.org/10.1126/sciadv.adk4321} {\bibfield  {journal} {\bibinfo  {journal} {Science Advances}\ }\textbf {\bibinfo {volume} {10}},\ \bibinfo {pages} {eadk4321} (\bibinfo {year} {2024})}\BibitemShut {NoStop}%
\bibitem [{\citenamefont {Patra}\ \emph {et~al.}(2024)\citenamefont {Patra}, \citenamefont {Jahromi}, \citenamefont {Singh},\ and\ \citenamefont {Or\'us}}]{PhysRevResearch.6.013326}%
  \BibitemOpen
  \bibfield  {author} {\bibinfo {author} {\bibfnamefont {S.}~\bibnamefont {Patra}}, \bibinfo {author} {\bibfnamefont {S.~S.}\ \bibnamefont {Jahromi}}, \bibinfo {author} {\bibfnamefont {S.}~\bibnamefont {Singh}},\ and\ \bibinfo {author} {\bibfnamefont {R.}~\bibnamefont {Or\'us}},\ }\bibfield  {title} {\bibinfo {title} {Efficient tensor network simulation of ibm's largest quantum processors},\ }\href {https://doi.org/10.1103/PhysRevResearch.6.013326} {\bibfield  {journal} {\bibinfo  {journal} {Phys. Rev. Res.}\ }\textbf {\bibinfo {volume} {6}},\ \bibinfo {pages} {013326} (\bibinfo {year} {2024})}\BibitemShut {NoStop}%
\bibitem [{\citenamefont {Cirac}\ \emph {et~al.}(2021)\citenamefont {Cirac}, \citenamefont {P\'erez-Garc\'{\i}a}, \citenamefont {Schuch},\ and\ \citenamefont {Verstraete}}]{RevModPhys.93.045003}%
  \BibitemOpen
  \bibfield  {author} {\bibinfo {author} {\bibfnamefont {J.~I.}\ \bibnamefont {Cirac}}, \bibinfo {author} {\bibfnamefont {D.}~\bibnamefont {P\'erez-Garc\'{\i}a}}, \bibinfo {author} {\bibfnamefont {N.}~\bibnamefont {Schuch}},\ and\ \bibinfo {author} {\bibfnamefont {F.}~\bibnamefont {Verstraete}},\ }\bibfield  {title} {\bibinfo {title} {Matrix product states and projected entangled pair states: Concepts, symmetries, theorems},\ }\href {https://doi.org/10.1103/RevModPhys.93.045003} {\bibfield  {journal} {\bibinfo  {journal} {Rev. Mod. Phys.}\ }\textbf {\bibinfo {volume} {93}},\ \bibinfo {pages} {045003} (\bibinfo {year} {2021})}\BibitemShut {NoStop}%
\bibitem [{\citenamefont {Schollwöck}(2011)}]{Schoellwock2011}%
  \BibitemOpen
  \bibfield  {author} {\bibinfo {author} {\bibfnamefont {U.}~\bibnamefont {Schollwöck}},\ }\bibfield  {title} {\bibinfo {title} {The density-matrix renormalization group in the age of matrix product states},\ }\href {https://doi.org/https://doi.org/10.1016/j.aop.2010.09.012} {\bibfield  {journal} {\bibinfo  {journal} {Annals of Physics}\ }\textbf {\bibinfo {volume} {326}},\ \bibinfo {pages} {96} (\bibinfo {year} {2011})},\ \bibinfo {note} {january 2011 Special Issue}\BibitemShut {NoStop}%
\bibitem [{\citenamefont {Crosswhite}\ and\ \citenamefont {Bacon}(2008)}]{Crosswhite2008}%
  \BibitemOpen
  \bibfield  {author} {\bibinfo {author} {\bibfnamefont {G.~M.}\ \bibnamefont {Crosswhite}}\ and\ \bibinfo {author} {\bibfnamefont {D.}~\bibnamefont {Bacon}},\ }\bibfield  {title} {\bibinfo {title} {Finite automata for caching in matrix product algorithms},\ }\href {https://doi.org/10.1103/PhysRevA.78.012356} {\bibfield  {journal} {\bibinfo  {journal} {Phys. Rev. A}\ }\textbf {\bibinfo {volume} {78}},\ \bibinfo {pages} {012356} (\bibinfo {year} {2008})}\BibitemShut {NoStop}%
\bibitem [{\citenamefont {McCulloch}(2007)}]{McCulloch2007}%
  \BibitemOpen
  \bibfield  {author} {\bibinfo {author} {\bibfnamefont {I.~P.}\ \bibnamefont {McCulloch}},\ }\bibfield  {title} {\bibinfo {title} {From density-matrix renormalization group to matrix product states},\ }\href {https://doi.org/10.1088/1742-5468/2007/10/P10014} {\bibfield  {journal} {\bibinfo  {journal} {Journal of Statistical Mechanics: Theory and Experiment}\ }\textbf {\bibinfo {volume} {2007}},\ \bibinfo {pages} {P10014} (\bibinfo {year} {2007})}\BibitemShut {NoStop}%
\bibitem [{\citenamefont {Ba\~nuls}\ \emph {et~al.}(2009)\citenamefont {Ba\~nuls}, \citenamefont {Hastings}, \citenamefont {Verstraete},\ and\ \citenamefont {Cirac}}]{PhysRevLett.102.240603}%
  \BibitemOpen
  \bibfield  {author} {\bibinfo {author} {\bibfnamefont {M.~C.}\ \bibnamefont {Ba\~nuls}}, \bibinfo {author} {\bibfnamefont {M.~B.}\ \bibnamefont {Hastings}}, \bibinfo {author} {\bibfnamefont {F.}~\bibnamefont {Verstraete}},\ and\ \bibinfo {author} {\bibfnamefont {J.~I.}\ \bibnamefont {Cirac}},\ }\bibfield  {title} {\bibinfo {title} {Matrix product states for dynamical simulation of infinite chains},\ }\href {https://doi.org/10.1103/PhysRevLett.102.240603} {\bibfield  {journal} {\bibinfo  {journal} {Phys. Rev. Lett.}\ }\textbf {\bibinfo {volume} {102}},\ \bibinfo {pages} {240603} (\bibinfo {year} {2009})}\BibitemShut {NoStop}%
\bibitem [{\citenamefont {Hastings}\ and\ \citenamefont {Mahajan}(2015)}]{PhysRevA.91.032306}%
  \BibitemOpen
  \bibfield  {author} {\bibinfo {author} {\bibfnamefont {M.~B.}\ \bibnamefont {Hastings}}\ and\ \bibinfo {author} {\bibfnamefont {R.}~\bibnamefont {Mahajan}},\ }\bibfield  {title} {\bibinfo {title} {Connecting entanglement in time and space: Improving the folding algorithm},\ }\href {https://doi.org/10.1103/PhysRevA.91.032306} {\bibfield  {journal} {\bibinfo  {journal} {Phys. Rev. A}\ }\textbf {\bibinfo {volume} {91}},\ \bibinfo {pages} {032306} (\bibinfo {year} {2015})}\BibitemShut {NoStop}%
\bibitem [{\citenamefont {Oseledets}(2011)}]{Oseledets2011}%
  \BibitemOpen
  \bibfield  {author} {\bibinfo {author} {\bibfnamefont {I.~V.}\ \bibnamefont {Oseledets}},\ }\bibfield  {title} {\bibinfo {title} {Tensor-train decomposition},\ }\href {https://doi.org/10.1137/090752286} {\bibfield  {journal} {\bibinfo  {journal} {SIAM Journal on Scientific Computing}\ }\textbf {\bibinfo {volume} {33}},\ \bibinfo {pages} {2295} (\bibinfo {year} {2011})}\BibitemShut {NoStop}%
\bibitem [{\citenamefont {Vidal}(2003)}]{vidal_efficient_2003}%
  \BibitemOpen
  \bibfield  {author} {\bibinfo {author} {\bibfnamefont {G.}~\bibnamefont {Vidal}},\ }\bibfield  {title} {\bibinfo {title} {Efficient {{Classical Simulation}} of {{Slightly Entangled Quantum Computations}}},\ }\href {https://doi.org/10.1103/PhysRevLett.91.147902} {\bibfield  {journal} {\bibinfo  {journal} {Physical Review Letters}\ }\textbf {\bibinfo {volume} {91}},\ \bibinfo {pages} {147902} (\bibinfo {year} {2003})}\BibitemShut {NoStop}%
\bibitem [{\citenamefont {García}\ and\ \citenamefont {Márquez~Romero}(2024)}]{ReviewGarcia2024}%
  \BibitemOpen
  \bibfield  {author} {\bibinfo {author} {\bibfnamefont {M.~D.}\ \bibnamefont {García}}\ and\ \bibinfo {author} {\bibfnamefont {A.}~\bibnamefont {Márquez~Romero}},\ }\bibfield  {title} {\bibinfo {title} {Survey on computational applications of tensor-network simulations},\ }\href {https://doi.org/10.1109/ACCESS.2024.3519676} {\bibfield  {journal} {\bibinfo  {journal} {IEEE Access}\ }\textbf {\bibinfo {volume} {12}},\ \bibinfo {pages} {193212} (\bibinfo {year} {2024})}\BibitemShut {NoStop}%
\bibitem [{\citenamefont {Khoromskij}(2011)}]{khoromskij_odlog_2011-1}%
  \BibitemOpen
  \bibfield  {author} {\bibinfo {author} {\bibfnamefont {B.~N.}\ \bibnamefont {Khoromskij}},\ }\bibfield  {title} {\bibinfo {title} {O(dlog{\thinspace}n)-quantics approximation of n-d tensors in high-dimensional numerical modeling},\ }\href {https://doi.org/10.1007/s00365-011-9131-1} {\bibfield  {journal} {\bibinfo  {journal} {Constructive Approximation}\ }\textbf {\bibinfo {volume} {34}},\ \bibinfo {pages} {257} (\bibinfo {year} {2011})}\BibitemShut {NoStop}%
\bibitem [{\citenamefont {Waintal}\ \emph {et~al.}(2026)\citenamefont {Waintal}, \citenamefont {Huang},\ and\ \citenamefont {Groth}}]{SciPostPhysLectNotes.133}%
  \BibitemOpen
  \bibfield  {author} {\bibinfo {author} {\bibfnamefont {X.}~\bibnamefont {Waintal}}, \bibinfo {author} {\bibfnamefont {C.-H.}\ \bibnamefont {Huang}},\ and\ \bibinfo {author} {\bibfnamefont {C.~W.}\ \bibnamefont {Groth}},\ }\bibfield  {title} {\bibinfo {title} {Who can compete with quantum computers? lecture notes on quantum inspired tensor networks computational techniques},\ }\href {https://doi.org/10.21468/SciPostPhysLectNotes.133} {\bibfield  {journal} {\bibinfo  {journal} {SciPost Phys. Lect. Notes}\ ,\ \bibinfo {pages} {133}} (\bibinfo {year} {2026})}\BibitemShut {NoStop}%
\bibitem [{\citenamefont {Oseledets}\ and\ \citenamefont {Tyrtyshnikov}(2010)}]{oseledets_tt-cross_2010}%
  \BibitemOpen
  \bibfield  {author} {\bibinfo {author} {\bibfnamefont {I.}~\bibnamefont {Oseledets}}\ and\ \bibinfo {author} {\bibfnamefont {E.}~\bibnamefont {Tyrtyshnikov}},\ }\bibfield  {title} {\bibinfo {title} {{{TT-cross}} approximation for multidimensional arrays},\ }\href {https://doi.org/10.1016/j.laa.2009.07.024} {\bibfield  {journal} {\bibinfo  {journal} {Linear Algebra and its Applications}\ }\textbf {\bibinfo {volume} {432}},\ \bibinfo {pages} {70} (\bibinfo {year} {2010})}\BibitemShut {NoStop}%
\bibitem [{\citenamefont {Ritter}\ \emph {et~al.}(2024)\citenamefont {Ritter}, \citenamefont {N\'u\~nez Fern\'andez}, \citenamefont {Wallerberger}, \citenamefont {von Delft}, \citenamefont {Shinaoka},\ and\ \citenamefont {Waintal}}]{Ritter2024}%
  \BibitemOpen
  \bibfield  {author} {\bibinfo {author} {\bibfnamefont {M.~K.}\ \bibnamefont {Ritter}}, \bibinfo {author} {\bibfnamefont {Y.}~\bibnamefont {N\'u\~nez Fern\'andez}}, \bibinfo {author} {\bibfnamefont {M.}~\bibnamefont {Wallerberger}}, \bibinfo {author} {\bibfnamefont {J.}~\bibnamefont {von Delft}}, \bibinfo {author} {\bibfnamefont {H.}~\bibnamefont {Shinaoka}},\ and\ \bibinfo {author} {\bibfnamefont {X.}~\bibnamefont {Waintal}},\ }\bibfield  {title} {\bibinfo {title} {Quantics tensor cross interpolation for high-resolution parsimonious representations of multivariate functions},\ }\href {https://doi.org/10.1103/PhysRevLett.132.056501} {\bibfield  {journal} {\bibinfo  {journal} {Phys. Rev. Lett.}\ }\textbf {\bibinfo {volume} {132}},\ \bibinfo {pages} {056501} (\bibinfo {year} {2024})}\BibitemShut {NoStop}%
\bibitem [{\citenamefont {Núñez~Fernández}\ \emph {et~al.}(2025)\citenamefont {Núñez~Fernández}, \citenamefont {Ritter}, \citenamefont {Jeannin}, \citenamefont {Li}, \citenamefont {Kloss}, \citenamefont {Louvet}, \citenamefont {Terasaki}, \citenamefont {Parcollet}, \citenamefont {von Delft}, \citenamefont {Shinaoka},\ and\ \citenamefont {Waintal}}]{Núñez2025}%
  \BibitemOpen
  \bibfield  {author} {\bibinfo {author} {\bibfnamefont {Y.}~\bibnamefont {Núñez~Fernández}}, \bibinfo {author} {\bibfnamefont {M.~K.}\ \bibnamefont {Ritter}}, \bibinfo {author} {\bibfnamefont {M.}~\bibnamefont {Jeannin}}, \bibinfo {author} {\bibfnamefont {J.-W.}\ \bibnamefont {Li}}, \bibinfo {author} {\bibfnamefont {T.}~\bibnamefont {Kloss}}, \bibinfo {author} {\bibfnamefont {T.}~\bibnamefont {Louvet}}, \bibinfo {author} {\bibfnamefont {S.}~\bibnamefont {Terasaki}}, \bibinfo {author} {\bibfnamefont {O.}~\bibnamefont {Parcollet}}, \bibinfo {author} {\bibfnamefont {J.}~\bibnamefont {von Delft}}, \bibinfo {author} {\bibfnamefont {H.}~\bibnamefont {Shinaoka}},\ and\ \bibinfo {author} {\bibfnamefont {X.}~\bibnamefont {Waintal}},\ }\bibfield  {title} {\bibinfo {title} {Learning tensor networks with tensor cross interpolation: New algorithms and libraries},\ }\href {https://doi.org/10.21468/SciPostPhys.18.3.104} {\bibfield  {journal} {\bibinfo  {journal} {SciPost Phys.}\ }\textbf {\bibinfo {volume} {18}},\
  \bibinfo {pages} {104} (\bibinfo {year} {2025})}\BibitemShut {NoStop}%
\bibitem [{\citenamefont {N\'u\~nez Fern\'andez}\ \emph {et~al.}(2022)\citenamefont {N\'u\~nez Fern\'andez}, \citenamefont {Jeannin}, \citenamefont {Dumitrescu}, \citenamefont {Kloss}, \citenamefont {Kaye}, \citenamefont {Parcollet},\ and\ \citenamefont {Waintal}}]{Nunez2022}%
  \BibitemOpen
  \bibfield  {author} {\bibinfo {author} {\bibfnamefont {Y.}~\bibnamefont {N\'u\~nez Fern\'andez}}, \bibinfo {author} {\bibfnamefont {M.}~\bibnamefont {Jeannin}}, \bibinfo {author} {\bibfnamefont {P.~T.}\ \bibnamefont {Dumitrescu}}, \bibinfo {author} {\bibfnamefont {T.}~\bibnamefont {Kloss}}, \bibinfo {author} {\bibfnamefont {J.}~\bibnamefont {Kaye}}, \bibinfo {author} {\bibfnamefont {O.}~\bibnamefont {Parcollet}},\ and\ \bibinfo {author} {\bibfnamefont {X.}~\bibnamefont {Waintal}},\ }\bibfield  {title} {\bibinfo {title} {Learning feynman diagrams with tensor trains},\ }\href {https://doi.org/10.1103/PhysRevX.12.041018} {\bibfield  {journal} {\bibinfo  {journal} {Phys. Rev. X}\ }\textbf {\bibinfo {volume} {12}},\ \bibinfo {pages} {041018} (\bibinfo {year} {2022})}\BibitemShut {NoStop}%
\bibitem [{\citenamefont {Ishida}\ \emph {et~al.}(2025)\citenamefont {Ishida}, \citenamefont {Okada}, \citenamefont {Hoshino},\ and\ \citenamefont {Shinaoka}}]{Ishida2025}%
  \BibitemOpen
  \bibfield  {author} {\bibinfo {author} {\bibfnamefont {H.}~\bibnamefont {Ishida}}, \bibinfo {author} {\bibfnamefont {N.}~\bibnamefont {Okada}}, \bibinfo {author} {\bibfnamefont {S.}~\bibnamefont {Hoshino}},\ and\ \bibinfo {author} {\bibfnamefont {H.}~\bibnamefont {Shinaoka}},\ }\bibfield  {title} {\bibinfo {title} {Low-rank quantics tensor train representations of feynman diagrams for multiorbital electron-phonon models},\ }\href {https://doi.org/10.1103/tkcp-p5br} {\bibfield  {journal} {\bibinfo  {journal} {Phys. Rev. Lett.}\ }\textbf {\bibinfo {volume} {135}},\ \bibinfo {pages} {046502} (\bibinfo {year} {2025})}\BibitemShut {NoStop}%
\bibitem [{\citenamefont {H\"olscher}\ \emph {et~al.}(2025)\citenamefont {H\"olscher}, \citenamefont {Rao}, \citenamefont {M\"uller}, \citenamefont {Klepsch}, \citenamefont {Luckow}, \citenamefont {Stollenwerk},\ and\ \citenamefont {Wilhelm}}]{Holscher2025}%
  \BibitemOpen
  \bibfield  {author} {\bibinfo {author} {\bibfnamefont {L.}~\bibnamefont {H\"olscher}}, \bibinfo {author} {\bibfnamefont {P.}~\bibnamefont {Rao}}, \bibinfo {author} {\bibfnamefont {L.}~\bibnamefont {M\"uller}}, \bibinfo {author} {\bibfnamefont {J.}~\bibnamefont {Klepsch}}, \bibinfo {author} {\bibfnamefont {A.}~\bibnamefont {Luckow}}, \bibinfo {author} {\bibfnamefont {T.}~\bibnamefont {Stollenwerk}},\ and\ \bibinfo {author} {\bibfnamefont {F.~K.}\ \bibnamefont {Wilhelm}},\ }\bibfield  {title} {\bibinfo {title} {Quantum-inspired fluid simulation of two-dimensional turbulence with gpu acceleration},\ }\href {https://doi.org/10.1103/PhysRevResearch.7.013112} {\bibfield  {journal} {\bibinfo  {journal} {Phys. Rev. Res.}\ }\textbf {\bibinfo {volume} {7}},\ \bibinfo {pages} {013112} (\bibinfo {year} {2025})}\BibitemShut {NoStop}%
\bibitem [{\citenamefont {Pisoni}\ \emph {et~al.}(2026)\citenamefont {Pisoni}, \citenamefont {Peddinti}, \citenamefont {Tiunov}, \citenamefont {Guzman},\ and\ \citenamefont {Aolita}}]{pisoni2026compressionsimulationsynthesisturbulent}%
  \BibitemOpen
  \bibfield  {author} {\bibinfo {author} {\bibfnamefont {S.}~\bibnamefont {Pisoni}}, \bibinfo {author} {\bibfnamefont {R.~D.}\ \bibnamefont {Peddinti}}, \bibinfo {author} {\bibfnamefont {E.}~\bibnamefont {Tiunov}}, \bibinfo {author} {\bibfnamefont {S.~E.}\ \bibnamefont {Guzman}},\ and\ \bibinfo {author} {\bibfnamefont {L.}~\bibnamefont {Aolita}},\ }\href {https://arxiv.org/abs/2506.05477} {\bibinfo {title} {Compression, simulation, and synthesis of turbulent flows with tensor trains}} (\bibinfo {year} {2026}),\ \Eprint {https://arxiv.org/abs/2506.05477} {arXiv:2506.05477} \BibitemShut {NoStop}%
\bibitem [{\citenamefont {Pham}\ \emph {et~al.}(2026)\citenamefont {Pham}, \citenamefont {Moon},\ and\ \citenamefont {Kim}}]{Pham2026}%
  \BibitemOpen
  \bibfield  {author} {\bibinfo {author} {\bibfnamefont {K.~T.}\ \bibnamefont {Pham}}, \bibinfo {author} {\bibfnamefont {G.~E.}\ \bibnamefont {Moon}},\ and\ \bibinfo {author} {\bibfnamefont {J.}~\bibnamefont {Kim}},\ }\bibfield  {title} {\bibinfo {title} {Randomized tensor-train truncation on gpus for quantum-inspired computational fluid dynamics},\ }\href {https://doi.org/https://doi.org/10.1016/j.cpc.2026.110259} {\bibfield  {journal} {\bibinfo  {journal} {Computer Physics Communications}\ }\textbf {\bibinfo {volume} {327}},\ \bibinfo {pages} {110259} (\bibinfo {year} {2026})}\BibitemShut {NoStop}%
\bibitem [{\citenamefont {Gavrilyuk}\ and\ \citenamefont {Khoromskij}(2011)}]{gavrilyuk_quantized-tt-cayley_2011}%
  \BibitemOpen
  \bibfield  {author} {\bibinfo {author} {\bibfnamefont {I.}~\bibnamefont {Gavrilyuk}}\ and\ \bibinfo {author} {\bibfnamefont {B.}~\bibnamefont {Khoromskij}},\ }\bibfield  {title} {\bibinfo {title} {Quantized-{{TT-Cayley Transform}} for {{Computing}} the {{Dynamics}} and the {{Spectrum}} of {{High-Dimensional Hamiltonians}}},\ }\href {https://doi.org/10.2478/cmam-2011-0015} {\bibfield  {journal} {\bibinfo  {journal} {Comput. Methods Appl. Math.}\ }\textbf {\bibinfo {volume} {11}},\ \bibinfo {pages} {273} (\bibinfo {year} {2011})}\BibitemShut {NoStop}%
\bibitem [{\citenamefont {Dolgov}\ \emph {et~al.}(2012)\citenamefont {Dolgov}, \citenamefont {Khoromskij},\ and\ \citenamefont {Savostyanov}}]{dolgov_superfast_2012}%
  \BibitemOpen
  \bibfield  {author} {\bibinfo {author} {\bibfnamefont {S.}~\bibnamefont {Dolgov}}, \bibinfo {author} {\bibfnamefont {B.}~\bibnamefont {Khoromskij}},\ and\ \bibinfo {author} {\bibfnamefont {D.}~\bibnamefont {Savostyanov}},\ }\bibfield  {title} {\bibinfo {title} {Superfast {{Fourier Transform Using QTT Approximation}}},\ }\href {https://doi.org/10.1007/s00041-012-9227-4} {\bibfield  {journal} {\bibinfo  {journal} {Journal of Fourier Analysis and Applications}\ }\textbf {\bibinfo {volume} {18}},\ \bibinfo {pages} {915} (\bibinfo {year} {2012})}\BibitemShut {NoStop}%
\bibitem [{\citenamefont {Savostyanov}(2012)}]{SAVOSTYANOV20123215}%
  \BibitemOpen
  \bibfield  {author} {\bibinfo {author} {\bibfnamefont {D.}~\bibnamefont {Savostyanov}},\ }\bibfield  {title} {\bibinfo {title} {Qtt-rank-one vectors with qtt-rank-one and full-rank fourier images},\ }\href {https://doi.org/https://doi.org/10.1016/j.laa.2011.11.008} {\bibfield  {journal} {\bibinfo  {journal} {Linear Algebra and its Applications}\ }\textbf {\bibinfo {volume} {436}},\ \bibinfo {pages} {3215} (\bibinfo {year} {2012})}\BibitemShut {NoStop}%
\bibitem [{\citenamefont {Rohshap}\ \emph {et~al.}(2025{\natexlab{a}})\citenamefont {Rohshap}, \citenamefont {Ritter}, \citenamefont {Shinaoka}, \citenamefont {von Delft}, \citenamefont {Wallerberger},\ and\ \citenamefont {Kauch}}]{PhysRevResearch.7.023087}%
  \BibitemOpen
  \bibfield  {author} {\bibinfo {author} {\bibfnamefont {S.}~\bibnamefont {Rohshap}}, \bibinfo {author} {\bibfnamefont {M.~K.}\ \bibnamefont {Ritter}}, \bibinfo {author} {\bibfnamefont {H.}~\bibnamefont {Shinaoka}}, \bibinfo {author} {\bibfnamefont {J.}~\bibnamefont {von Delft}}, \bibinfo {author} {\bibfnamefont {M.}~\bibnamefont {Wallerberger}},\ and\ \bibinfo {author} {\bibfnamefont {A.}~\bibnamefont {Kauch}},\ }\bibfield  {title} {\bibinfo {title} {Two-particle calculations with quantics tensor trains: Solving the parquet equations},\ }\href {https://doi.org/10.1103/PhysRevResearch.7.023087} {\bibfield  {journal} {\bibinfo  {journal} {Phys. Rev. Res.}\ }\textbf {\bibinfo {volume} {7}},\ \bibinfo {pages} {023087} (\bibinfo {year} {2025}{\natexlab{a}})}\BibitemShut {NoStop}%
\bibitem [{\citenamefont {Shinaoka}\ \emph {et~al.}(2023)\citenamefont {Shinaoka}, \citenamefont {Wallerberger}, \citenamefont {Murakami}, \citenamefont {Nogaki}, \citenamefont {Sakurai}, \citenamefont {Werner},\ and\ \citenamefont {Kauch}}]{PhysRevX.13.021015}%
  \BibitemOpen
  \bibfield  {author} {\bibinfo {author} {\bibfnamefont {H.}~\bibnamefont {Shinaoka}}, \bibinfo {author} {\bibfnamefont {M.}~\bibnamefont {Wallerberger}}, \bibinfo {author} {\bibfnamefont {Y.}~\bibnamefont {Murakami}}, \bibinfo {author} {\bibfnamefont {K.}~\bibnamefont {Nogaki}}, \bibinfo {author} {\bibfnamefont {R.}~\bibnamefont {Sakurai}}, \bibinfo {author} {\bibfnamefont {P.}~\bibnamefont {Werner}},\ and\ \bibinfo {author} {\bibfnamefont {A.}~\bibnamefont {Kauch}},\ }\bibfield  {title} {\bibinfo {title} {Multiscale space-time ansatz for correlation functions of quantum systems based on quantics tensor trains},\ }\href {https://doi.org/10.1103/PhysRevX.13.021015} {\bibfield  {journal} {\bibinfo  {journal} {Phys. Rev. X}\ }\textbf {\bibinfo {volume} {13}},\ \bibinfo {pages} {021015} (\bibinfo {year} {2023})}\BibitemShut {NoStop}%
\bibitem [{\citenamefont {Frankenbach}\ \emph {et~al.}(2025)\citenamefont {Frankenbach}, \citenamefont {Ritter}, \citenamefont {Pelz}, \citenamefont {Ritz}, \citenamefont {von Delft},\ and\ \citenamefont {Ge}}]{jx7h-lsqk}%
  \BibitemOpen
  \bibfield  {author} {\bibinfo {author} {\bibfnamefont {M.}~\bibnamefont {Frankenbach}}, \bibinfo {author} {\bibfnamefont {M.~K.}\ \bibnamefont {Ritter}}, \bibinfo {author} {\bibfnamefont {M.}~\bibnamefont {Pelz}}, \bibinfo {author} {\bibfnamefont {N.}~\bibnamefont {Ritz}}, \bibinfo {author} {\bibfnamefont {J.}~\bibnamefont {von Delft}},\ and\ \bibinfo {author} {\bibfnamefont {A.}~\bibnamefont {Ge}},\ }\bibfield  {title} {\bibinfo {title} {Compressing local vertex functions from the multipoint numerical renormalization group using quantics tensor cross interpolation},\ }\href {https://doi.org/10.1103/jx7h-lsqk} {\bibfield  {journal} {\bibinfo  {journal} {Phys. Rev. Res.}\ }\textbf {\bibinfo {volume} {7}},\ \bibinfo {pages} {043032} (\bibinfo {year} {2025})}\BibitemShut {NoStop}%
\bibitem [{\citenamefont {Sakaue}\ \emph {et~al.}(2025)\citenamefont {Sakaue}, \citenamefont {Shinaoka},\ and\ \citenamefont {Sakurai}}]{SciPostPhys.19.2.038}%
  \BibitemOpen
  \bibfield  {author} {\bibinfo {author} {\bibfnamefont {K.}~\bibnamefont {Sakaue}}, \bibinfo {author} {\bibfnamefont {H.}~\bibnamefont {Shinaoka}},\ and\ \bibinfo {author} {\bibfnamefont {R.}~\bibnamefont {Sakurai}},\ }\bibfield  {title} {\bibinfo {title} {Adaptive sampling-based optimization of quantics tensor trains for noisy functions: Applications to quantum simulations},\ }\href {https://doi.org/10.21468/SciPostPhys.19.2.038} {\bibfield  {journal} {\bibinfo  {journal} {SciPost Phys.}\ }\textbf {\bibinfo {volume} {19}},\ \bibinfo {pages} {038} (\bibinfo {year} {2025})}\BibitemShut {NoStop}%
\bibitem [{\citenamefont {Antão}\ \emph {et~al.}(2026)\citenamefont {Antão}, \citenamefont {Moustaj}, \citenamefont {Sun},\ and\ \citenamefont {Lado}}]{antao2026tensornetworksolversultralarge}%
  \BibitemOpen
  \bibfield  {author} {\bibinfo {author} {\bibfnamefont {T.~V.~C.}\ \bibnamefont {Antão}}, \bibinfo {author} {\bibfnamefont {A.}~\bibnamefont {Moustaj}}, \bibinfo {author} {\bibfnamefont {Y.}~\bibnamefont {Sun}},\ and\ \bibinfo {author} {\bibfnamefont {J.~L.}\ \bibnamefont {Lado}},\ }\href {https://arxiv.org/abs/2607.00991} {\bibinfo {title} {Tensor network solvers for ultra-large tight-binding hamiltonians: algorithms and applications}} (\bibinfo {year} {2026}),\ \Eprint {https://arxiv.org/abs/2607.00991} {arXiv:2607.00991} \BibitemShut {NoStop}%
\bibitem [{\citenamefont {Sun}\ \emph {et~al.}(2025)\citenamefont {Sun}, \citenamefont {Niedermeier}, \citenamefont {Ant\~ao}, \citenamefont {Fumega},\ and\ \citenamefont {Lado}}]{Yitao2025}%
  \BibitemOpen
  \bibfield  {author} {\bibinfo {author} {\bibfnamefont {Y.}~\bibnamefont {Sun}}, \bibinfo {author} {\bibfnamefont {M.}~\bibnamefont {Niedermeier}}, \bibinfo {author} {\bibfnamefont {T.~V.~C.}\ \bibnamefont {Ant\~ao}}, \bibinfo {author} {\bibfnamefont {A.~O.}\ \bibnamefont {Fumega}},\ and\ \bibinfo {author} {\bibfnamefont {J.~L.}\ \bibnamefont {Lado}},\ }\bibfield  {title} {\bibinfo {title} {Self-consistent tensor network method for correlated super-moir\'e matter beyond one billion sites},\ }\href {https://doi.org/10.1103/krjp-mn4v} {\bibfield  {journal} {\bibinfo  {journal} {Phys. Rev. Res.}\ }\textbf {\bibinfo {volume} {7}},\ \bibinfo {pages} {043288} (\bibinfo {year} {2025})}\BibitemShut {NoStop}%
\bibitem [{\citenamefont {Rohshap}\ \emph {et~al.}(2025{\natexlab{b}})\citenamefont {Rohshap}, \citenamefont {Li}, \citenamefont {Lorenz}, \citenamefont {Hasil}, \citenamefont {Held}, \citenamefont {Kauch},\ and\ \citenamefont {Wallerberger}}]{qlrs-6f8t}%
  \BibitemOpen
  \bibfield  {author} {\bibinfo {author} {\bibfnamefont {S.}~\bibnamefont {Rohshap}}, \bibinfo {author} {\bibfnamefont {J.-W.}\ \bibnamefont {Li}}, \bibinfo {author} {\bibfnamefont {A.}~\bibnamefont {Lorenz}}, \bibinfo {author} {\bibfnamefont {S.}~\bibnamefont {Hasil}}, \bibinfo {author} {\bibfnamefont {K.}~\bibnamefont {Held}}, \bibinfo {author} {\bibfnamefont {A.}~\bibnamefont {Kauch}},\ and\ \bibinfo {author} {\bibfnamefont {M.}~\bibnamefont {Wallerberger}},\ }\bibfield  {title} {\bibinfo {title} {Entanglement across scales: Quantics tensor trains as a natural framework for renormalization},\ }\href {https://doi.org/10.1103/qlrs-6f8t} {\bibfield  {journal} {\bibinfo  {journal} {Phys. Rev. Res.}\ }\textbf {\bibinfo {volume} {7}},\ \bibinfo {pages} {043313} (\bibinfo {year} {2025}{\natexlab{b}})}\BibitemShut {NoStop}%
\bibitem [{\citenamefont {Oseledets}(2010)}]{oseledets_approximation_2010}%
  \BibitemOpen
  \bibfield  {author} {\bibinfo {author} {\bibfnamefont {I.~V.}\ \bibnamefont {Oseledets}},\ }\bibfield  {title} {\bibinfo {title} {Approximation of $2^d \times 2^d$ {{Matrices Using Tensor Decomposition}}},\ }\href {https://doi.org/10.1137/090757861} {\bibfield  {journal} {\bibinfo  {journal} {SIAM Journal on Matrix Analysis and Applications}\ }\textbf {\bibinfo {volume} {31}},\ \bibinfo {pages} {2130} (\bibinfo {year} {2010})}\BibitemShut {NoStop}%
\bibitem [{\citenamefont {Kazeev}\ and\ \citenamefont {Khoromskij}(2012)}]{doi:10.1137/100820479}%
  \BibitemOpen
  \bibfield  {author} {\bibinfo {author} {\bibfnamefont {V.~A.}\ \bibnamefont {Kazeev}}\ and\ \bibinfo {author} {\bibfnamefont {B.~N.}\ \bibnamefont {Khoromskij}},\ }\bibfield  {title} {\bibinfo {title} {Low-rank explicit qtt representation of the laplace operator and its inverse},\ }\href {https://doi.org/10.1137/100820479} {\bibfield  {journal} {\bibinfo  {journal} {SIAM Journal on Matrix Analysis and Applications}\ }\textbf {\bibinfo {volume} {33}},\ \bibinfo {pages} {742} (\bibinfo {year} {2012})}\BibitemShut {NoStop}%
\bibitem [{\citenamefont {Kazeev}\ \emph {et~al.}(2013)\citenamefont {Kazeev}, \citenamefont {Khoromskij},\ and\ \citenamefont {Tyrtyshnikov}}]{kazeev_multilevel_2013}%
  \BibitemOpen
  \bibfield  {author} {\bibinfo {author} {\bibfnamefont {V.~A.}\ \bibnamefont {Kazeev}}, \bibinfo {author} {\bibfnamefont {B.~N.}\ \bibnamefont {Khoromskij}},\ and\ \bibinfo {author} {\bibfnamefont {E.~E.}\ \bibnamefont {Tyrtyshnikov}},\ }\bibfield  {title} {\bibinfo {title} {Multilevel toeplitz matrices generated by tensor-structured vectors and convolution with logarithmic complexity},\ }\href {https://doi.org/10.1137/110844830} {\bibfield  {journal} {\bibinfo  {journal} {SIAM Journal on Scientific Computing}\ }\textbf {\bibinfo {volume} {35}},\ \bibinfo {pages} {A1511} (\bibinfo {year} {2013})}\BibitemShut {NoStop}%
\bibitem [{\citenamefont {Gelß}\ and\ \citenamefont {Schütte}(2018)}]{gelss2018tensorgeneratedfractalsusingtensor}%
  \BibitemOpen
  \bibfield  {author} {\bibinfo {author} {\bibfnamefont {P.}~\bibnamefont {Gelß}}\ and\ \bibinfo {author} {\bibfnamefont {C.}~\bibnamefont {Schütte}},\ }\href {https://arxiv.org/abs/1812.00814} {\bibinfo {title} {Tensor-generated fractals: Using tensor decompositions for creating self-similar patterns}} (\bibinfo {year} {2018}),\ \Eprint {https://arxiv.org/abs/1812.00814} {arXiv:1812.00814} \BibitemShut {NoStop}%
\bibitem [{\citenamefont {Villalba-D{\'i}ez}\ \emph {et~al.}(2026)\citenamefont {Villalba-D{\'i}ez}, \citenamefont {Gonz{\'a}lez-Marcos}, \citenamefont {Losada-Gonz{\'a}lez},\ and\ \citenamefont {Ordieres-Mer{\'e}}}]{Villalba-Diez2026}%
  \BibitemOpen
  \bibfield  {author} {\bibinfo {author} {\bibfnamefont {J.}~\bibnamefont {Villalba-D{\'i}ez}}, \bibinfo {author} {\bibfnamefont {A.}~\bibnamefont {Gonz{\'a}lez-Marcos}}, \bibinfo {author} {\bibfnamefont {J.~C.}\ \bibnamefont {Losada-Gonz{\'a}lez}},\ and\ \bibinfo {author} {\bibfnamefont {J.}~\bibnamefont {Ordieres-Mer{\'e}}},\ }\bibfield  {title} {\bibinfo {title} {Fractal photonic crystals with controlled disorder for robust 3d-integrated on-chip quantum mode localization},\ }\href {https://doi.org/10.1007/s11082-026-08741-3} {\bibfield  {journal} {\bibinfo  {journal} {Optical and Quantum Electronics}\ }\textbf {\bibinfo {volume} {58}},\ \bibinfo {pages} {166} (\bibinfo {year} {2026})}\BibitemShut {NoStop}%
\bibitem [{\citenamefont {Wei\ss{}e}\ \emph {et~al.}(2006)\citenamefont {Wei\ss{}e}, \citenamefont {Wellein}, \citenamefont {Alvermann},\ and\ \citenamefont {Fehske}}]{RevModPhys.78.275}%
  \BibitemOpen
  \bibfield  {author} {\bibinfo {author} {\bibfnamefont {A.}~\bibnamefont {Wei\ss{}e}}, \bibinfo {author} {\bibfnamefont {G.}~\bibnamefont {Wellein}}, \bibinfo {author} {\bibfnamefont {A.}~\bibnamefont {Alvermann}},\ and\ \bibinfo {author} {\bibfnamefont {H.}~\bibnamefont {Fehske}},\ }\bibfield  {title} {\bibinfo {title} {The kernel polynomial method},\ }\href {https://doi.org/10.1103/RevModPhys.78.275} {\bibfield  {journal} {\bibinfo  {journal} {Rev. Mod. Phys.}\ }\textbf {\bibinfo {volume} {78}},\ \bibinfo {pages} {275} (\bibinfo {year} {2006})}\BibitemShut {NoStop}%
\bibitem [{\citenamefont {Kolehmainen}\ \emph {et~al.}(2026)\citenamefont {Kolehmainen}, \citenamefont {Lado},\ and\ \citenamefont {Moustaj}}]{kolehmainen2026onedimensionalquasicrystalstensornetworkfinitestate}%
  \BibitemOpen
  \bibfield  {author} {\bibinfo {author} {\bibfnamefont {M.}~\bibnamefont {Kolehmainen}}, \bibinfo {author} {\bibfnamefont {J.~L.}\ \bibnamefont {Lado}},\ and\ \bibinfo {author} {\bibfnamefont {A.}~\bibnamefont {Moustaj}},\ }\href {https://arxiv.org/abs/2609.06040} {\bibinfo {title} {One-dimensional quasicrystals with tensor-network finite-state automata}} (\bibinfo {year} {2026}),\ \Eprint {https://arxiv.org/abs/2609.06040} {arXiv:2609.06040} \BibitemShut {NoStop}%
\bibitem [{\citenamefont {Droste}\ \emph {et~al.}(2009)\citenamefont {Droste}, \citenamefont {Kuich},\ and\ \citenamefont {Vogler}}]{BookAutomaton}%
  \BibitemOpen
  \bibfield  {author} {\bibinfo {author} {\bibfnamefont {M.}~\bibnamefont {Droste}}, \bibinfo {author} {\bibfnamefont {W.}~\bibnamefont {Kuich}},\ and\ \bibinfo {author} {\bibfnamefont {H.}~\bibnamefont {Vogler}},\ }\href {https://doi.org/10.1007/978-3-642-01492-5} {\emph {\bibinfo {title} {Handbook of Weighted Automata}}}\ (\bibinfo {year} {2009})\BibitemShut {NoStop}%
\bibitem [{\citenamefont {Singh}\ \emph {et~al.}(2010)\citenamefont {Singh}, \citenamefont {Pfeifer},\ and\ \citenamefont {Vidal}}]{PhysRevA.82.050301}%
  \BibitemOpen
  \bibfield  {author} {\bibinfo {author} {\bibfnamefont {S.}~\bibnamefont {Singh}}, \bibinfo {author} {\bibfnamefont {R.~N.~C.}\ \bibnamefont {Pfeifer}},\ and\ \bibinfo {author} {\bibfnamefont {G.}~\bibnamefont {Vidal}},\ }\bibfield  {title} {\bibinfo {title} {Tensor network decompositions in the presence of a global symmetry},\ }\href {https://doi.org/10.1103/PhysRevA.82.050301} {\bibfield  {journal} {\bibinfo  {journal} {Phys. Rev. A}\ }\textbf {\bibinfo {volume} {82}},\ \bibinfo {pages} {050301(R)} (\bibinfo {year} {2010})}\BibitemShut {NoStop}%
\bibitem [{\citenamefont {Holzner}\ \emph {et~al.}(2011)\citenamefont {Holzner}, \citenamefont {Weichselbaum}, \citenamefont {McCulloch}, \citenamefont {Schollw\"ock},\ and\ \citenamefont {von Delft}}]{Holzner_chebyshev_2011}%
  \BibitemOpen
  \bibfield  {author} {\bibinfo {author} {\bibfnamefont {A.}~\bibnamefont {Holzner}}, \bibinfo {author} {\bibfnamefont {A.}~\bibnamefont {Weichselbaum}}, \bibinfo {author} {\bibfnamefont {I.~P.}\ \bibnamefont {McCulloch}}, \bibinfo {author} {\bibfnamefont {U.}~\bibnamefont {Schollw\"ock}},\ and\ \bibinfo {author} {\bibfnamefont {J.}~\bibnamefont {von Delft}},\ }\bibfield  {title} {\bibinfo {title} {Chebyshev matrix product state approach for spectral functions},\ }\href {https://doi.org/10.1103/PhysRevB.83.195115} {\bibfield  {journal} {\bibinfo  {journal} {Phys. Rev. B}\ }\textbf {\bibinfo {volume} {83}},\ \bibinfo {pages} {195115} (\bibinfo {year} {2011})}\BibitemShut {NoStop}%
\bibitem [{\citenamefont {Fishman}\ \emph {et~al.}(2022{\natexlab{a}})\citenamefont {Fishman}, \citenamefont {White},\ and\ \citenamefont {Stoudenmire}}]{ITensor}%
  \BibitemOpen
  \bibfield  {author} {\bibinfo {author} {\bibfnamefont {M.}~\bibnamefont {Fishman}}, \bibinfo {author} {\bibfnamefont {S.~R.}\ \bibnamefont {White}},\ and\ \bibinfo {author} {\bibfnamefont {E.~M.}\ \bibnamefont {Stoudenmire}},\ }\bibfield  {title} {\bibinfo {title} {{The ITensor Software Library for Tensor Network Calculations}},\ }\href {https://doi.org/10.21468/SciPostPhysCodeb.4} {\bibfield  {journal} {\bibinfo  {journal} {SciPost Phys. Codebases}\ ,\ \bibinfo {pages} {4}} (\bibinfo {year} {2022}{\natexlab{a}})}\BibitemShut {NoStop}%
\bibitem [{\citenamefont {Fishman}\ \emph {et~al.}(2022{\natexlab{b}})\citenamefont {Fishman}, \citenamefont {White},\ and\ \citenamefont {Stoudenmire}}]{ITensor-r0.3}%
  \BibitemOpen
  \bibfield  {author} {\bibinfo {author} {\bibfnamefont {M.}~\bibnamefont {Fishman}}, \bibinfo {author} {\bibfnamefont {S.~R.}\ \bibnamefont {White}},\ and\ \bibinfo {author} {\bibfnamefont {E.~M.}\ \bibnamefont {Stoudenmire}},\ }\bibfield  {title} {\bibinfo {title} {{Codebase release 0.3 for ITensor}},\ }\href {https://doi.org/10.21468/SciPostPhysCodeb.4-r0.3} {\bibfield  {journal} {\bibinfo  {journal} {SciPost Phys. Codebases}\ ,\ \bibinfo {pages} {4}} (\bibinfo {year} {2022}{\natexlab{b}})}\BibitemShut {NoStop}%
\bibitem [{\citenamefont {Fujiwara}\ \emph {et~al.}(1989)\citenamefont {Fujiwara}, \citenamefont {Kohmoto},\ and\ \citenamefont {Tokihiro}}]{PhysRevB.40.7413}%
  \BibitemOpen
  \bibfield  {author} {\bibinfo {author} {\bibfnamefont {T.}~\bibnamefont {Fujiwara}}, \bibinfo {author} {\bibfnamefont {M.}~\bibnamefont {Kohmoto}},\ and\ \bibinfo {author} {\bibfnamefont {T.}~\bibnamefont {Tokihiro}},\ }\bibfield  {title} {\bibinfo {title} {Multifractal wave functions on a fibonacci lattice},\ }\href {https://doi.org/10.1103/PhysRevB.40.7413} {\bibfield  {journal} {\bibinfo  {journal} {Phys. Rev. B}\ }\textbf {\bibinfo {volume} {40}},\ \bibinfo {pages} {7413(R)} (\bibinfo {year} {1989})}\BibitemShut {NoStop}%
\bibitem [{\citenamefont {Mac\'e}\ \emph {et~al.}(2016)\citenamefont {Mac\'e}, \citenamefont {Jagannathan},\ and\ \citenamefont {Pi\'echon}}]{PhysRevB.93.205153}%
  \BibitemOpen
  \bibfield  {author} {\bibinfo {author} {\bibfnamefont {N.}~\bibnamefont {Mac\'e}}, \bibinfo {author} {\bibfnamefont {A.}~\bibnamefont {Jagannathan}},\ and\ \bibinfo {author} {\bibfnamefont {F.}~\bibnamefont {Pi\'echon}},\ }\bibfield  {title} {\bibinfo {title} {Fractal dimensions of wave functions and local spectral measures on the fibonacci chain},\ }\href {https://doi.org/10.1103/PhysRevB.93.205153} {\bibfield  {journal} {\bibinfo  {journal} {Phys. Rev. B}\ }\textbf {\bibinfo {volume} {93}},\ \bibinfo {pages} {205153} (\bibinfo {year} {2016})}\BibitemShut {NoStop}%
\bibitem [{\citenamefont {S{\"u}t{\H{o}}}(1989)}]{Suto1989}%
  \BibitemOpen
  \bibfield  {author} {\bibinfo {author} {\bibfnamefont {A.}~\bibnamefont {S{\"u}t{\H{o}}}},\ }\bibfield  {title} {\bibinfo {title} {Singular continuous spectrum on a cantor set of zero lebesgue measure for the fibonacci hamiltonian},\ }\href {https://doi.org/10.1007/BF01044450} {\bibfield  {journal} {\bibinfo  {journal} {Journal of Statistical Physics}\ }\textbf {\bibinfo {volume} {56}},\ \bibinfo {pages} {525} (\bibinfo {year} {1989})}\BibitemShut {NoStop}%
\bibitem [{\citenamefont {Ilan}\ \emph {et~al.}(2004)\citenamefont {Ilan}, \citenamefont {Liberty}, \citenamefont {Mandel},\ and\ \citenamefont {Lifshitz}}]{ILAN01012004}%
  \BibitemOpen
  \bibfield  {author} {\bibinfo {author} {\bibfnamefont {R.}~\bibnamefont {Ilan}}, \bibinfo {author} {\bibfnamefont {E.}~\bibnamefont {Liberty}}, \bibinfo {author} {\bibfnamefont {S.~E.-D.}\ \bibnamefont {Mandel}},\ and\ \bibinfo {author} {\bibfnamefont {R.}~\bibnamefont {Lifshitz}},\ }\bibfield  {title} {\bibinfo {title} {Electrons and phonons on the square fibonacci tiling},\ }\href {https://doi.org/10.1080/00150190490462252} {\bibfield  {journal} {\bibinfo  {journal} {Ferroelectrics}\ }\textbf {\bibinfo {volume} {305}},\ \bibinfo {pages} {15} (\bibinfo {year} {2004})}\BibitemShut {NoStop}%
\bibitem [{\citenamefont {Lifshitz}(2002)}]{LIFSHITZ2002186}%
  \BibitemOpen
  \bibfield  {author} {\bibinfo {author} {\bibfnamefont {R.}~\bibnamefont {Lifshitz}},\ }\bibfield  {title} {\bibinfo {title} {The square fibonacci tiling},\ }\href {https://doi.org/https://doi.org/10.1016/S0925-8388(02)00169-X} {\bibfield  {journal} {\bibinfo  {journal} {Journal of Alloys and Compounds}\ }\textbf {\bibinfo {volume} {342}},\ \bibinfo {pages} {186} (\bibinfo {year} {2002})}\BibitemShut {NoStop}%
\bibitem [{\citenamefont {Zeckendorf}(1972)}]{Zeckendorf1972}%
  \BibitemOpen
  \bibfield  {author} {\bibinfo {author} {\bibfnamefont {{\'{E}}.}~\bibnamefont {Zeckendorf}},\ }\bibfield  {title} {\bibinfo {title} {Repr{\'{e}}sentation des nombres naturels par une somme de nombres de fibonacci ou de nombres de lucas},\ }\href@noop {} {\bibfield  {journal} {\bibinfo  {journal} {Bulletin de la Société Royale des Sciences de Liège}\ }\textbf {\bibinfo {volume} {41}},\ \bibinfo {pages} {179} (\bibinfo {year} {1972})}\BibitemShut {NoStop}%
\bibitem [{\citenamefont {Jagannathan}(2026)}]{7rhl-z7f3}%
  \BibitemOpen
  \bibfield  {author} {\bibinfo {author} {\bibfnamefont {A.}~\bibnamefont {Jagannathan}},\ }\bibfield  {title} {\bibinfo {title} {Metallic mean quasicrystals and their topological invariants},\ }\href {https://doi.org/10.1103/7rhl-z7f3} {\bibfield  {journal} {\bibinfo  {journal} {Phys. Rev. B}\ } (\bibinfo {year} {2026})}\BibitemShut {NoStop}%
\bibitem [{\citenamefont {Krebbekx}\ \emph {et~al.}(2023)\citenamefont {Krebbekx}, \citenamefont {Moustaj}, \citenamefont {Dajani},\ and\ \citenamefont {Morais~Smith}}]{PhysRevB.108.104204}%
  \BibitemOpen
  \bibfield  {author} {\bibinfo {author} {\bibfnamefont {J.~P.~J.}\ \bibnamefont {Krebbekx}}, \bibinfo {author} {\bibfnamefont {A.}~\bibnamefont {Moustaj}}, \bibinfo {author} {\bibfnamefont {K.}~\bibnamefont {Dajani}},\ and\ \bibinfo {author} {\bibfnamefont {C.}~\bibnamefont {Morais~Smith}},\ }\bibfield  {title} {\bibinfo {title} {Multifractal properties of tribonacci chains},\ }\href {https://doi.org/10.1103/PhysRevB.108.104204} {\bibfield  {journal} {\bibinfo  {journal} {Phys. Rev. B}\ }\textbf {\bibinfo {volume} {108}},\ \bibinfo {pages} {104204} (\bibinfo {year} {2023})}\BibitemShut {NoStop}%
\bibitem [{\citenamefont {Ashraff}\ \emph {et~al.}(1990)\citenamefont {Ashraff}, \citenamefont {Luck},\ and\ \citenamefont {Stinchcombe}}]{PhysRevB.41.4314}%
  \BibitemOpen
  \bibfield  {author} {\bibinfo {author} {\bibfnamefont {J.~A.}\ \bibnamefont {Ashraff}}, \bibinfo {author} {\bibfnamefont {J.-M.}\ \bibnamefont {Luck}},\ and\ \bibinfo {author} {\bibfnamefont {R.~B.}\ \bibnamefont {Stinchcombe}},\ }\bibfield  {title} {\bibinfo {title} {Dynamical properties of two-dimensional quasicrystals},\ }\href {https://doi.org/10.1103/PhysRevB.41.4314} {\bibfield  {journal} {\bibinfo  {journal} {Phys. Rev. B}\ }\textbf {\bibinfo {volume} {41}},\ \bibinfo {pages} {4314} (\bibinfo {year} {1990})}\BibitemShut {NoStop}%
\bibitem [{\citenamefont {Mandel}\ and\ \citenamefont {Lifshitz}(2006)}]{Even-DarMandel21022006}%
  \BibitemOpen
  \bibfield  {author} {\bibinfo {author} {\bibfnamefont {S.~E.-D.}\ \bibnamefont {Mandel}}\ and\ \bibinfo {author} {\bibfnamefont {R.}~\bibnamefont {Lifshitz}},\ }\bibfield  {title} {\bibinfo {title} {Electronic energy spectra and wave functions on the square fibonacci tiling},\ }\href {https://doi.org/10.1080/14786430500313846} {\bibfield  {journal} {\bibinfo  {journal} {Philosophical Magazine}\ }\textbf {\bibinfo {volume} {86}},\ \bibinfo {pages} {759} (\bibinfo {year} {2006})}\BibitemShut {NoStop}%
\bibitem [{\citenamefont {Fu}\ and\ \citenamefont {Liu}(1993)}]{Fu1993}%
  \BibitemOpen
  \bibfield  {author} {\bibinfo {author} {\bibfnamefont {X.}~\bibnamefont {Fu}}\ and\ \bibinfo {author} {\bibfnamefont {Y.}~\bibnamefont {Liu}},\ }\bibfield  {title} {\bibinfo {title} {Renormalization-group approach for the local density of states of two-dimensional fibonacci quasilattices},\ }\href {https://doi.org/10.1103/PhysRevB.47.3026} {\bibfield  {journal} {\bibinfo  {journal} {Phys. Rev. B}\ }\textbf {\bibinfo {volume} {47}},\ \bibinfo {pages} {3026} (\bibinfo {year} {1993})}\BibitemShut {NoStop}%
\bibitem [{\citenamefont {{Mosseri, R.}}\ and\ \citenamefont {{Sadoc, J.F.}}(1982)}]{refId0}%
  \BibitemOpen
  \bibfield  {author} {\bibinfo {author} {\bibnamefont {{Mosseri, R.}}}\ and\ \bibinfo {author} {\bibnamefont {{Sadoc, J.F.}}},\ }\bibfield  {title} {\bibinfo {title} {The bethe lattice : a regular tiling of the hyperbolic plane},\ }\href {https://doi.org/10.1051/jphyslet:01982004308024900} {\bibfield  {journal} {\bibinfo  {journal} {J. Physique Lett.}\ }\textbf {\bibinfo {volume} {43}},\ \bibinfo {pages} {249} (\bibinfo {year} {1982})}\BibitemShut {NoStop}%
\bibitem [{\citenamefont {S\"oderberg}(1993)}]{PhysRevE.47.4582}%
  \BibitemOpen
  \bibfield  {author} {\bibinfo {author} {\bibfnamefont {B.}~\bibnamefont {S\"oderberg}},\ }\bibfield  {title} {\bibinfo {title} {Bethe lattices in hyperbolic space},\ }\href {https://doi.org/10.1103/PhysRevE.47.4582} {\bibfield  {journal} {\bibinfo  {journal} {Phys. Rev. E}\ }\textbf {\bibinfo {volume} {47}},\ \bibinfo {pages} {4582} (\bibinfo {year} {1993})}\BibitemShut {NoStop}%
\end{thebibliography}%
\end{document}